\pdfoutput=1
\documentclass[cernpreprint, atlasdraft=false, texlive=2023, UKenglish, texmf, orcidlogo]{atlasdoc}
 
\usepackage[subfigure=true]{atlaspackage}
\usepackage{atlasbiblatex}
\usepackage{dcolumn,siunitx}
\usepackage{atlasphysics}
 
\graphicspath{{logos/}{figures/}}
\usepackage{graphbox}
\usepackage{multirow}
 
\usepackage{ANA-HDBS-2019-31-PAPER-defs}

\AtlasTitle{Search for Higgs boson pair production in association with a vector boson in $pp$ collisions at $\sqrt{s}=13\,\ensuremath{\text{TeV}}$ with the ATLAS detector}
 
\PreprintIdNumber{CERN-EP-2022-159}
\AtlasJournalRef{Eur. Phys. J. C 83 (2023) 519}
\AtlasDOI{10.1140/epjc/s10052-023-11559-y}
\AtlasAbstract{
This paper reports a search for Higgs boson pair ($hh$) production in association with a vector boson ($W\; {\textrm or}\; Z$) using 139\xspace\mbox{fb\(^{-1}\)} of proton--proton collision data at $\sqrt{s}=13\,\ensuremath{\text{TeV}}$ recorded with the ATLAS detector at the Large Hadron Collider. The search is performed in final states in which the vector boson decays leptonically ($W\to\ell\nu,\, Z\to\ell\ell,\nu\nu$ with $\ell=e, \mu$) and the Higgs bosons each decay into a pair of $b$-quarks. It targets $Vhh$ signals from both non-resonant $hh$ production, present in the Standard Model (SM), and resonant $hh$ production, as predicted in some SM extensions.
A 95\% confidence-level upper limit of 183~(87) times the SM cross-section is observed (expected) for non-resonant $Vhh$ production when assuming the kinematics are as expected in the SM.
Constraints are also placed on Higgs boson coupling modifiers.
For the resonant search, upper limits on the production cross-sections are derived for two specific models: one is the production of a vector boson along with a neutral heavy scalar resonance $H$, in the mass range 260--1000\,GeV, that decays into $hh$, and the other is the production of a heavier neutral pseudoscalar resonance $A$ that decays into a $Z$ boson and $H$ boson, where the $A$ boson mass is 360--800\,GeV and the $H$ boson mass is 260--400\,GeV. Constraints are also derived in the parameter space of two-Higgs-doublet models.
}
 
\author{The ATLAS Collaboration}
 
\AtlasRefCode{HDBS-2019-31}
 
\AtlasJournal{EPJC}

\hypersetup{pdftitle={ATLAS document},pdfauthor={The ATLAS Collaboration}}
 
\begin{document}
 
\maketitle
 
\tableofcontents

\newpage

% The next lines are included from the .//tex/introduction.tex input file
\section{Introduction}
\label{sec:intro}

Since the discovery of the Higgs boson ($h$) with a mass ($m_h$) of approximately $125\,\GeV$ in 2012~\cite{HIGG-2012-27,CMS-HIG-12-028}, its couplings to vector bosons and fermions have been found to be consistent with Standard Model~(SM) predictions within the current measurement precision~\cite{HIGG-2015-07,HIGG-2021-23,CMS-HIG-22-001}, providing strong evidence that the Higgs boson has SM properties. The SM also predicts the Higgs boson self-coupling (or trilinear coupling) as well as quartic couplings with itself and with massive vector bosons. These couplings are direct consequences of electroweak symmetry breaking~(EWSB)~\cite{Englert:1964et,Higgs:1964pj,Guralnik:1964eu} and are yet to be confirmed experimentally.

The Higgs boson self-coupling and quartic coupling to vector bosons can be probed through studies of Higgs boson pair ($hh$) production. In proton--proton ($pp$) collisions, SM production of $hh$ is dominated by the gluon--gluon fusion (ggF) process~\cite{Baglio:2012np,deFlorian:2016spz}.
Extensive searches for this process have led to significant bounds on the Higgs boson self-coupling~\cite{HDBS-2018-58,CMS-HIG-17-030,HDBS-2018-34,CMS-HIG-19-018}. Production through the vector-boson fusion~(VBF) process has the second largest cross-section~\cite{Baglio:2012np,deFlorian:2016spz}.
Searches for VBF $hh$ production have resulted in additional constraints on the Higgs boson self-coupling and are also sensitive to the SM prediction of the Higgs boson quartic coupling to vector bosons~\cite{HDBS-2018-18,CMS-HIG-20-005,CMS:2022nmn}. While produced through non-resonant processes in the SM, $hh$ production can also occur in resonant processes in scenarios beyond the Standard Model~(BSM) through the decays of heavy resonances, such as the heavy Higgs boson predicted in two-Higgs-doublet models~(2HDM)~\cite{Lee:1973iz,Branco:2011iw,Kling:2016opi} or the spin-2 Kaluza--Klein gravitons in Randall--Sundrum models~\cite{Randall:1999ee,Fitzpatrick:2007qr,Agashe:2007zd}. Searches for resonant $hh$ production in ggF and VBF processes have also led to constraints in the parameter spaces of these models~\cite{ATLAS:2022hwc,CMS-HIG-17-009,CMS:2021yci}.

This paper reports a search for Higgs boson pairs produced in association with a vector boson, $\Vhh~(V=W,Z)$, a process previously unexplored. The search targets both non-resonant $hh$ production, which occurs in the SM, and BSM-inspired resonant $hh$ production. It is performed on a dataset of $pp$ collisions at a centre-of-mass energy of $\sqrt{s}=13$~\TeV\ collected between 2015 and 2018 with the ATLAS detector at the Large Hadron Collider (LHC), corresponding to an integrated luminosity of $139\pm 2.4$\,\fbi~\cite{LUCID2}.
The search considers vector bosons decaying into leptons ($W\to\ell\nu$, $Z\to\ell\ell,\nu\nu$) and Higgs bosons decaying into a pair of $b$-quarks ($h\to bb$), leading to three distinct leptonic channels: $Zhh\to \vvbbbb$ (denoted by 0L), $Whh\to \lvbbbb$ (denoted by 1L), and $Zhh\to \llbbbb$ (denoted by 2L). Here $\ell$ denotes either an electron ($e$) or a muon ($\mu$).\footnote{Although $W\to\tau\nu$ and $Z\to\tau\tau$ are not considered explicitly, final states with electrons or muons from $\tau$ decays are included.}

\begin{figure}[!htb]
\begin{center}
\subfigure[]{\includegraphics[align=c,width=0.32\textwidth]{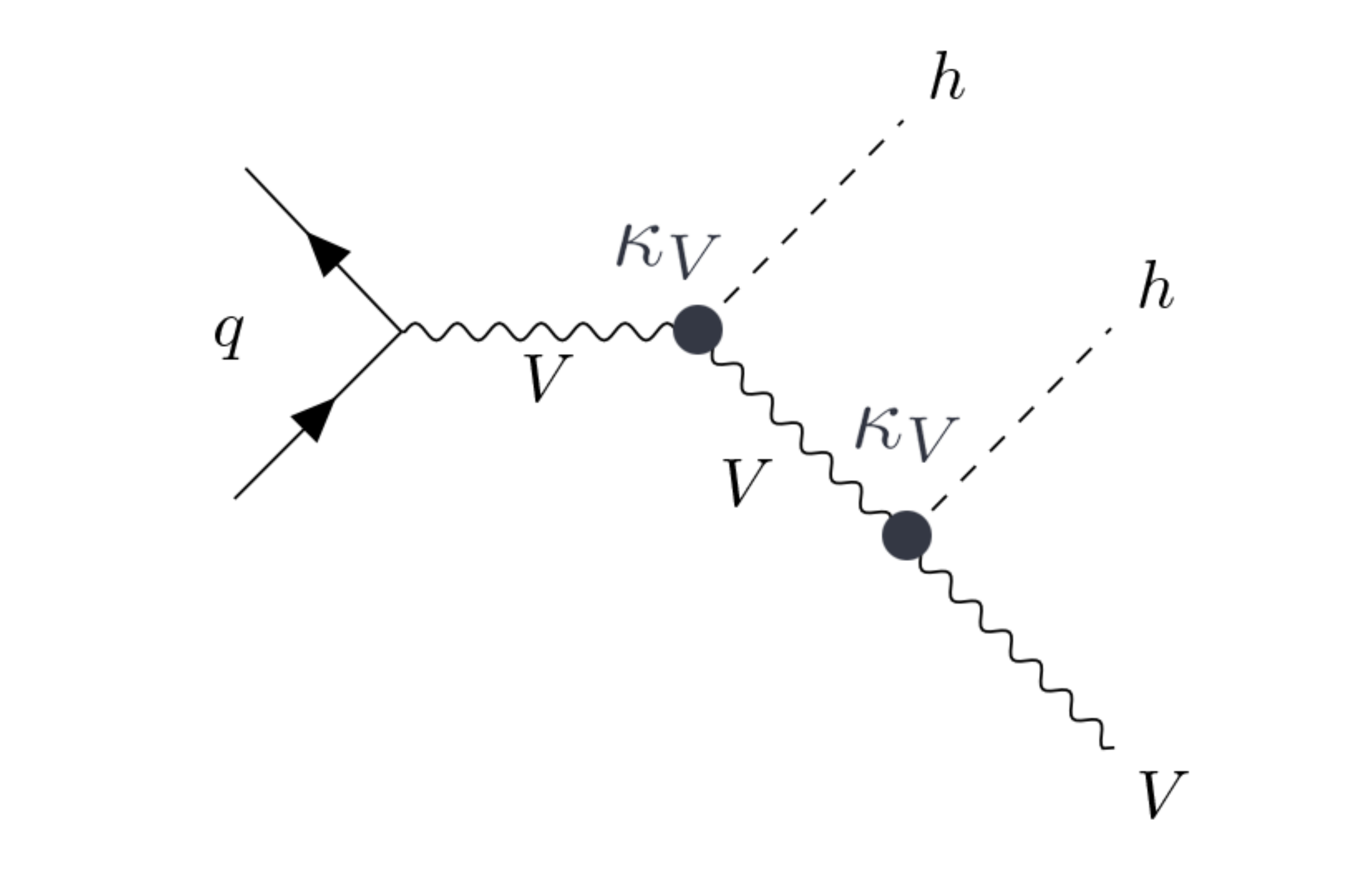}\label{fig:Diagrams_CV}}\hspace*{0.3cm}
\subfigure[]{\includegraphics[align=c,width=0.32\textwidth]{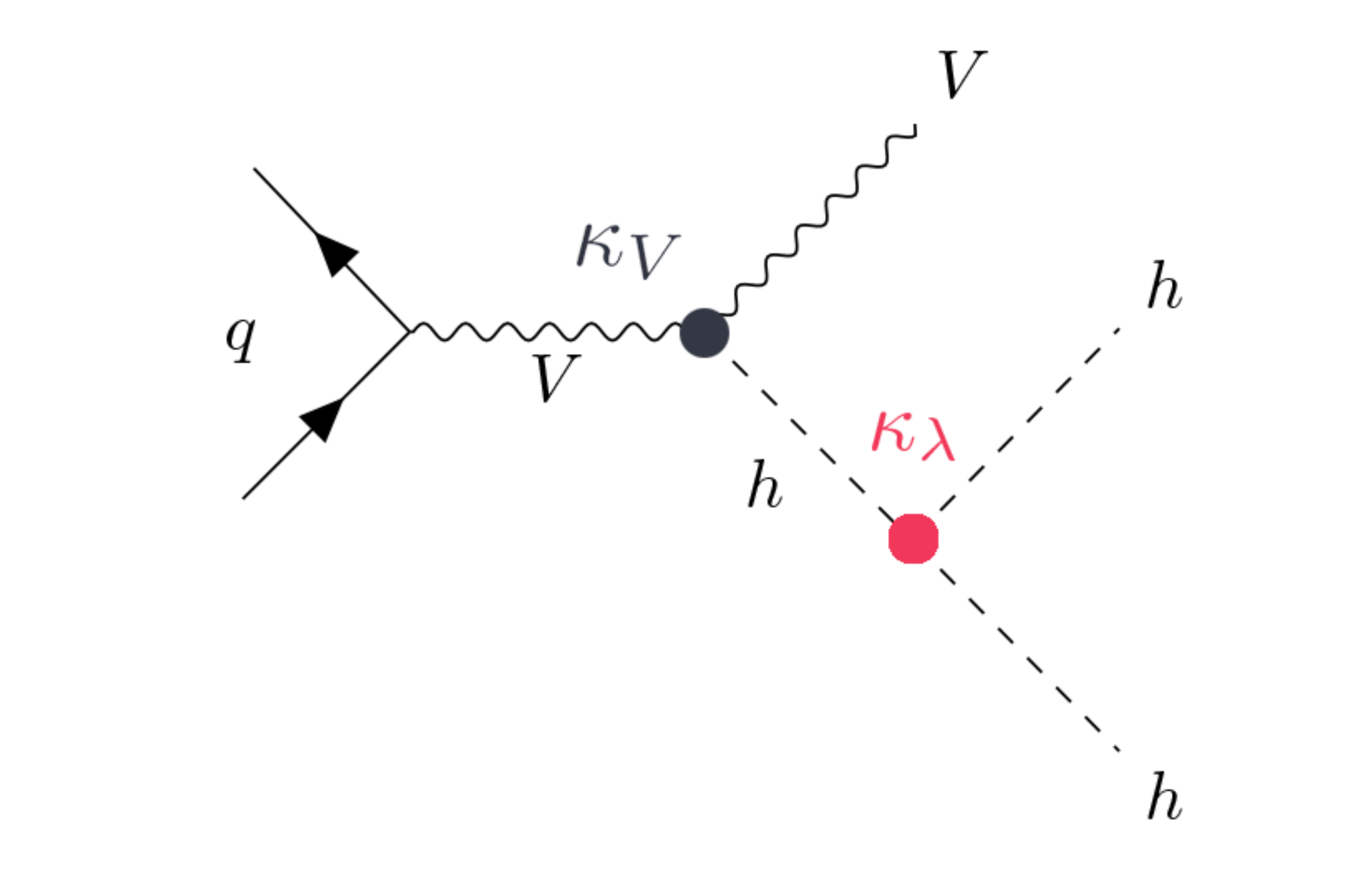}\label{fig:Diagrams_Chhh}}\hspace*{0.3cm}
\subfigure[]{\includegraphics[align=c,width=0.32\textwidth]{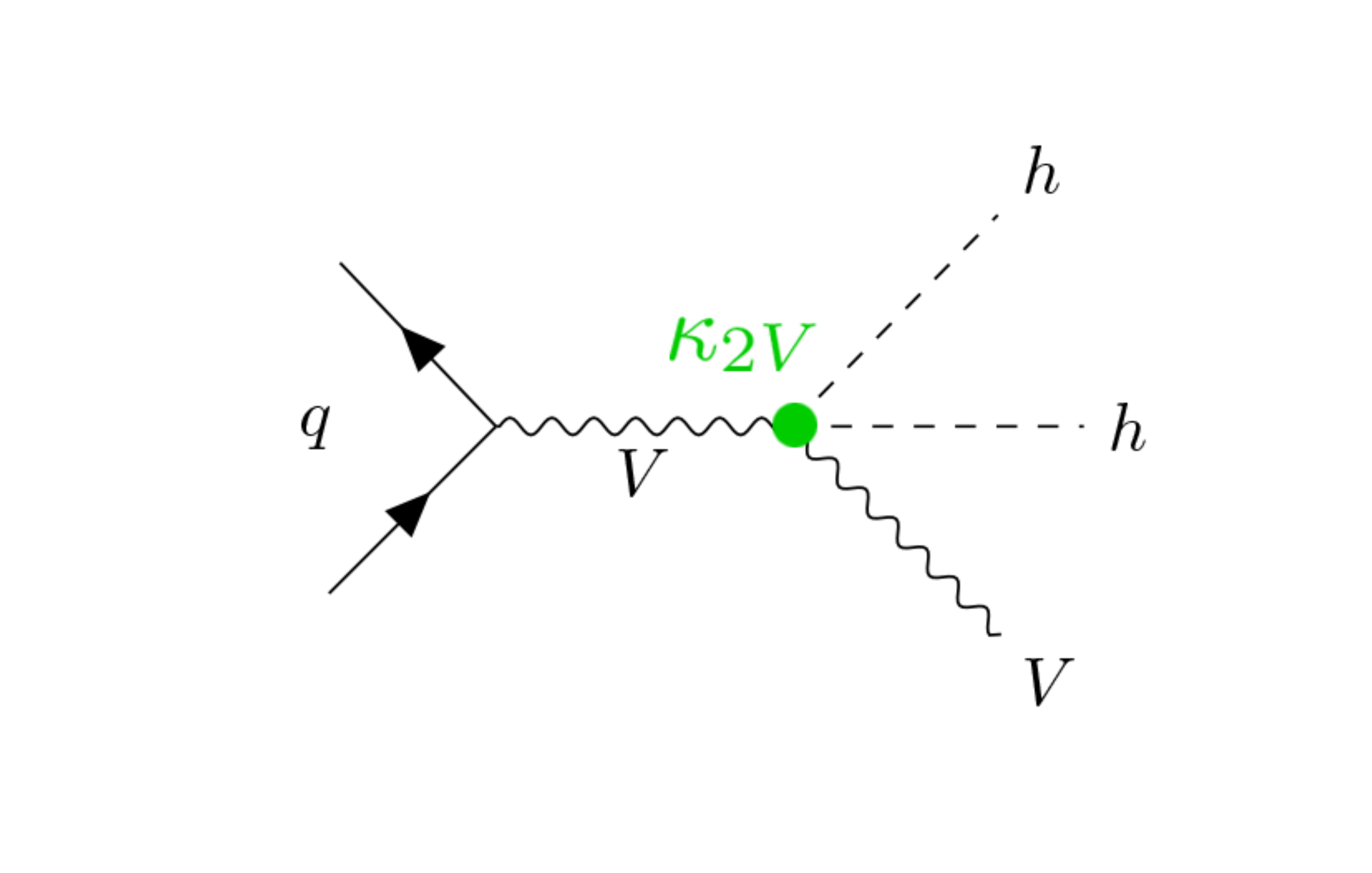}\label{fig:Diagrams_ChhVV}}
\end{center}
\caption{Leading-order Feynman diagrams of non-resonant $hh$ production in association with a vector boson $V$ expected in the SM from (a) Higgs boson coupling to vector bosons, (b) Higgs boson self-coupling, and (c) Higgs boson quartic coupling to vector bosons. The coupling modifiers \CV, \Chhh, and \ChhVV are discussed in Section~\ref{sec:samples}.}
\label{fig:NonResDiagrams}
\end{figure}

Non-resonant $hh$ production in association with a $V$ boson arises in the SM from three distinct Higgs boson couplings: coupling to vector bosons, self-coupling, and quartic coupling to vector bosons.
The leading-order processes are depicted in Figures~\ref{fig:Diagrams_CV}, \ref{fig:Diagrams_Chhh}, and \ref{fig:Diagrams_ChhVV}, respectively. In the SM, the cross-sections of these processes at the LHC are small compared with those of the ggF and VBF processes, $0.50\pm 0.01$~fb for \Whh ($W^+hh$: $0.329\pm 0.007$~fb, and $W^-hh$: $0.173\pm 0.005$~fb) and $0.36\pm 0.01$~fb for \Zhh at $\sqrt{s}=13\,\TeV$ with $m_h=125\,\GeV$~\cite{Baglio:2012np,deFlorian:2016spz}, computed at next-to-next-to-leading-order~(NNLO) accuracy in QCD.
 
\FloatBarrier
 
\begin{figure}[!htb]
\begin{center}
\subfigure[]{\includegraphics[align=c,width=0.4\textwidth]{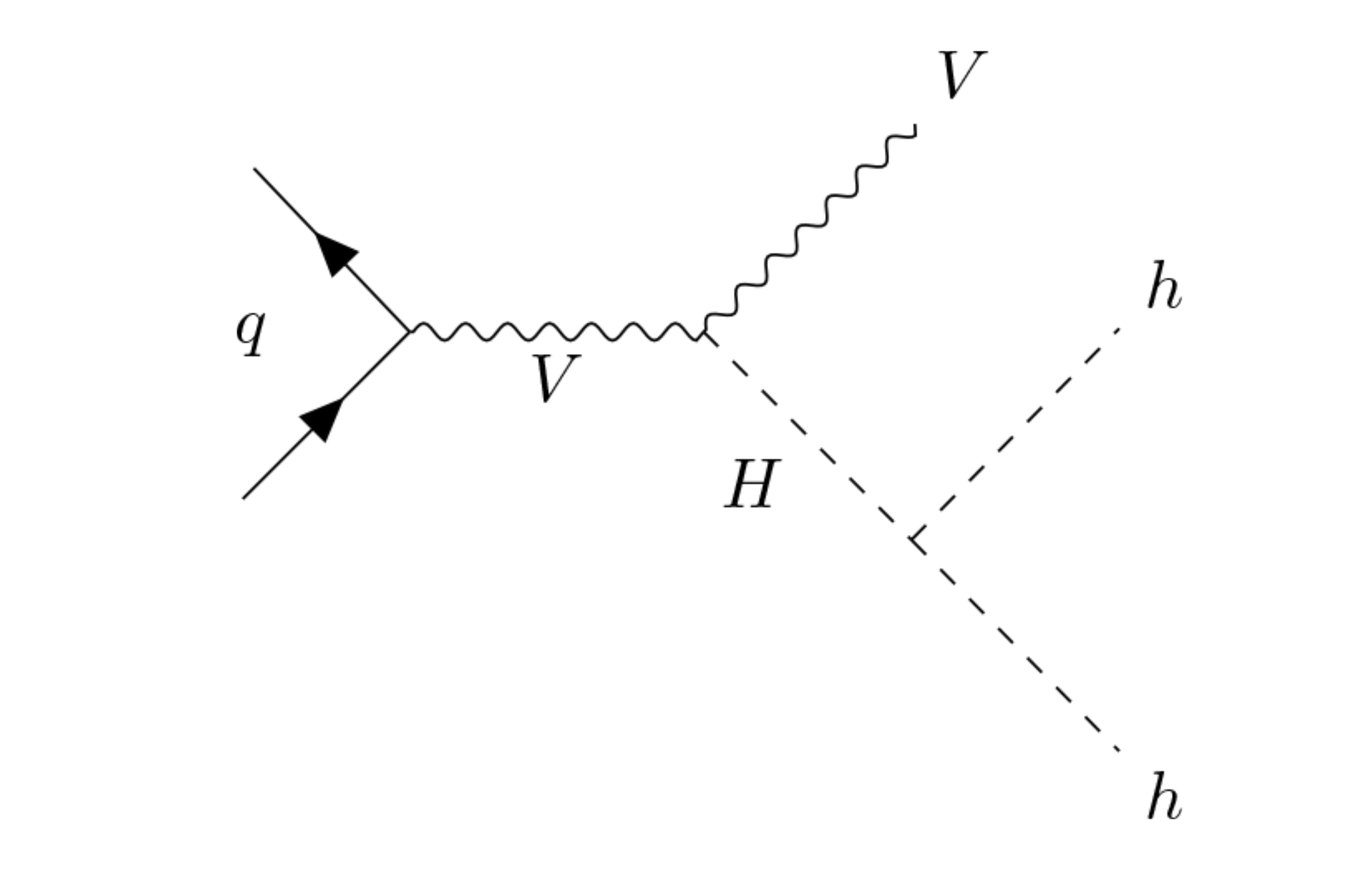}\label{fig:Diagrams_VH}} \hspace*{2.0cm}
\subfigure[]{\includegraphics[align=c,width=0.4\textwidth]{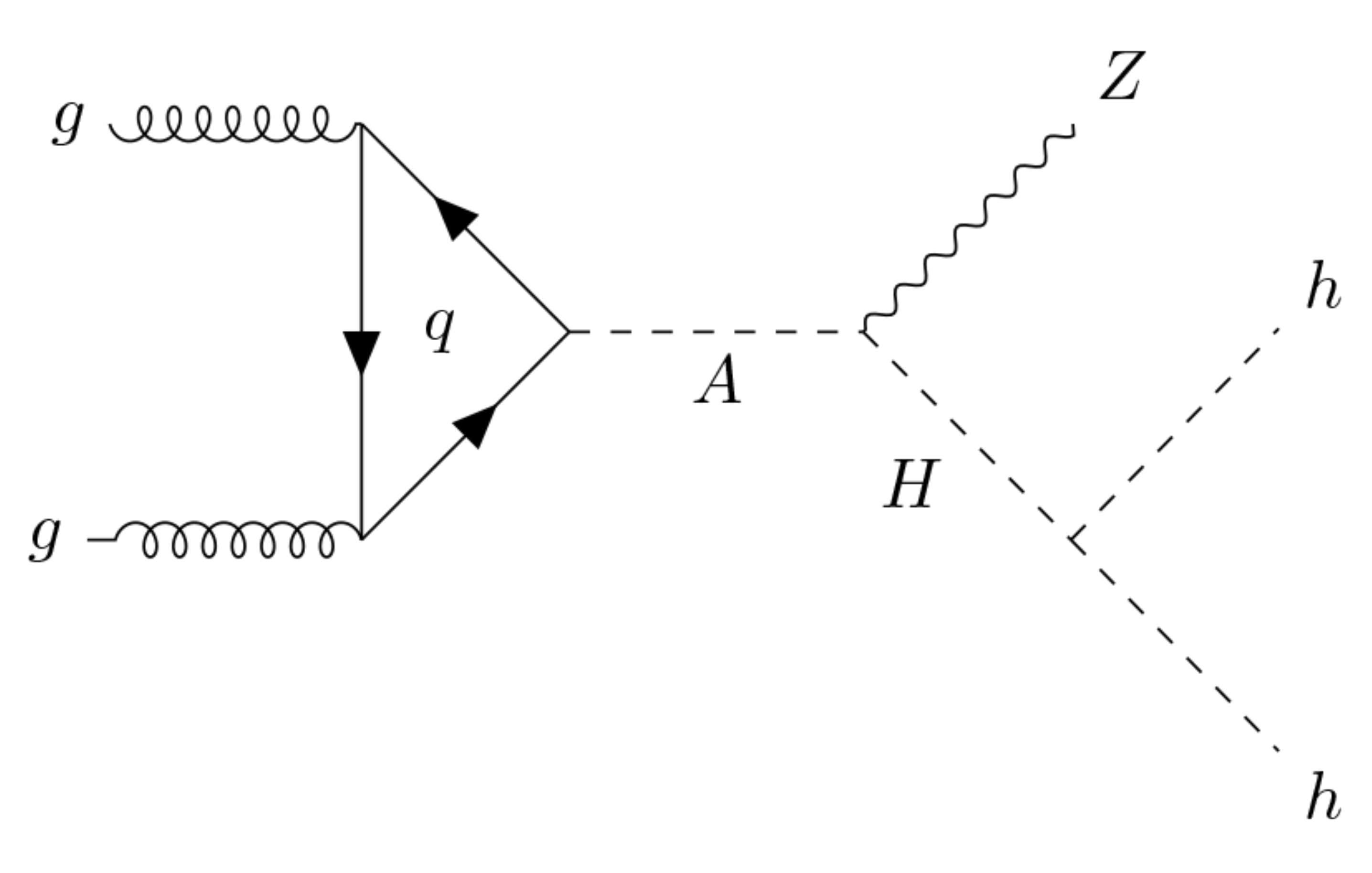}\label{fig:Diagrams_AZH}}
\end{center}
\caption{Leading-order Feynman diagrams of resonant $hh$ production in association with a vector boson $V$ predicted in some BSM scenarios from the decay of a heavy scalar $H$ originating from (a) an off-shell vector boson and (b) the decay of a neutral heavy pseudoscalar $A$.}
\end{figure}

Two BSM scenarios are considered for resonant $hh$ production as illustrated in Figures~\ref{fig:Diagrams_VH} and~\ref{fig:Diagrams_AZH}. The first scenario, labelled as \VH, is the `Higgstrahlung' production of a generic neutral CP-even scalar $H$ boson which couples directly to vector bosons and decays into $hh$, i.e.\ $VH\to\Vhh$. Examples of such a scalar resonance are the CP-even heavy Higgs boson predicted in the electroweak singlet model~\cite{Muhlleitner:2017dkd} or in the type-II 2HDM~\cite{Coleppa:2013dya}.  In this search, the scalar $H$ is assumed to be a narrow resonance; i.e.\ its natural width is much smaller than the expected experimental relative mass resolution of approximately 3\%. This scenario was explored previously by ATLAS in a VBF $hh$ search~\cite{HDBS-2018-18}. The $VH$ search is complementary because it is sensitive to the $HWW$ and $HZZ$  couplings separately, while the VBF search is sensitive only to their combination.
The second scenario, labelled as \AZH, is a specific process in the 2HDM which predicts three neutral Higgs bosons: two CP-even scalars $h$ and $H$ (with mass hierarchy $m_H>m_h$), and one CP-odd scalar $A$. In parts of the 2HDM parameter space where the light Higgs boson $h$ is similar to the SM Higgs boson and has a mass $m_h\sim 125\,\GeV$ favourable for electroweak baryogenesis, the $A$ boson has a mass below about 800~\GeV\ but is heavier than the $H$ boson~\cite{Dorsch:2014qja}. If $m_H$ is in the range $2m_h <m_H<2m_t$, the $H\to hh$ decay branching ratio could be substantial, leading to a sizeable rate for $gg\to A\to ZH\to Zhh$. Here $m_t$ is the mass of the top quark. For this search, natural widths up to 20\% of its mass are considered for $A$, and a narrow width is assumed for $H$.
Searches for $A\to ZH$ were performed previously by ATLAS and CMS in the $H\to bb,\tau\tau$, and $WW$ decay channels~\cite{CMS-HIG-15-001,EXOT-2016-34,CMS-HIG-18-012,HDBS-2018-13}.

% End of text imported from the .//tex/introduction.tex input file

% The next lines are included from the .//tex/detector.tex input file
\section{ATLAS detector}
 
\newcommand{\AtlasCoordFootnote}{
ATLAS uses a right-handed coordinate system with its origin at the nominal interaction point (IP) in the centre of the detector and the \(z\)-axis along the beam pipe. The \(x\)-axis points from the IP to the centre of the LHC ring, and the \(y\)-axis points upwards. Cylindrical coordinates \((r,\phi)\) are used in the transverse plane, \(\phi\) being the azimuthal angle around the \(z\)-axis. The pseudorapidity is defined in terms of the polar angle \(\theta\) as \(\eta = -\ln \tan(\theta/2)\). Angular distance is measured in units of \(\Delta R \equiv \sqrt{(\Delta\eta)^{2} + (\Delta\phi)^{2}}\).}
 
The ATLAS detector~\cite{PERF-2007-01} at the LHC covers nearly the entire solid angle around the collision point.\footnote{\AtlasCoordFootnote} It consists of an inner tracking detector surrounded by a thin superconducting solenoid, electromagnetic and hadron calorimeters, and a muon spectrometer incorporating three large superconducting air-core toroidal magnets.
 
The inner-detector system (ID) is immersed in a \SI{2}{\tesla} axial magnetic field and provides charged-particle tracking in the range \(|\eta| < 2.5\). The high-granularity silicon pixel detector covers the vertex region and typically provides four measurements per track, the first hit normally being in the insertable B-layer (IBL) installed before Run~2~\cite{ATLAS-TDR-19,PIX-2018-001}. It is followed by the silicon microstrip tracker (SCT), which usually provides eight measurements per track. These silicon detectors are complemented by the transition radiation tracker (TRT), which enables radially extended track reconstruction up to \(|\eta| = 2.0\). The TRT also provides electron identification information based on the fraction of hits (typically 30 in total) above a higher energy-deposit threshold corresponding to transition radiation.

The calorimeter system covers the pseudorapidity range \(|\eta| < 4.9\). Within the region \(|\eta|< 3.2\), the electromagnetic calorimeter (ECAL) consists of barrel and endcap high-granularity lead/liquid-argon (LAr) calorimeters, with an additional thin LAr presampler covering \(|\eta| < 1.8\) to correct for energy loss in material upstream of the calorimeters. Hadron calorimetry is provided by the steel/scintillator-tile calorimeter,  segmented into three barrel structures within \(|\eta| < 1.7\), and two copper/LAr hadron endcap calorimeters. The solid angle coverage is completed with forward copper/LAr and tungsten/LAr calorimeter modules optimised for electromagnetic and hadronic energy measurements respectively.

The muon spectrometer (MS) comprises separate trigger and high-precision tracking chambers measuring the deflection of muons in a magnetic field generated by the superconducting air-core toroidal magnets. The field integral of the toroids ranges between \num{2.0} and \SI{6.0}{\tesla\metre} across most of the detector. Three layers of precision chambers, each consisting of layers of monitored drift tubes, covers the region \(|\eta| < 2.7\), complemented by cathode-strip chambers in the forward region, where the background is highest. The muon trigger system covers the range \(|\eta| < 2.4\) with resistive-plate chambers in the barrel, and thin-gap chambers in the endcap regions.

Interesting events are selected by the first-level trigger system implemented in custom hardware, followed by selections made by algorithms implemented in software in the high-level trigger~\cite{TRIG-2016-01}.  The first-level trigger accepts events from the \SI{40}{\MHz} bunch crossings at a rate below \SI{100}{\kHz}, which the high-level trigger further reduces in order to record events to disk at about \SI{1}{\kHz}.

An extensive software suite~\cite{ATL-SOFT-PUB-2021-001} is used in data simulation, in the reconstruction and analysis of real and simulated data, in detector operations, and in the trigger and data acquisition systems of the experiment.

% End of text imported from the .//tex/detector.tex input file

% The next lines are included from the .//tex/samples.tex input file
\section{Data and Monte Carlo samples}
\label{sec:samples}

The data used in this analysis were collected using unprescaled single-lepton, missing transverse momentum (\met), or single-photon triggers.
The single-lepton trigger requirements are applied as a logical OR of single-electron or single-muon triggers~\cite{TRIG-2018-05,TRIG-2018-01}, with transverse momentum (\pT) thresholds that started at 20~\GeV\ or 24~\GeV\ in 2015 for muons or electrons, respectively, and increased to 26~\GeV\ in 2016--2018.
The \met trigger~\cite{TRIG-2019-01} threshold was raised from 70~\GeV\ to 110~\GeV\ between the 2015 and 2018 data-taking periods.
The single-photon trigger~\cite{TRIG-2018-05} was only used in this analysis for background estimation and had a threshold of 140~\GeV\ for the full data-taking period.
Events are selected for analysis only if they are of good quality and if all the relevant
detector components are known to have been in good operating condition~\cite{DAPR-2018-01}, which corresponds to a total integrated luminosity of $139.0$\,\ifb.
The uncertainty in the combined 2015--2018 integrated luminosity is 1.7\%~\cite{ATLAS-CONF-2019-021}, obtained using the LUCID-2 detector~\cite{LUCID2} for the primary luminosity measurements.
The recorded events contain an average of 34 inelastic $pp$ collisions per bunch-crossing.

Monte Carlo (MC) simulations are used to optimise the search sensitivities and to estimate background contributions.
MC samples were produced with various event generators, interfaced to different programs for parton showers, hadronisation, and underlying-event simulations. No potential interference between signal and background processes is considered.
Table~\ref{tab:mc} summarises the simulation of the signal and background processes relevant for the searches described in this paper.
All MC samples are passed through the ATLAS detector simulation program~\cite{SOFT-2010-01} based on \GEANT~\cite{Agostinelli:2002hh}.
Simulated processes are normalised to the most accurate theoretical cross-section predictions currently available.
The effects of multiple interactions in the same or nearby bunch crossings (pile-up) were modelled by overlaying minimum-bias events, simulated using the soft QCD processes of \Pythia[8.186]~\cite{Pythia8} with the A3~\cite{ATL-PHYS-PUB-2016-017} set of tuned parameters and the \NNPDF[2.3lo]~\cite{Ball:2012cx} PDF. For all samples of simulated events, except for those generated using \Sherpa~\cite{sherpa}, the \EVTGEN[1.6.0] program~\cite{Lange:2001uf} was used to describe the decays of bottom and charm hadrons.

\begin{table}[!htb]
\begin{center}
\caption{List of MC event generators, PDFs, and parton shower, hadronisation and underlying event (UE) models used to simulate signal and background processes. Different versions are the results of matching different calculations and models at the time the samples were produced.
Here $\Vl$ and $\Vh$, respectively, denote $V$ decaying leptonically or hadronically.
The last column shows the calculation orders of cross-sections used.
Cross-section orders in the last column can be leading order (LO), next-to-leading order (NLO), or next-to-next-to-leading order (NNLO) in terms of QCD or electroweak (EW) accuracy. The mass of the Higgs boson $h$ is set to $125\,\GeV$ in the simulation.}
\label{tab:mc}\vspace*{-0.2cm}{\scriptsize
\hspace*{-1.1cm}\begin{tabular}{lllllll} \toprule
Process             &  \multicolumn{3}{c}{Matrix element}    &  Parton   &  UE   & Cross-section      \\  \cline{2-4}
&  generator &  order & PDF  &  shower   &  model &  order             \\
\midrule
\multicolumn{7}{l}{Signal samples} \\
\midrule
Non-resonant \Vhh   & \MGNLO~\cite{Alwall:2014hca}  &  LO  &  \NNPDF[2.3lo]~\cite{Ball:2012cx} &  \Pythia[8.244]~\cite{Sjostrand:2014zea}  &  A14~\cite{ATL-PHYS-PUB-2014-021}  &  NNLO~\cite{Baglio:2012np}  \\
Resonant \VH        & \MGNLO                        &  LO  &  \NNPDF[2.3lo]      &  \Pythia[8.244]  &  A14      &  LO   \\
Resonant \AZH       & \MGNLO                        &  LO  &  \NNPDF[2.3lo]      &  \Pythia[8.244]  &  A14      &  LO  \\[1mm]
\midrule
\multicolumn{7}{l}{Top quark processes} \\
\midrule
\ttbar              &  \POWHEGBOX[v2]~\cite{Frixione:2007nw}  &  NLO & \NNPDF[3.0nlo]~\cite{Ball:2014uwa}  &  \Pythia[8.230]  &  A14      & NNLO  \\
&                                         &      &                                  &                 &           & +NNLL~\cite{Czakon:2011xx}            \\
Single-$t$: s-channel &  \POWHEGBOX[v2]~\cite{Alioli:2009je}    &  NLO & \NNPDF[3.0nlo]  &  \Pythia[8.230]  &  A14      & NLO~\cite{Kidonakis:2010tc}    \\
Single-$t$: t-channel &  \POWHEGBOX[v2]~\cite{Alioli:2009je}    &  NLO & \NNPDF[3.0nlo]  &  \Pythia[8.230]  &  A14      & NLO~\cite{Kidonakis:2011wy}    \\
Single-$t$: $Wt$      &  \POWHEGBOX[v2]~\cite{Re:2010bp}        &  NLO & \NNPDF[3.0nlo]  &  \Pythia[8.230]  &  A14      & NNLO~\cite{Kidonakis:2010ux}    \\[2mm]
$ttH$               &  \POWHEGBOX[v2]                         &  NLO & \NNPDF[3.0nlo]      &  \Pythia[8.230]  &  A14      & NLO  \\
&                                         &      &              &                 &           & (QCD+EW)~\cite{deFlorian:2016spz}        \\
$ttV$      &  \MGNLO  &  NLO  &  \NNPDF[3.0nlo]   &  \Pythia[8.230]  &  A14      & NLO         \\
&             &       &            &                 &           & (QCD+EW)    \\
$tt\ell\ell^{\S}$        &  \MGNLO  &  NLO  &  \NNPDF[3.0nlo]   &  \Pythia[8.230]  &  A14      & NLO       \\
$tttt$              &  \MGNLO  &  NLO  &  \NNPDF[3.1lo]   &  \Pythia[8.230]  &  A14      & NLO        \\
&             &       &            &                 &           & (QCD+EW)              \\
$ttt$               &  \MGNLO  &   LO  &  \NNPDF[2.3lo]   &  \Pythia[8.230]  &  A14      & LO    \\
$ttWW$              &  \MGNLO  &   LO  &  \NNPDF[2.3lo]   &  \Pythia[8.230]  &  A14      & LO    \\
$tWZ$               &  \MGNLO  &  NLO  &  \NNPDF[2.3lo]   &  \Pythia[8.212]  &  A14      & NLO   \\
$tZ$                &  \MGNLO  &   LO  &  \NNPDF[2.3lo]   &  \Pythia[8.212]  &  A14      & LO    \\[1mm]
\midrule
\multicolumn{7}{l}{Single- and multi-boson production} \\
\midrule
\Vjets              &  \Sherpa[2.2.1]~\cite{Bothmann:2019yzt,Gleisberg:2008fv,Cascioli:2011va} & NLO$\dagger$ &  \NNPDF[3.0nnlo]~\cite{Ball:2014uwa}  &  \Sherpa 2.2.1~\cite{Schumann:2007mg,Hoeche:2012yf}  &  Default  & NNLO~\cite{Catani:2009sm} \\
\Gjets              &  \Sherpa[2.2.2]  &  NLO$\dagger$  & \NNPDF[3.0nnlo]  &  \Sherpa[2.2.2]  &  Default   & NNLO  \\[2mm]
$qq\to \Vl\Vlh$          &  \Sherpa[2.2.1]  &  NLO           & \NNPDF[3.0nnlo]  &  \Sherpa 2.2.1  &  Default   & NLO$^{\ddagger}$   \\
$gg\to \Vl\Vlh$          &  \Sherpa[2.2.2]  &  NLO           & \NNPDF[3.0nnlo]  &  \Sherpa 2.2.2  &  Default   & NLO$^{\ddagger}$   \\
$\Vl\Vl V $         &  \Sherpa[2.2.2]  &  NLO           & \NNPDF[3.0nnlo]  &  \Sherpa 2.2.2  &  Default   & LO   \\
$\Vl\Vh\Vh$         &  \MGNLO      &   LO           &  \NNPDF[2.3lo]      &  \Pythia 8.243  &  A14      & LO  \\[2mm]
$qq\to VH$          &  \POWHEGBOX[v2]~\cite{Alioli:2010xd} & NLO    &  \NNPDF[3.0nlo]   & \Pythia[8.212]   &  AZNLO~\cite{STDM-2012-23}    & NNLO(QCD)            \\
&                                     &        &               &                 &                               & +NLO(EW)             \\
$gg\to ZH$          &  \POWHEGBOX[v2]  & NLO           & \NNPDF[3.0nlo]   & \Pythia[8.212]   &  AZNLO    & NLO+NLL               \\[1mm]
\toprule
\multicolumn{7}{l}{$\S$ Here, the same-flavour $\ell\ell$ pairs are from $Z^*/\gamma^*\to\ell\ell$.} \\
\multicolumn{7}{l}{$\dagger$ Produced with up to two extra jets at NLO and up to four extra jets at LO.} \\
\multicolumn{7}{l}{$\ddagger$ For the diboson samples the cross-sections are calculated by the Monte Carlo generator at NLO accuracy in QCD.} \\
\end{tabular}}
\end{center}
\end{table}

\subsection{Non-resonant signal samples}
 
As shown in Figure~\ref{fig:NonResDiagrams}, three types of the Higgs boson couplings are responsible for non-resonant $hh$ production in the SM. Although these couplings are predicted in the SM once the Higgs boson mass $m_h$ is known, their values should be determined experimentally. Deviations from their SM values are traditionally parameterised using the coupling  modifiers~\cite{LHCHiggsCrossSectionWorkingGroup:2013rie} denoted, in this paper, by \CV, \Chhh, and \ChhVV for the $hVV$, $hhh$, and $hhVV$ vertices, respectively, assuming the same coupling modifiers for the $W$ and $Z$ bosons.\footnote{In the SM, $\CV=1,\, \Chhh=1$, and $\ChhVV=1$.}
 
While it is impractical to produce signal samples for arbitrary values of the coupling modifiers probed by this analysis, such samples can be constructed from six independent samples composed of different combinations of the leading-order (LO) processes shown in Figure~\ref{fig:NonResDiagrams}.
These component samples were produced with $\CV=\Chhh=\ChhVV=1$  for the following couplings and their combinations:
\begin{itemize}
\setlength\itemsep{-0.5em}
\item[(1)] \CV: diagram without the $hhh$ and $hhVV$ vertices, Figure~\ref{fig:Diagrams_CV};
\item[(2)] \Chhh: diagram with the $hhh$ vertex, Figure~\ref{fig:Diagrams_Chhh};
\item[(3)] \ChhVV: diagram with the $hhVV$ vertex, Figure~\ref{fig:Diagrams_ChhVV};
\item[(4)] $\CV+\Chhh$:  all diagrams except those with an $hhVV$ vertex, Figures~\ref{fig:Diagrams_CV} and~\ref{fig:Diagrams_Chhh};
\item[(5)] $\CV+\ChhVV$: all diagrams except those with an $hhh$ vertex, Figures~\ref{fig:Diagrams_CV} and~\ref{fig:Diagrams_ChhVV};
\item[(6)] $\Chhh+\ChhVV$: all diagrams with either an $hhh$ or an $hhVV$ vertex, Figures~\ref{fig:Diagrams_Chhh} and~\ref{fig:Diagrams_ChhVV}.
\end{itemize}
 
Samples 1--3 determine the contributions from their respective diagrams, while samples 4--6 allow the determination of the contributions from interference between the three diagrams.
Non-resonant signal samples for any coupling modifier deviations from the SM values can be built from the combinations of these component samples, weighted by the coupling modifiers.
A SM \Vhh sample with $\CV=\Chhh=\ChhVV=1$ was also produced to validate this procedure. Kinematic distributions of the reweighted sample are found to agree with those of the directly produced sample.
Since the $hVV$ coupling has been constrained through the single Higgs boson measurements~\cite{HIGG-2021-23}, $\CV=1$ is assumed in this paper.

In addition to the quark-initiated diagrams shown in Figure~\ref{fig:NonResDiagrams}, \Zhh can also be produced via the gluon-initiated  $gg\to Zhh$ diagram. Although $gg\to Zhh$ is technically a higher-order process for \Zhh production, its cross-section is predicted to be approximately 24\% of that of the $qq\to Zhh$ process in the SM~\cite{deFlorian:2016spz}.
A correction is applied to the non-resonant $qq\to Zhh$ cross-section, as a function of \Chhh and \ChhVV, to account for the $gg\to Zhh$ contribution.
The generator-level difference, in both normalisation and shape, between the $qq\to Zhh$ sample and the sum of the $qq\to Zhh$ and $gg\to Zhh$ samples is taken as an uncertainty.
For a signal as expected in the SM, the normalisation component of this uncertainty has a magnitude of 24\%.
It tends to be larger than 24\% for large \ChhVV and smaller than 24\% for large \Chhh.

\subsection{Resonant signal samples}

The production of a generic CP-even scalar resonance $H$ in association with a $V$ boson, where the $H$ decays into a pair of the $h$ Higgs bosons, is modelled using the $qq\to VH\to Vhh$ process in the 2HDM as shown in Figure~\ref{fig:Diagrams_VH}. The heavy Higgs boson is assumed to have a narrow width, i.e.\ its total decay width is far smaller than the experimental \mhh resolution of approximately 3\%, but to decay promptly nevertheless. Ten signal samples were produced for each leptonic channel, corresponding to $m_H$ values of 260, 280, 300, 400, 500, 600, 700, 800, 900, and 1000~\GeV. The lower bound is dictated by $2m_h$, while the upper bound is limited by the ability to reconstruct two separate jets from the highly boosted $h\to bb$ decay.
 
The $gg\to A\to ZH\to Zhh$ signal samples were produced for mass combinations of  $m_A=360,\, 400,\, 500,\, 600,\, 700,\, \mathrm{and}~800\,\GeV$ and $m_H=260,\, 300,\, \mathrm{and}~400\,\GeV$ subject to the kinematic bound of $m_A>m_H+m_Z$, leading to a total of 15 $(m_A,m_H)$ grid points for each of the 0L and 2L channels.
These mass combinations cover the 2HDM parameter space relevant for this search, but unexplored by previous $gg\to A\to ZH$ searches~\cite{CMS-HIG-15-001,EXOT-2016-34,CMS-HIG-18-012,HDBS-2018-13}. The CP-odd $A$ boson could have a substantial total decay width, depending on the 2HDM parameter values.
Two sets of samples were produced, one for a narrow-width (NW) $A$ boson  and the other for a large-width~(LW) $A$ boson.
In both scenarios, the total decay width of the heavy CP-even $H$ boson is assumed to be narrow.
For the LW samples, the $A$ boson width is assumed to be 20\% of its mass.
To perform searches for $A$ bosons with different widths, MC distributions for an $A$ boson with width equal to 5\%, 10\%, or 15\% of its mass are derived through reweighting. The generator-level $A$ boson mass distribution for each of these intermediate widths is compared with that of the LW $A$ boson to derive reweighting factors which are then applied to the distributions of the simulated LW MC samples to obtain the corresponding distributions for the intermediate widths.

% End of text imported from the .//tex/samples.tex input file

% The next lines are included from the .//tex/objects.tex input file
\section{Object reconstruction and identification}
\label{sec:objects}
 
As discussed in Section~\ref{sec:intro}, the signal events are characterised by the products of the targeted decays of the vector and Higgs bosons:  electrons or muons, jets ($b$-jets in particular), and missing transverse momentum. Additionally, events with identified hadronically decaying $\tau$-leptons are vetoed to reduce backgrounds. Events with photons are used for background estimation. The reconstruction and identification of these physics objects are described in this section.

Electrons are reconstructed by matching topological energy clusters in the ECAL with tracks in the ID~\cite{EGAM-2018-01} and are required to have $\pt>7\,\GeV$ within the pseudorapidity range $|\eta|<2.47$.
They are identified using likelihood-based identification criteria that combined the requirements of calorimeter shower shape, track-to-cluster matching, and associated track qualities. Electron candidates are required to satisfy the \tight criterion in the 1L channel and the \loose criterion for the rest.  All candidates must satisfy a loose track- and calorimeter-based isolation criterion to minimise the number of jets misidentified as electrons.

Similarly to electrons, photon reconstruction starts with topological energy clusters in the ECAL~\cite{EGAM-2018-01}, but the clusters are required to have either no matching tracks (unconverted photons) or one or two matching tracks consistent with a conversion vertex (converted photons).  Photon candidates are required to have $\pt>10\,\GeV$ within the pseudorapidity range $|\eta|<1.37$  or $1.52<|\eta|<2.37$ and satisfy the \tight  identification criterion as well as a calorimeter-based isolation requirement called `{\em TightCaloOnly}'~\cite{EGAM-2018-01} to suppress backgrounds.

Muons are reconstructed by matching tracks in the ID to either full tracks or track segments in the MS or, for $|\eta|<0.1$ only, to energy deposits in the calorimeter~\cite{MUON-2018-03}.\footnote{Designed to compensate for the low muon reconstruction efficiency due to a gap in the MS coverage.} They must have $\pt>7\,\GeV$ and be within $|\eta|<2.5$, the combined acceptance of the ID and MS.  Muons must satisfy the \medium quality criterion in the 1L channel and the \loose criterion in others. The same isolation requirement used for the electron selection is applied to all muon candidates to reduce the rate of muons from heavy-flavour decays.

Electrons (muons) are required to have associated tracks satisfying $|d_0/\sigma_{d_0}|<5\,(3)$ and $|z_0\sin\theta|<0.5\,\textrm{mm}$, where $d_0$ is the transverse impact parameter relative to the beam line, $\sigma_{d_0}$ is its uncertainty, and $z_0$ is the distance between the longitudinal position of the track along the beam line at the point where $d_0$ is measured and the longitudinal position of the primary $pp$ collision vertex.
 
Jets are reconstructed from particle-flow objects using the anti-$k_t$ algorithm~\cite{Cacciari:2008gp,Fastjet} with a radius parameter of $R=0.4$ and are calibrated as described in Ref.~\cite{PERF-2016-04}. They are required to have $\pt>20\,\GeV$ and $|\eta|<4.5$. Jets with $\pt<60$\,\GeV\ and $|\eta|<2.4$ originating from pile-up are suppressed with the jet-vertex-tagger~\cite{PERF-2014-03}, a likelihood discriminant based on matching enough of the jet's tracks to the primary vertex.
 
Jets containing $b$-hadrons ($b$-jets) are identified with the DL1r algorithm~\cite{FTAG-2018-01}. The algorithm is based on information such as properties of displaced tracks and reconstructed secondary vertices in the jet. A jet is $b$-tagged if the DL1r value is above a preset threshold. Four thresholds, referred to as working points~(WP), are defined with average efficiencies of 60\%, 70\%, 77\%, and 85\% for tagging $b$-jets from simulated $t\bar{t}$ events. A pseudo-continuous $b$-tagging score is defined for each jet as the number of WPs it satisfies, with zero being failing, and four passing, all WPs.
In some cases, instead of directly applying $b$-tagging in MC simulation,
events with $c$- and light-flavor jets are weighted by the probability that these jets pass the $b$-tagging
requirement~\cite{HIGG-2016-29} and a $b$-tagging score is chosen for each jet based on its $b$-tagging probability.
This procedure, referred to as `truth'-tagging, increases the effective MC sample size for subdominant backgrounds.
A momentum correction is applied to $b$-tagged jets to account for the energy lost to soft radiation and to muons and neutrinos in semileptonic $b$-hadron decays, following the procedure used in Ref.~\cite{HIGG-2016-29}. Furthermore, correction factors are applied to the simulated samples to compensate for the differences between the $b$-tagging efficiencies in data and simulation~\cite{FTAG-2018-01}.

Hadronically decaying $\tau$-leptons, \tauh, are identified from the reconstructed jets~\cite{PERF-2013-06,ATL-PHYS-PUB-2019-033,ATLAS-CONF-2017-029}. A Recurrent Neural Network multivariate discriminant is used to select jets with energy-deposit profiles consistent with those expected from \tauh decay products and to match tracks in the ID to the \tauh candidates, using the `medium' \tauh criteria.
The \tauh candidates must have one or three associated tracks and satisfy the requirements of $\pt>20\,\GeV$ and $|\eta|<2.5$, excluding the region $1.37<|\eta|<1.52$.
 
The missing transverse momentum $\ptmissvec$, with magnitude $\met$, is calculated as the negative vectorial sum of the transverse momenta of reconstructed physics objects, namely  electrons, muons, photons, hadronically decaying $\tau$-leptons, and jets~\cite{PERF-2016-07,ATLAS-CONF-2018-023}. A component, called the `soft-term', from energy deposits due to the underlying event and other soft radiation not included in the physics objects is added in the $\ptmissvec$ calculation. A $\ptmissvec$ significance, $\Smet$, is defined in order to test whether the measured $\ptmissvec$ is incompatible with zero real $\ptmissvec$. It is calculated from the measured $\ptmissvec$, its resolution, and the correlation between measurements parallel and perpendicular to the $\ptmissvec$ direction~\cite{ATLAS-CONF-2018-038}.
 
A sequential overlap-removal procedure is applied to ensure that energy deposits in the calorimeter and tracks in the ID are not included in two or more different reconstructed objects.
If two electrons share a track, the electron with lower $\pt$ is removed.
If an electron and muon share an inner-detector track, the muon is removed if it is calorimeter-tagged, and the electron is removed otherwise.
The closest jet within $\Delta R = 0.2$ of a selected electron is removed.
If the nearest jet surviving that selection is within $\Delta R = 0.4$ of the electron, the electron is discarded.
Muons are usually removed if they are
separated from the nearest jet by $\Delta R < 0.4+10\,\GeV/\pt^\mu$, since this reduces the background from heavy-flavour decays
inside jets. However, if this jet has fewer than three associated tracks, the muon is kept and the jet is
removed instead; this avoids an inefficiency for high-energy muons undergoing significant energy loss in
the calorimeter. A $\tau_{h}$ candidate is rejected if it is separated by $\Delta R < 0.2$ from any selected electron,
muon, or jet.

% End of text imported from the .//tex/objects.tex input file

% The next lines are included from the .//tex/analysis.tex input file
\section{Analysis}
 
The non-resonant and resonant signal models targeted in this search share the same $Vhh$ final state but have different event kinematics.
The search starts with the selection of the $Vhh$ final states which are characterised by a leptonically decaying $V$ boson and four $b$-jets from the decays of the two Higgs bosons. Three leptonic channels (0L, 1L, 2L), one for each leptonic decay mode of the $W$ and $Z$ bosons ($Z\to\nu\nu$, $W\to\ell\nu$, $Z\to\ell\ell$), define the signal regions (SRs) of the search. Multivariate techniques based on boosted decision trees (BDT), trained to distinguish between signal and background events in each SR and for each signal model, provide the final discriminants to extract potential signal contributions. This is achieved through simultaneous fits to the BDT distributions observed in data with the hypotheses of signal plus background contributions. For the resonant models, mass requirements are applied to the resonance candidates before the fits to the BDT distributions. Background contributions are estimated using both data control regions~(CRs) and MC simulations.  The SR event selections, background estimations, the designs and trainings of the BDTs, and the mass requirements for resonant models are described in this section.

\FloatBarrier

% End of text imported from the .//tex/analysis.tex input file

% The next lines are included from the .//tex/selection.tex input file
\subsection{Signal region event selection}

The 0L channel is intended for the $Zhh\to\nu\nu\, bbbb$ final state. Candidate events are required to have $\met>150\,\GeV$ and $\Smet > 12$. The high \met criterion is necessitated by the trigger requirement. Events with identified \loose leptons or \tauh are vetoed. To suppress multi-jet backgrounds with mismeasured $\met$, the minimum azimuthal opening angle between $\ptmissvec$ and the Higgs boson candidates (see below) must satisfy $\min\left[\Delta\phi(\ptmissvec,\hvec)\right]>1$.
The ratio of the \met trigger efficiencies in data and simulation, measured in single-muon events as a function of $\ptmissvec+\boldsymbol{p}_\text{T}^{\mu}$, is applied as an event-weight  scale factor to simulated events~\cite{TRIG-2019-01}, with corresponding uncertainties described in Section~\ref{sec:sys-exp}.
 
The 1L channel is designed for the $Whh\to\ell\nu\, bbbb$ final state. In this channel, candidate events are selected by requiring exactly one \tight electron with $\pt>27\,\GeV$ or one \medium  muon with $\pt>25\,\GeV$. In addition, the events must have $\met>30\,\GeV$. Candidates with additional \loose leptons or identified \tauh are removed. The 1L channel is split into two SRs based on the charge of the lepton, 1L$+$ and 1L$-$, motivated by the expected large difference between the signal $W^+hh$ and $W^-hh$ cross-sections\footnote{The cross-section ratio of $W^+hh$ and $W^-hh$ is 1.9 in the SM~\cite{deFlorian:2016spz}. For the resonant signal models, the ratio varies between 1.8 and 2.3, depending on the resonance masses.} in contrast to the mostly charge-symmetric backgrounds.
 
The 2L channel targets the $Zhh\to\ell\ell\, bbbb$ final state. The $Z\to\ell\ell$ candidates are selected by requiring
exactly two oppositely charged \loose leptons with the same flavour, $e^+e^-$ or $\mu^+\mu^-$, at least one of which has $\pT>27\,\GeV$. The invariant mass of the lepton pair must satisfy $81<\mll<101\,\GeV$ for compatibility with the $Z$ boson mass.

Candidate $hh\to bbbb$ decays are selected by requiring at least four jets passing the 85\% $b$-tagging WP. Within the four-jet combination with the highest pseudo-continuous $b$-tagging scores (considering all combinations of four jets when more than one combination satisfies this requirement), the jets are paired to form the two $h\to bb$ candidates, $h_1$ and $h_2$,  by minimising the value of $|m_{h_1}-120\,\GeV|+|m_{h_2}-120\,\GeV|$. Here $m_{h_1}$ and $m_{h_2}$ are the invariant masses of the two candidates and  $120\,\GeV$ is their most probable value in simulation. The pair with the higher $\pT$ is labelled as $h_1$ and the other as $h_2$.
The efficiency for correctly identifying and pairing the four $b$-jets into two $h\to bb$ decays amongst the selected $b$-jets depends on the signal model and, if applicable, the resonance mass values. For the non-resonant SM \Vhh signal, the efficiency is 73\%. For the resonant \VH signal, the efficiency varies from 63\% at $m_H=260\,\GeV$ to 85\% at $m_H=1000\,\GeV$.
The efficiencies for the \AZH signals are similar to those for the \VH signals.

Table~\ref{tab:selection} summarises the selections that define the SRs for the three leptonic channels. Also included in the table are selections for control regions discussed below. The products of the acceptances and efficiencies of the SR selections, \AccEff, are shown in Figure~\ref{fig:AccEff} as functions of the model parameters for a few selected signal models.

\begin{table}[!htb]
\begin{center}
\caption{Selections for the 0L, 1L, and 2L signal regions, and \ttbar and \Vjets control regions.
The `---' symbol indicates no selection is applied.}
\label{tab:selection}
{\scriptsize
\begin{tabular}{ccccccc}\toprule
& \multicolumn{3}{c}{Signal regions}            && \multicolumn{2}{c}{Control regions} \\ \cline{2-4}\cline{6-7} \\
&     0L                      &  1L (1L$+$/1L$-$) &   2L                         &&   \ttbar          &  \Vjets  \\ \midrule \midrule
Trigger    &    \met                     &  single-lepton   &  single-lepton                &&   single-lepton    &  single-photon  \\ \midrule

&                      &   $=1$ \tight electron        &  $=2$ \loose leptons        &&  $=2$ \loose leptons   &  $=1$ photon with \\
Lepton    & 0 \loose leptons,    &  with $\pT>27\,\GeV$          &  ($e^+e^-$ or $\mu^+\mu^-$), &&  ($e^\pm\mu^\mp $),    &  $\pt>150\,\GeV$, \\
or photon & 0 identified \tauh   &  OR 1 \medium muon            &  $\ge 1$ lepton with         &&  $\ge 1$ lepton with   &  0 \loose leptons, \\
&                      &  with $\pT>25\,\GeV$,         &  $\pT>27\,\GeV$,             &&  $\pT>27\,\GeV$        &  0 identified \tauh \\
&                      &  0 additional \loose leptons, &  $81<m_{\ell\ell}<101\,\GeV$ \\
&                      &  0 identified \tauh           &                              \\ \midrule
 
& $\met>150$ GeV,           &  \multirow{3}{*}{$\met>30\,\GeV$}   & \multirow{3}{*}{---} && \multirow{3}{*}{---} & \multirow{3}{*}{---}   \\
\ptmissvec      & $\Smet>12$,               &                  &                              &&                    &                  \\
& $|\Delta\phi(\ptmissvec,h)|>1$  &                  &                              &&                    &                  \\ \midrule

Jets   & \multicolumn{6}{c}{$\ge 4$ jets with $\pt>20\,\GeV$ and passing the 85\% $b$-tagging WP} \\
 
\toprule
\end{tabular}}
\end{center}
\end{table}

\begin{figure}[!htb]
\begin{center}
\subfigure[]{\includegraphics[width=0.48\textwidth]{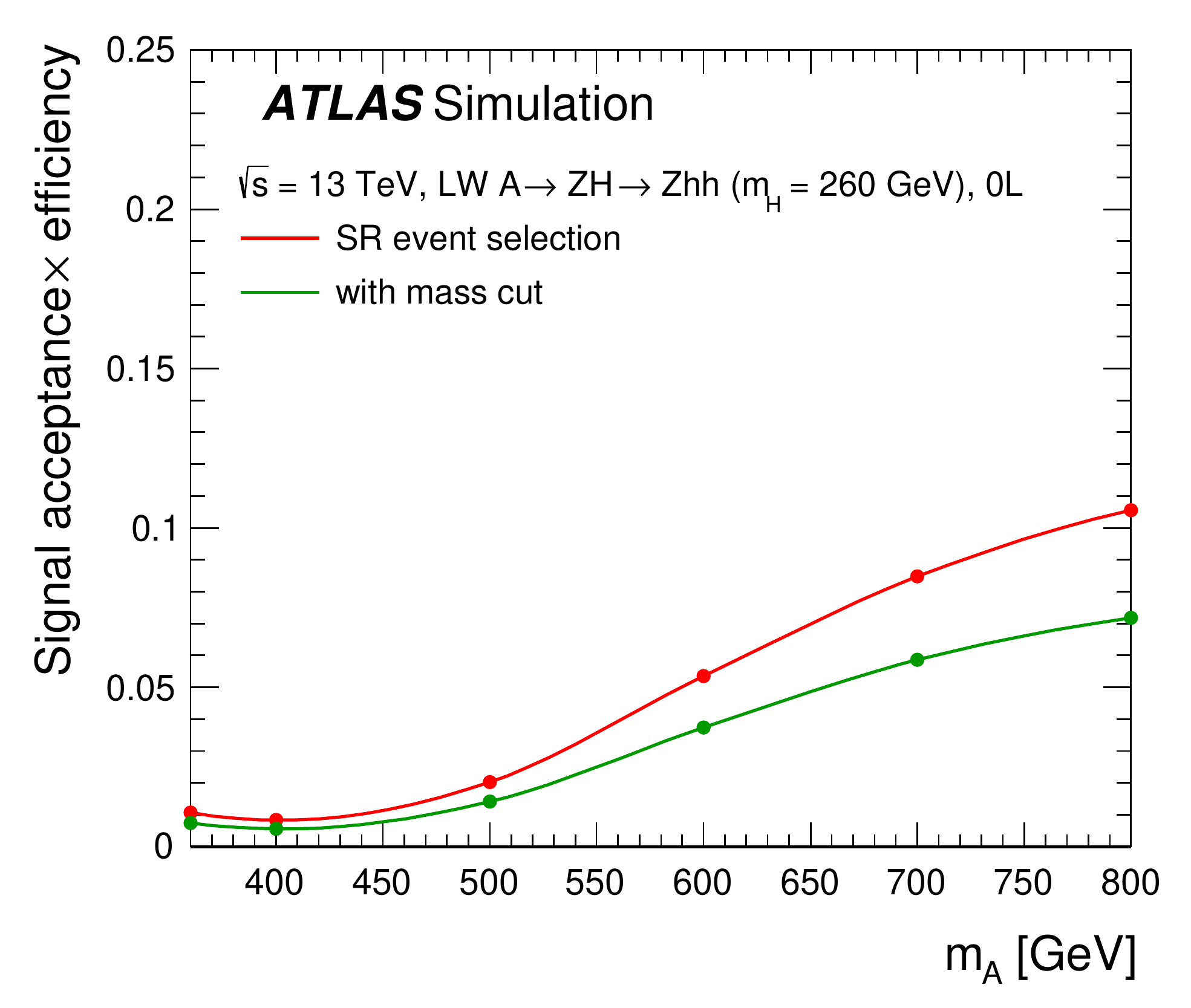}}
\subfigure[]{\includegraphics[width=0.48\textwidth]{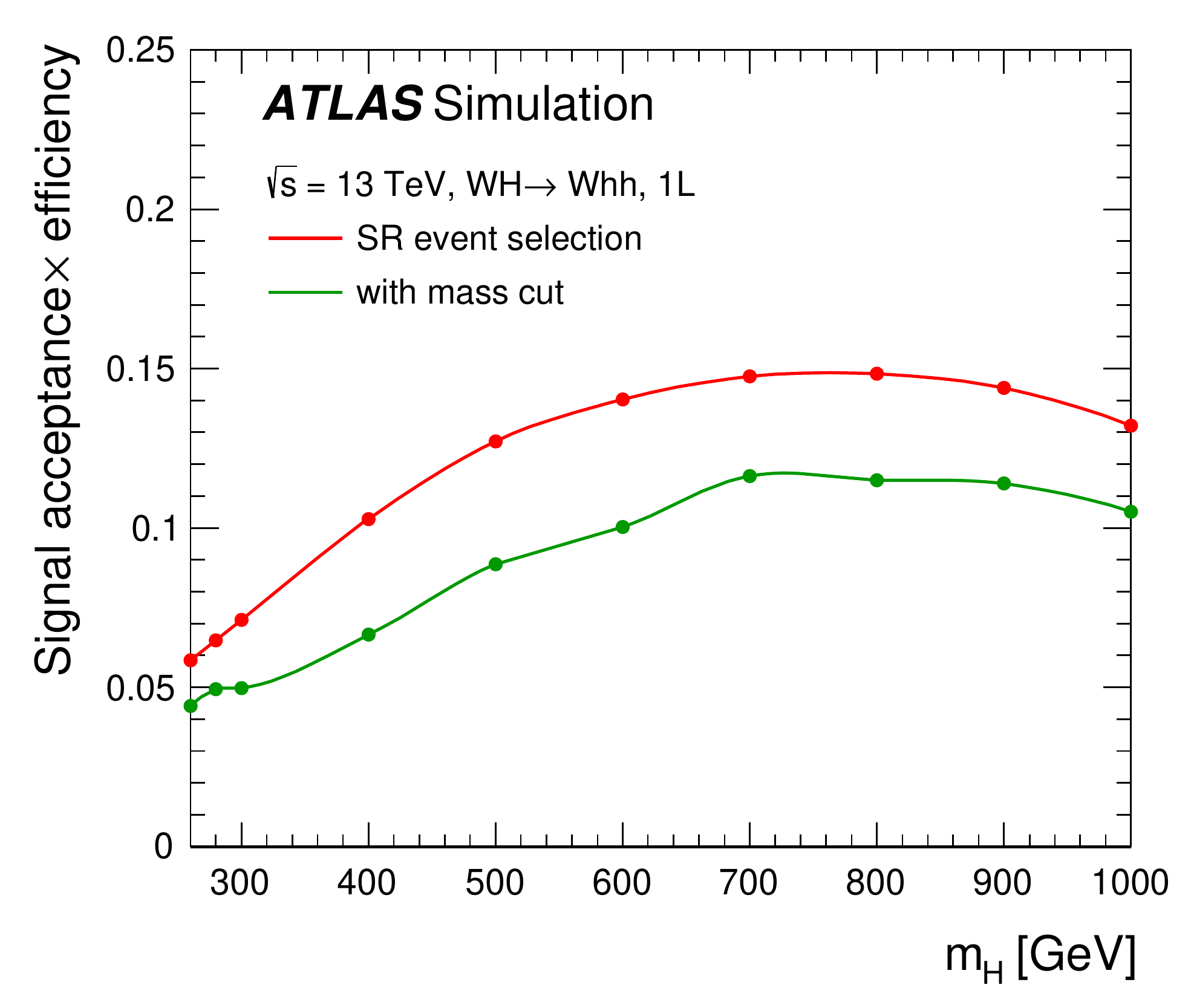}}  \\
\subfigure[]{\includegraphics[width=0.48\textwidth]{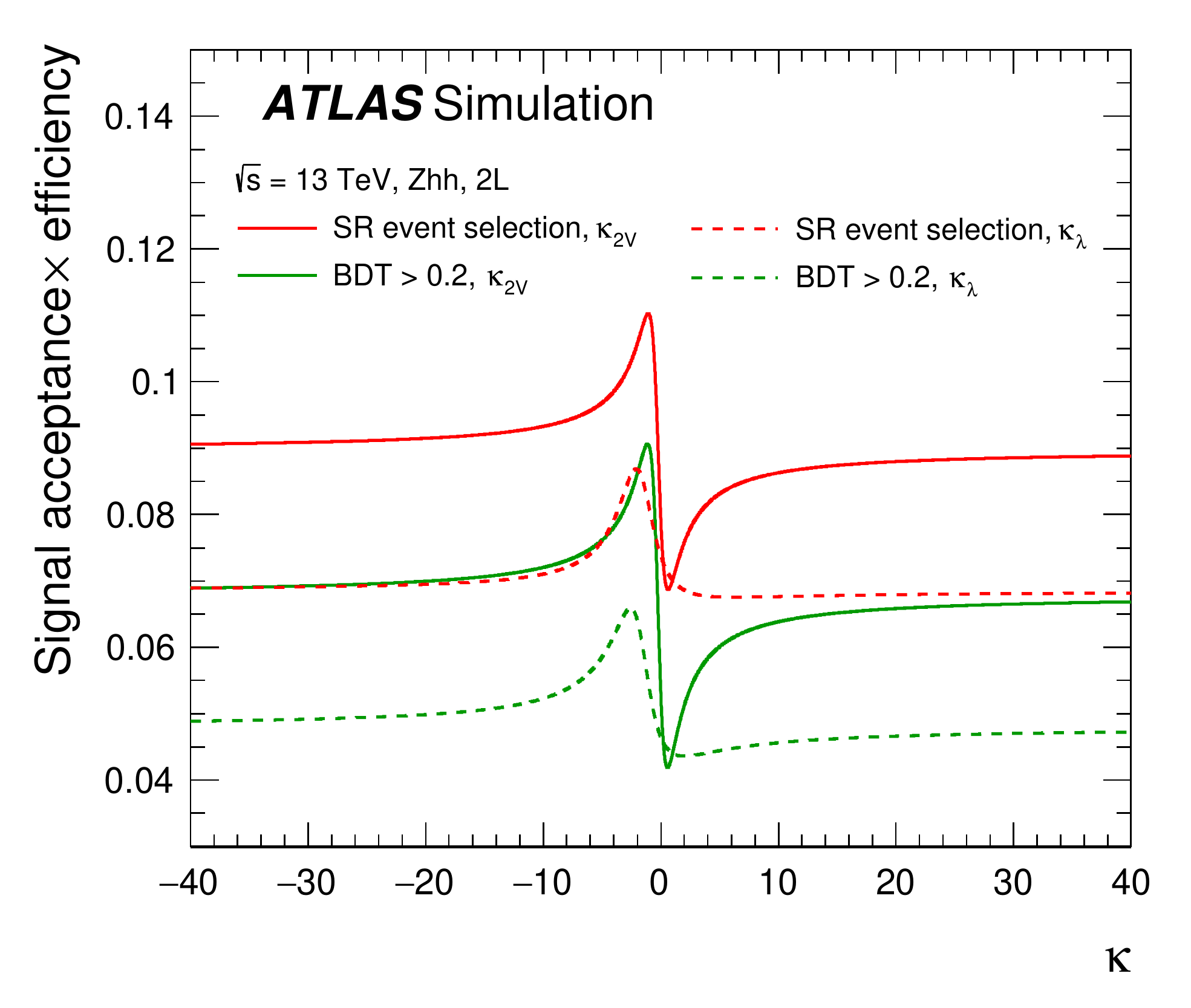}}
\end{center}
\caption{Example dependences of acceptance times efficiency (\AccEff) on signal model parameters: (a) $m_A$ for the \AZH search in the 0L channel for the case of $m_H=260\,\GeV$ and a LW $A$ boson, (b) $m_H$ for the \WH search in the 1L channel, and (c) \ChhVV and \Chhh for the non-resonant \Zhh search in the 2L channel. Each \AccEff value is calculated for the respective $\nu\nu bbbb$, $\ell\nu bbbb$, and $\ell\ell bbbb$ final state, with $\ell$ including the $e,\mu$, and $\tau$ leptons here.  For the resonant \VH and \AZH searches, the \AccEff value is shown both with and without the mass requirements on the reconstructed resonances. Similarly, for the \Zhh search, the \AccEff value is shown both with and without a BDT requirement. The small decreases in \AccEff at high $m_H$ values in (b) are due to the merging of $b$-jets from highly boosted $h\to bb$ decays. The structures in \AccEff around small values of $\kappa$ in (c) reflect large changes in the relative contributions of the three production diagrams shown in Figure~\ref{fig:NonResDiagrams}. }
\label{fig:AccEff}
\end{figure}

% End of text imported from the .//tex/selection.tex input file

% The next lines are included from the .//tex/backgrounds.tex input file
\subsection{Background estimations}
\label{sec:backgrounds}

Major background sources in the $Vhh$ search are the production of top-quark pairs (\ttbar), single top quarks, vector bosons in association with jets (\Vjets), diboson and multi-boson events, and multi-jet processes. Their relative contributions depend on the channel. In the 0L and 1L channels, the \ttbar background dominates, with the subleading contribution being from \Vjets (\Zjets for 0L and \Wjets for 1L). In the 2L channel, \Zjets and \ttbar are the two leading sources. A mixture of data-driven and simulation-based methods are used to estimate these backgrounds. MC simulations are used to model the kinematics of background processes except for the multi-jet backgrounds. Contributions from \ttbar and \Vjets processes are normalised to the data through the use of CRs, while the rest of the non-multi-jet sources are normalised to their theoretical cross-sections within the estimated uncertainties.  For the multi-jet backgrounds, their normalisations and kinematic models are derived from auxiliary data samples. These data-driven procedures are described below.

A \ttbar control region (\ttbarCR) is used to constrain the normalisations of the leptonic \ttbar events\footnote{Referring to \ttbar events with at least one $W$ boson decaying leptonically.} in the SRs.  The control region is defined by using a selection similar to that for the 2L SR, but requiring a different-flavour opposite-sign lepton pair ($e^\pm\mu^\mp$) instead of a same-flavour opposite-sign lepton pair ($e^+e^-$ or $\mu^+\mu^-$) and removing the dilepton mass requirement to increase the sample size. The jet requirements are the same as for the SR. These  selections are  summarised in Table~\ref{tab:selection}.
 
Top-quark-pair events have two $b$-jets from top-quark decays and can mimic signal events if there are two additional genuine $b$-jets from radiation or if there are misidentified $c$-jets or light-flavour jets ($j$) from radiation or hadronic $W$ boson decays.
The flavour content of this radiation is difficult to model, particularly for heavy flavours ($b$ or $c$). To allow for flavour-dependent variations of the \ttbar background contribution in the signal extraction, simulated \ttbar events are labelled according to the `truth' flavours of jets from the radiation: events with one or more $b$-hadrons (\ttbarB), events with one or more $c$-hadrons but no $b$-hadrons (\ttbarC), and events with no $b$-hadrons or $c$-hadrons (\ttbarL).
The `truth' flavour of a jet is determined from hadrons with $\pt>5\,\GeV$ found within a cone of size $\Delta R=0.3$ around the jet axis~\cite{FTAG-2018-01}.

The \Vjets process is a major background source in all three channels (\Zjets for 0L and 2L and \Wjets for 1L). A \Vjets CR (\VjetsCR), defined using \Gjets events, is used to assess their contributions. As summarised in Table~\ref{tab:selection}, events in \VjetsCR are selected by requiring exactly one photon with $\pt>150\,\GeV$ and no identified $e$, $\mu$, or $\tau_h$ from the data collected using a single-photon trigger. They must pass the same jet requirements as events in the SRs. At $\pt\gtrsim m_V$, \Gjets events and \Vjets events are expected to have similar kinematics because they originate from comparable diagrams and the impact of the finite mass of the $V$ bosons becomes small at high $\pt$.
Residual differences between \Gjets and \Vjets, driven mainly by lower-$\pt$ events, are taken into account by extrapolation uncertainties (see Section~\ref{sec:systematics}).
 
The \Vjets events can be selected as signal events if they have four jets passing the $b$-tagging requirement, either from genuine heavy-flavour jets or from misidentified light-flavour jets. Like the \ttbar events, modelling the jet flavour composition is challenging.
The simulated \Gjets events are categorised according to `truth' flavour and jet matching: events with three or more $b$-hadrons (\VjetsB), events with $\ge 1\, c$-hadron but with $\le 2\, b$-hadrons (\VjetsC), and events with zero $c$-hadrons and $\le 2\,b$-hadrons (\VjetsL).

The contribution from multi-jet events, including hadronic \ttbar events,\footnote{Referring to \ttbar events with both $W$ bosons decaying hadronically.} is negligible in the 0L and 2L SRs as well as in \ttbarCR, minor in the 1L SRs, but significant in \VjetsCR. Both its rate and kinematics are difficult to simulate. Auxiliary data samples rich in multi-jet events are selected to model the corresponding contributions in the 1L SRs and \VjetsCR.
For \VjetsCR, the auxiliary sample is defined by inverting the photon isolation requirement while keeping the rest of the selection the same.
The kinematic distributions of multi-jet events in \VjetsCR are modelled by the selected events in this auxiliary sample after subtracting approximately 10\% non-multi-jet contributions taken from simulation,
and an uncertainty in the shapes of these distributions is defined by varying the non-multi-jet contributions in the auxiliary sample by $\pm 100\%$.
These distributions are assigned a pre-fit normalisation equal to the difference between the number of data events and the non-multi-jet contributions in \VjetsCR, with an uncertainty of 100\%.
The multi-jet modelling in the 1L SRs follows the same approach, except that the auxiliary sample is selected by inverting the lepton isolation requirement and the multi-jet contribution in the 1L SRs is normalised through a fit to the distribution of the transverse mass of the lepton and \met system,\footnote{The transverse mass of a system is defined as $\sqrt{m^{2}+\pT^{2}}$, where $m$ and $\pT$ are the mass and transverse momentum of the system, respectively.} in events selected with the 1L SR requirements but with the \met criteria reversed to $\met<30\,\GeV$, which leads to negligible contamination of potential signals produced with cross-sections comparable to the search sensitivity.

Jet $b$-tagging scores offer the most sensitive information for separating the different flavour contributions of the \ttbar and \Vjets backgrounds. Thus the distributions of the sum of the pseudo-continuous $b$-tagging scores of the four jets with the highest scores, \pcbs, which ranges from 4 to 16, are used to disentangle contributions from the three flavour components. The \pcbs distributions observed in \ttbarCR and \VjetsCR, divided into three bins (4--9, 10--12, and 13--16), are included in the fits to determine the potential signal contributions as discussed in Section~\ref{sec:results}. In the fits, the flavour components of the \ttbar and \Vjets contributions are normalised with their separate normalisation factors. The \pcbs template, i.e.\ the shape of the \pcbs distribution, of each  component is taken from simulation except for the multi-jet contribution in \VjetsCR discussed above. Figure~\ref{fig:pcbs} compares the \pcbs distributions of the \ttbarCR and \VjetsCR observed in the data with their post-fit background expectations.
The \ttbar and \Vjets contributions comprise 92.4\% and 79.7\% of the total event yields in their respective CRs.
 
\begin{figure}[!htb]
\begin{center}
\subfigure[]{\includegraphics[width=0.45\textwidth]{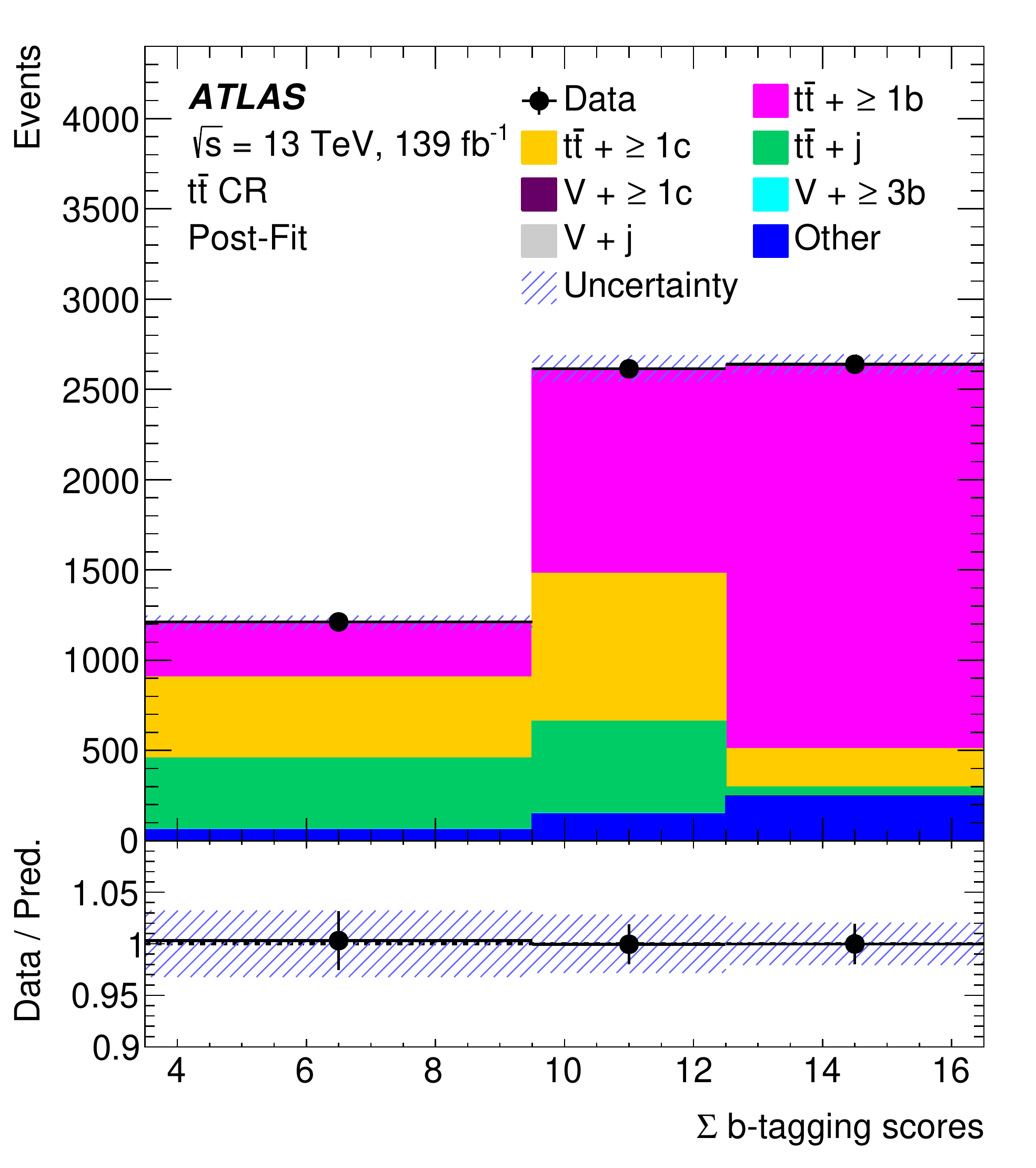}}\hspace*{1.0cm}
\subfigure[]{\includegraphics[width=0.45\textwidth]{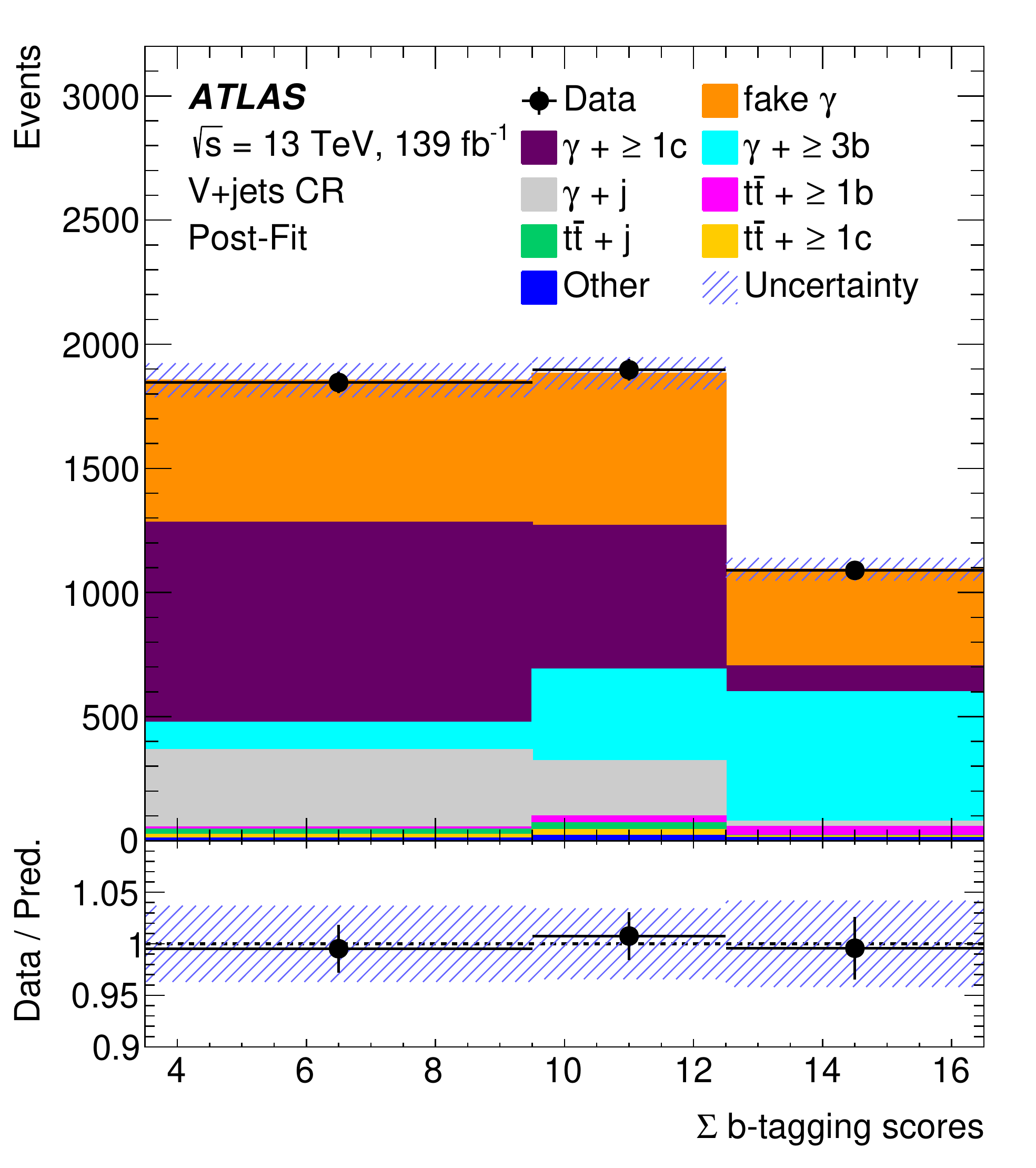}}
\end{center}
\caption{The background-only post-fit distributions of the sum of the pseudo-continuous $b$-tagging scores of the four jets with the highest scores in the (a) \ttbarCR and (b) \VjetsCR control regions.
The bottom panels show the ratios of the data to the total background expectations.  The hashed bands represent the combined statistical and systematic uncertainties in the total background predictions. }
\label{fig:pcbs}
\end{figure}

Validation regions~(VRs) are defined for each of the SRs and CRs (\ttbarCR and \VjetsCR) to study the background modelling uncertainties (see Section~\ref{sec:systematics}). Events in VRs are selected in the same way as those for SRs and CRs but with different jet requirements. Instead of four $b$-tagged jets, events are required to have at least four jets, exactly three of which pass the 85\% $b$-tagging WP. The highest-\pt non-$b$-tagged jet in  $|\eta|<2.5$ is taken as the fourth jet to build $hh$ candidates and is assigned a pseudo-continuous $b$-tagging score of 4, which is found to best reproduce the multivariate discriminants in the signal regions.

% End of text imported from the .//tex/backgrounds.tex input file

% The next lines are included from the .//tex/bdt.tex input file
\subsection{Multivariate discriminant}
\label{sec:bdt}

BDTs are used to exploit the kinematic differences between signal and background events passing the SR event selection. One BDT is constructed for each channel (0L, 1L, 2L) and each signal model (\Vhh, \VH, \AZH), resulting in eight BDTs in total (three each for 0L and 2L, two for 1L). To minimise the complexity of the analysis, these BDTs are built in the same way where possible.  They differ in having channel-dependent and signal-model-dependent variables.  BDT input variables are chosen through extensive comparisons between their distributions in signal and background events while minimising correlations among the variables. Moreover, for the resonant \VH and $\AZH$ searches, only variables with weak correlations with the resonance mass are considered so as to lessen the dependence of the BDTs on any particular hypothesised signal mass value. Requirements on the reconstructed resonance masses are applied separately (Section~\ref{sec:signal}). Table~\ref{tab:BDT} summarises the input variables used for the eight BDTs.

A common set of seven input variables are used for all eight BDTs. These variables are
\begin{itemize}
\item the sum ($m_{h_1}+m_{h_2}$) and the difference ($m_{h_1}-m_{h_2}$) of the masses of the Higgs boson candidates $h_1$ and $h_2$,
\item the sum of the $b$-tagging scores \pcbs,
\item the number of jets ($\njets$),
\item the sum of the transverse energy of jets excluding the four selected $b$-jets ($\HTex$),
\item  $m_{h_1}^\mathrm{FSR}$ and $m_{h_2}^\mathrm{FSR}$, the masses of the two Higgs boson candidates calculated by  including additional jets in a cone of size $\Delta R=0.8$ around each $b$-jet's axis. This calculation is intended to recover jet energy lost due to final-state radiation~(FSR).
\end{itemize}
 
Input variables specific to channels or signal models are
the invariant mass \mhh and transverse momentum \pthh of the reconstructed Higgs boson pair,
\met, the transverse momentum of the vector boson (\ptV),  the transverse mass of the lepton--\met system (\mtW),
and four variables exploiting differences between the angular distributions of signal and background events.
Because of its strong correlation with the resonance mass, \mhh is not used in the BDTs for the resonant searches. Instead, requirements on \mhh are applied afterwards (see Section~\ref{sec:signal}).  The variable \pthh is used in all BDTs except those for the \AZH search, due to its correlation with $m_A$. For the same reason, \met (\ptV) is not used in the 0L (2L) channel of the \AZH search.
The variables \coshOne and \coshTwo~\footnote{$\Delta\eta$ and $\Delta\phi$ are, respectively, the differences in pseudorapidity and azimuth of the two $b$-jets of the $h\to bb$ candidate.} of the $h_1$ and $h_2$ candidates exploit the  angular differences between the signal $h\to bb$ decay and the background gluon-splitting $g^*\to bb$ process and are motivated by $m_{h1}^{2}\propto\coshOne$ and $m_{h2}^{2}\propto\coshTwo$ in the approximation that jets are massless. Similarly, the Higgs bosons and vector bosons are expected to be produced more centrally for signal events and more forwardly for background events. The two rapidity difference variables, between $h_1$ and $h_2$ (\Draph) and between $V$ and $hh$ (\DrapV), are designed to take advantage of these subtle differences.
 
Figure~\ref{fig:BDTVars} compares the \mhh, $m_{h_1}-m_{h_2}$, and $\HTex$ distributions in data with those expected from background processes for events passing the SR selections in the three leptonic channels.  The background distributions are obtained from the fit to the background-only hypothesis discussed in Section~\ref{sec:results}. Also shown are the expected distributions from example signal models.

\begin{table}[!htb]
\begin{center}
\caption{Variables used in the BDT discriminant in each of the channels and for each signal model; see text for the variable definitions.  The \cm symbol indicates the inclusion of the variable. The BDTs for the \VH and \AZH searches exclude variables strongly correlated with the resonance mass.}
\label{tab:BDT}
\begin{tabular}{lccclcclccc}\toprule
& \multicolumn{10}{c}{Channel and signal model} \\ \cline{2-11}
& \multicolumn{3}{c}{0L} && \multicolumn{2}{c}{1L}  && \multicolumn{3}{c}{2L} \\ \cline{2-4}\cline{6-7}\cline{9-11}
Variable            &  \Vhh &  \VH  & \AZH  &&  \Vhh &  \VH  &&  \Vhh & \VH  & \AZH  \\  \midrule
$m_{h_1}+m_{h_2}$   &  \cm &  \cm   & \cm   &&  \cm & \cm    &&  \cm & \cm   & \cm    \\[1mm]
$m_{h_1}-m_{h_2}$   &  \cm &  \cm   & \cm   &&  \cm & \cm    &&  \cm & \cm   & \cm    \\[1mm]
$\njets$            &  \cm &  \cm   & \cm   &&  \cm & \cm    &&  \cm & \cm   & \cm    \\[1mm]
$\HTex$             &  \cm &  \cm   & \cm   &&  \cm & \cm    &&  \cm & \cm   & \cm    \\[1mm]
\pcbs               &  \cm &  \cm   & \cm   &&  \cm & \cm    &&  \cm & \cm   & \cm    \\[1mm]
$m_{h_1}^\mathrm{FSR}$    &  \cm &  \cm   & \cm   &&  \cm & \cm    &&  \cm & \cm   & \cm    \\[1mm]
$m_{h_2}^\mathrm{FSR}$    &  \cm &  \cm   & \cm   &&  \cm & \cm    &&  \cm & \cm   & \cm    \\[2mm]
 
$\mhh$              &  \cm &        &       &&  \cm &        &&  \cm &       &        \\[1mm]
$\pthh$             &  \cm &  \cm   &       &&  \cm & \cm    &&  \cm & \cm   &        \\[1mm]
 
$\met$              &  \cm &  \cm   &       &&  \cm & \cm    &&  \cm & \cm   & \cm    \\ [1mm]
$\ptV$              &      &        &       &&  \cm & \cm    &&  \cm & \cm   &        \\[1mm]
$\mtW$             &      &        &       &&  \cm &        &&      &       &        \\[1mm]
 
\coshOne            &  \cm &  \cm   &       &&  \cm & \cm    &&  \cm & \cm   &        \\[1mm]
\coshTwo            &  \cm &  \cm   &       &&  \cm & \cm    &&  \cm & \cm   &        \\[1mm]
\Draph              &  \cm &  \cm   &       &&  \cm & \cm    &&  \cm & \cm   &        \\[1mm]
\DrapV              &      &        &       &&      &        &&  \cm & \cm   &        \\
\toprule
\end{tabular}
\end{center}
\end{table}

\begin{figure}[!htbp]
\begin{center}
\subfigure[]{\includegraphics[width=0.30\textwidth]{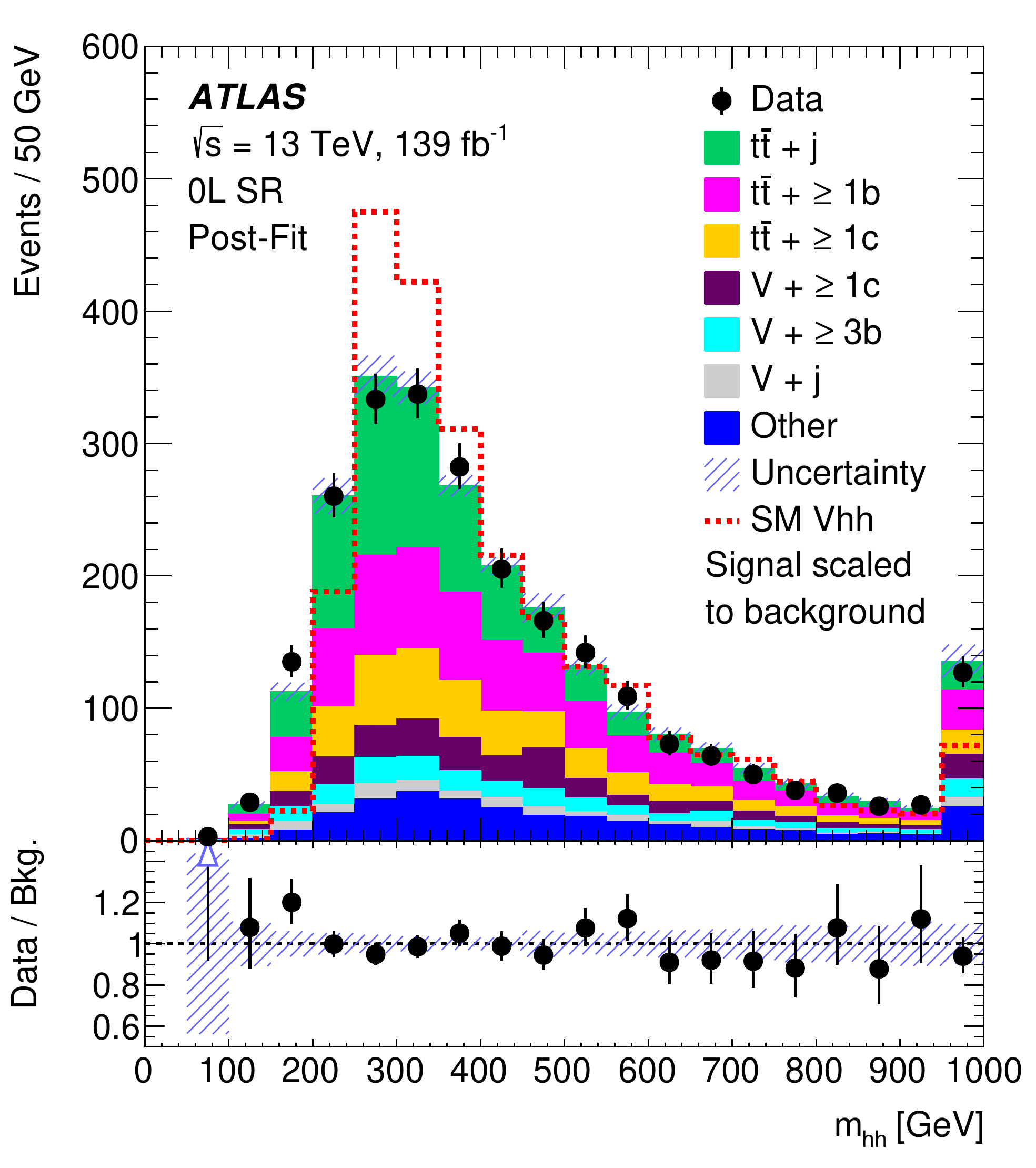}}
\subfigure[]{\includegraphics[width=0.30\textwidth]{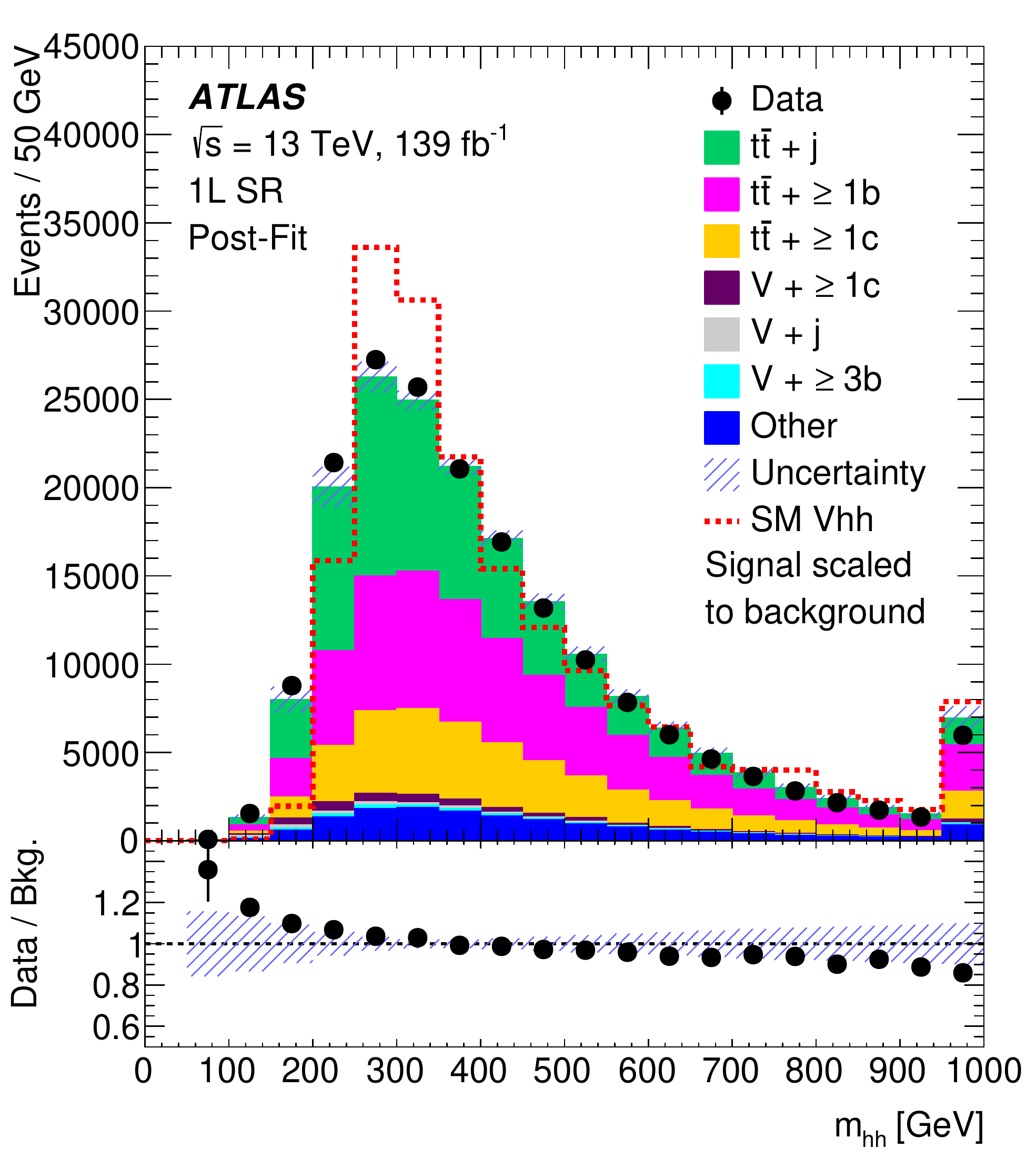}}
\subfigure[]{\includegraphics[width=0.30\textwidth]{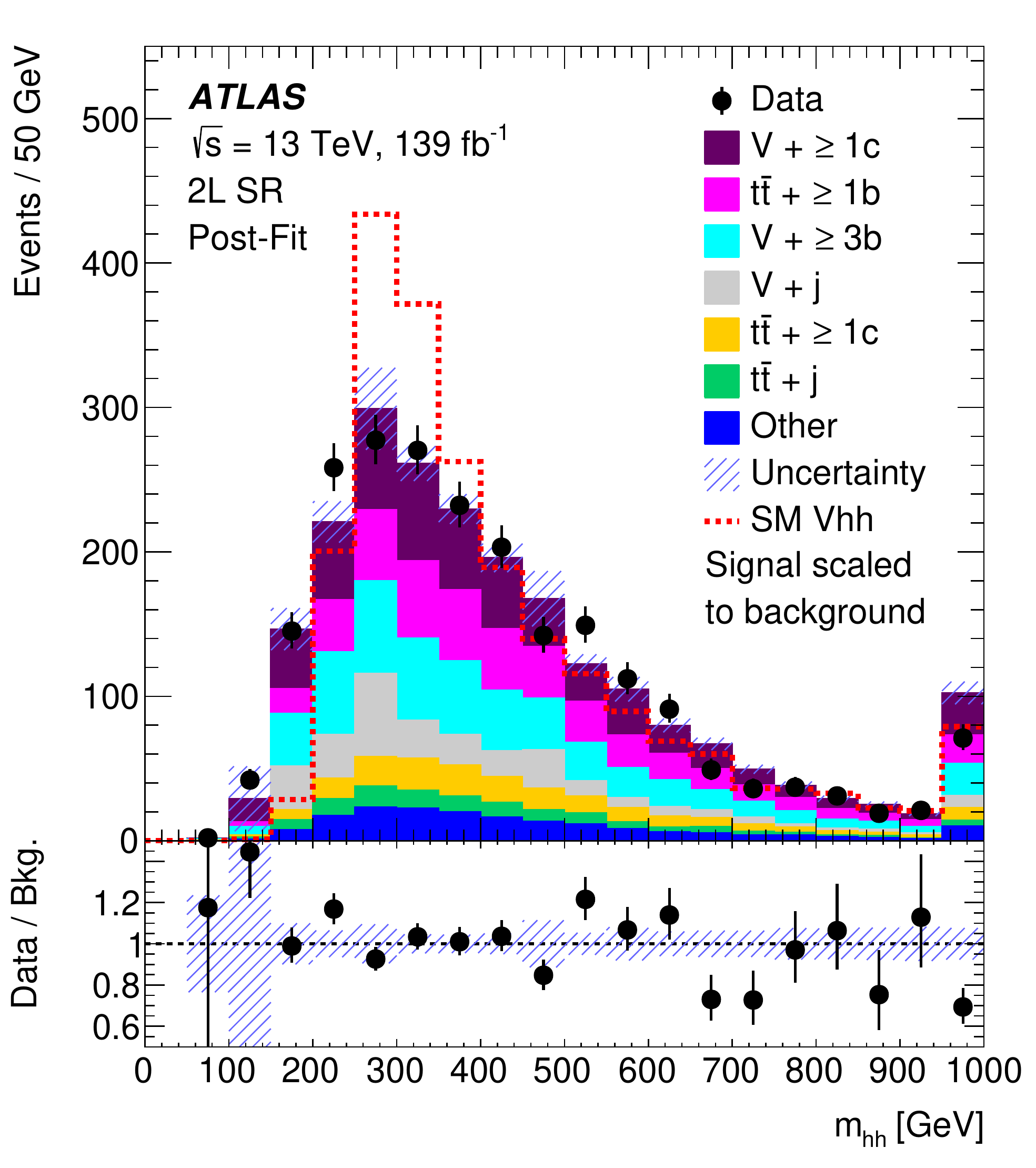}} \\
\subfigure[]{\includegraphics[width=0.30\textwidth]{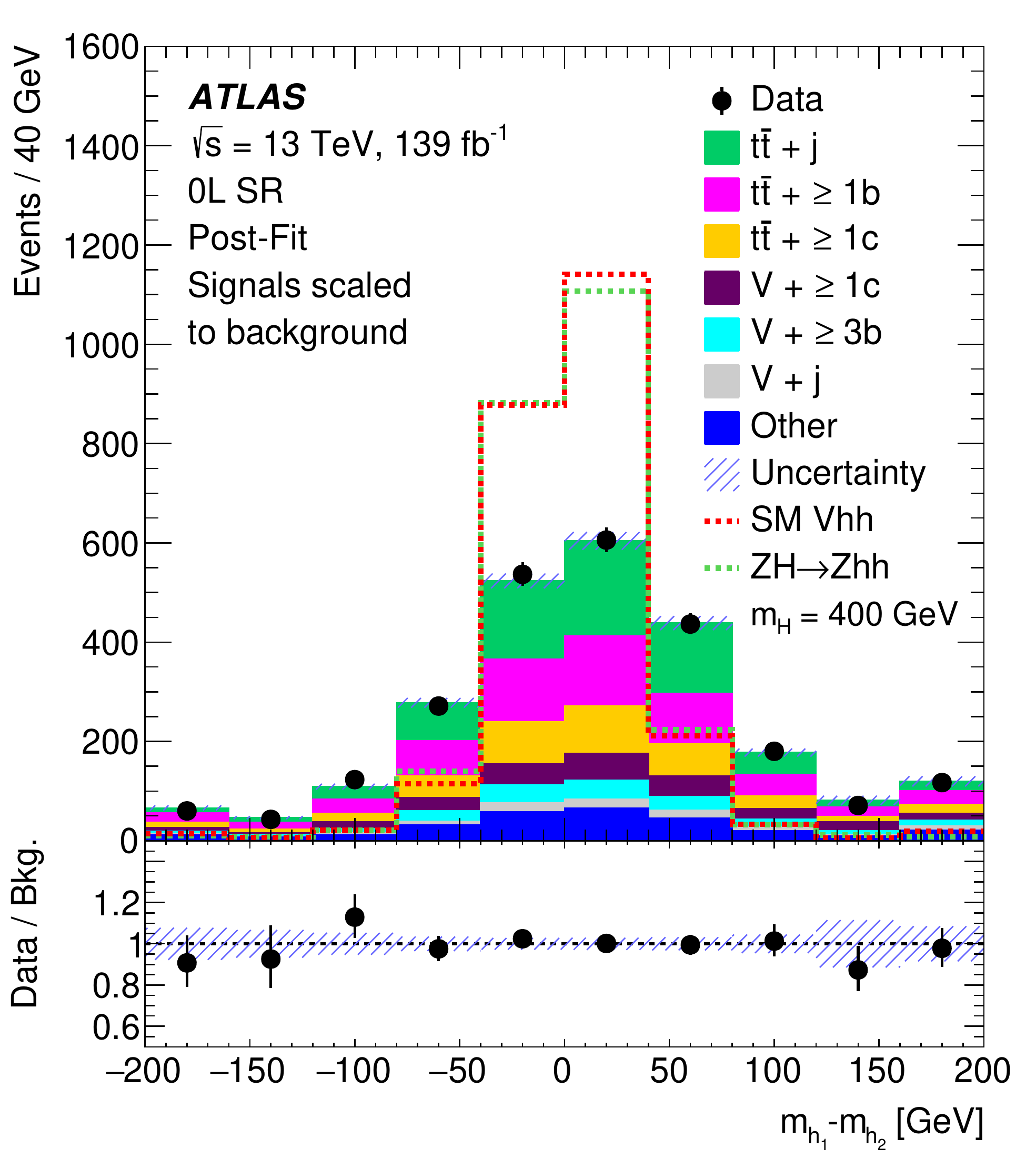}}
\subfigure[]{\includegraphics[width=0.30\textwidth]{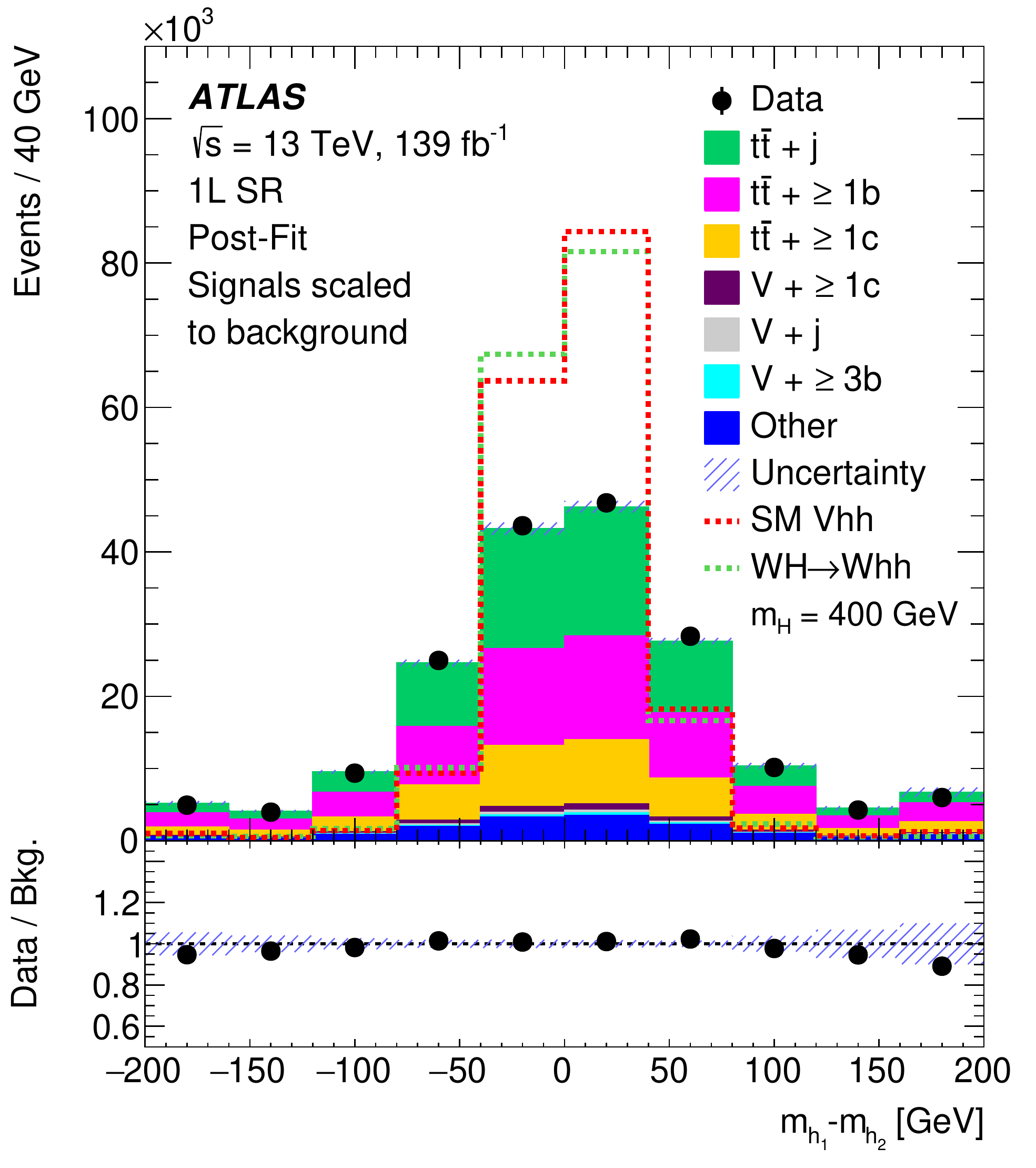}}
\subfigure[]{\includegraphics[width=0.30\textwidth]{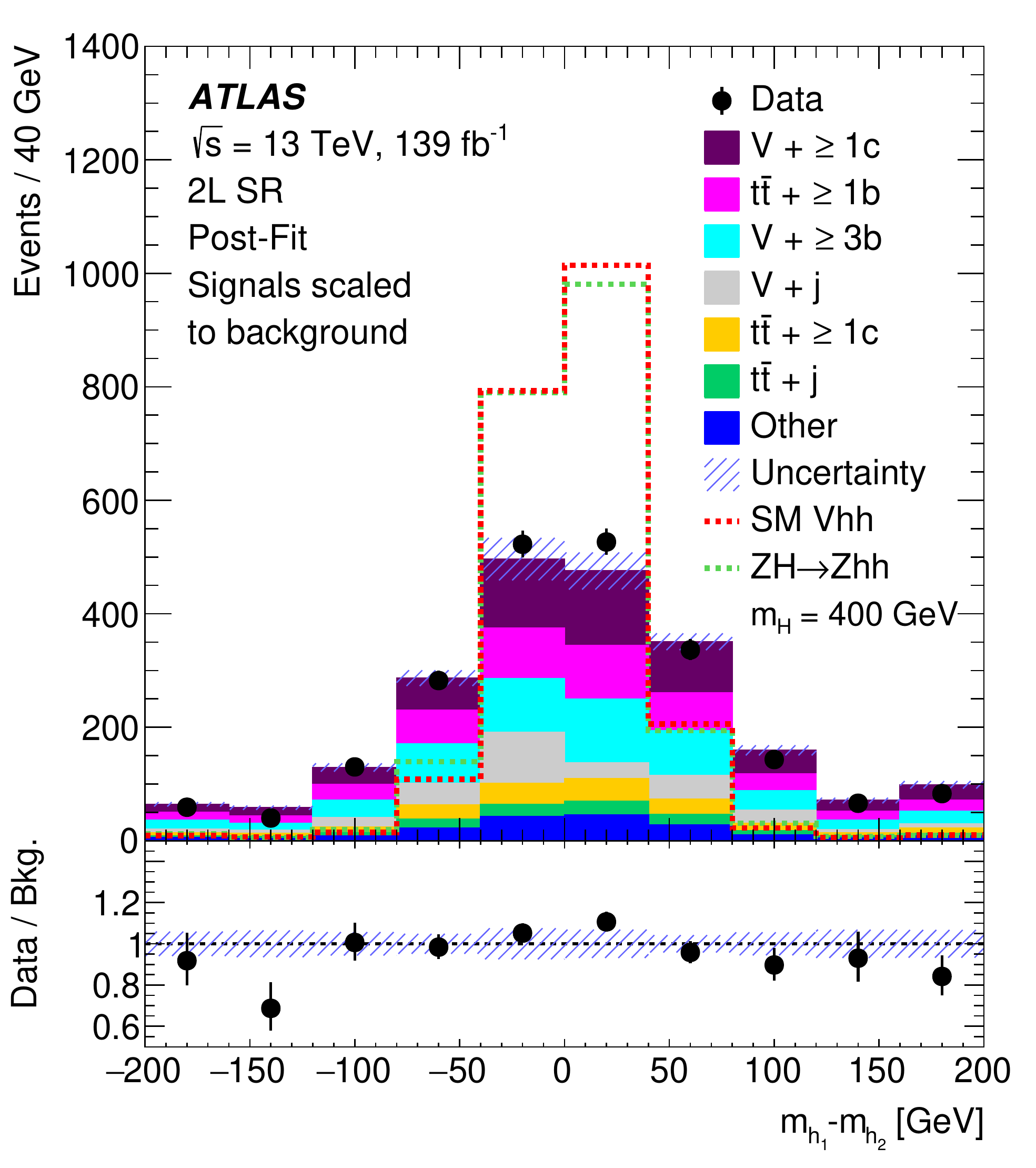}} \\
\subfigure[]{\includegraphics[width=0.30\textwidth]{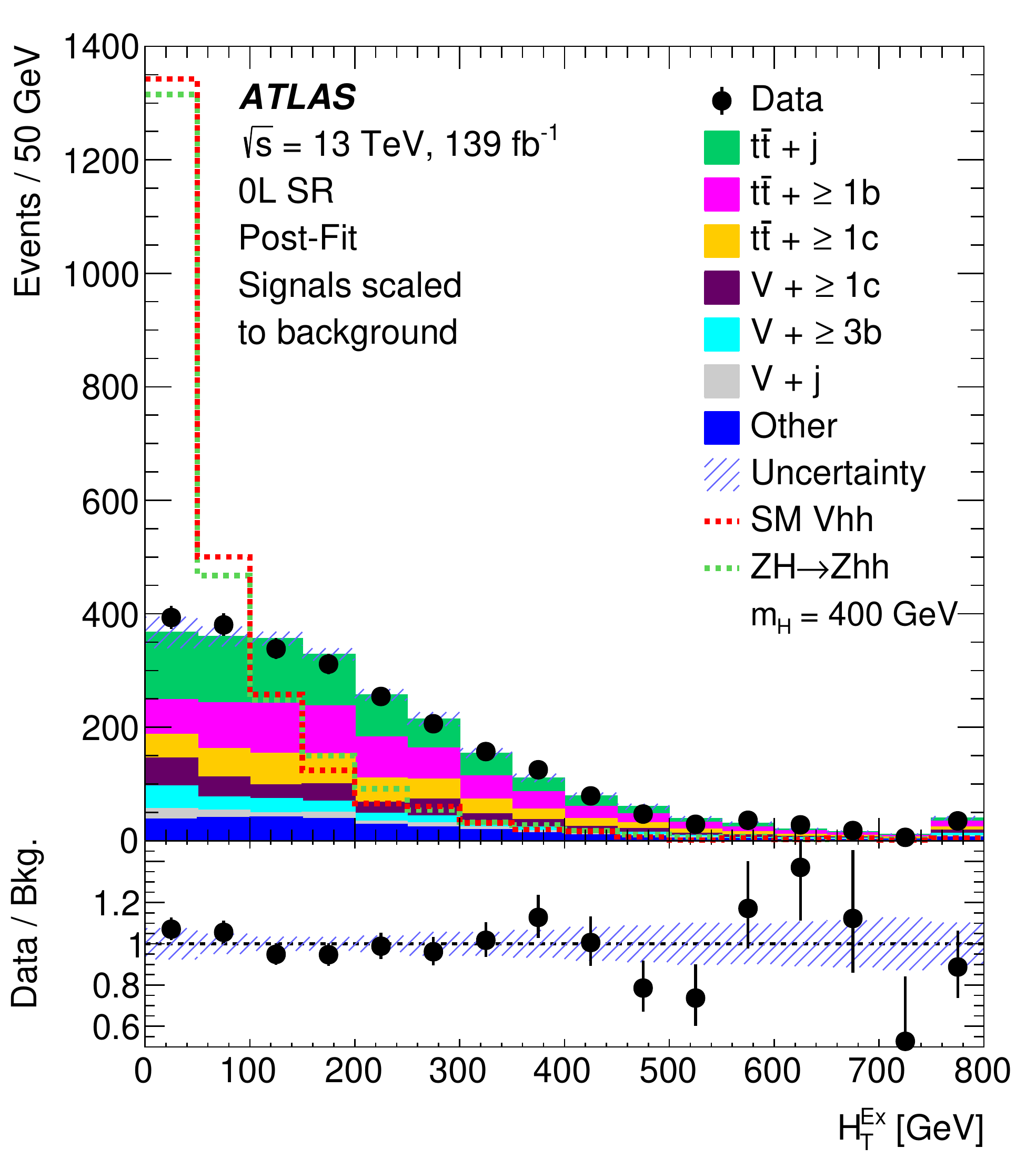}}
\subfigure[]{\includegraphics[width=0.30\textwidth]{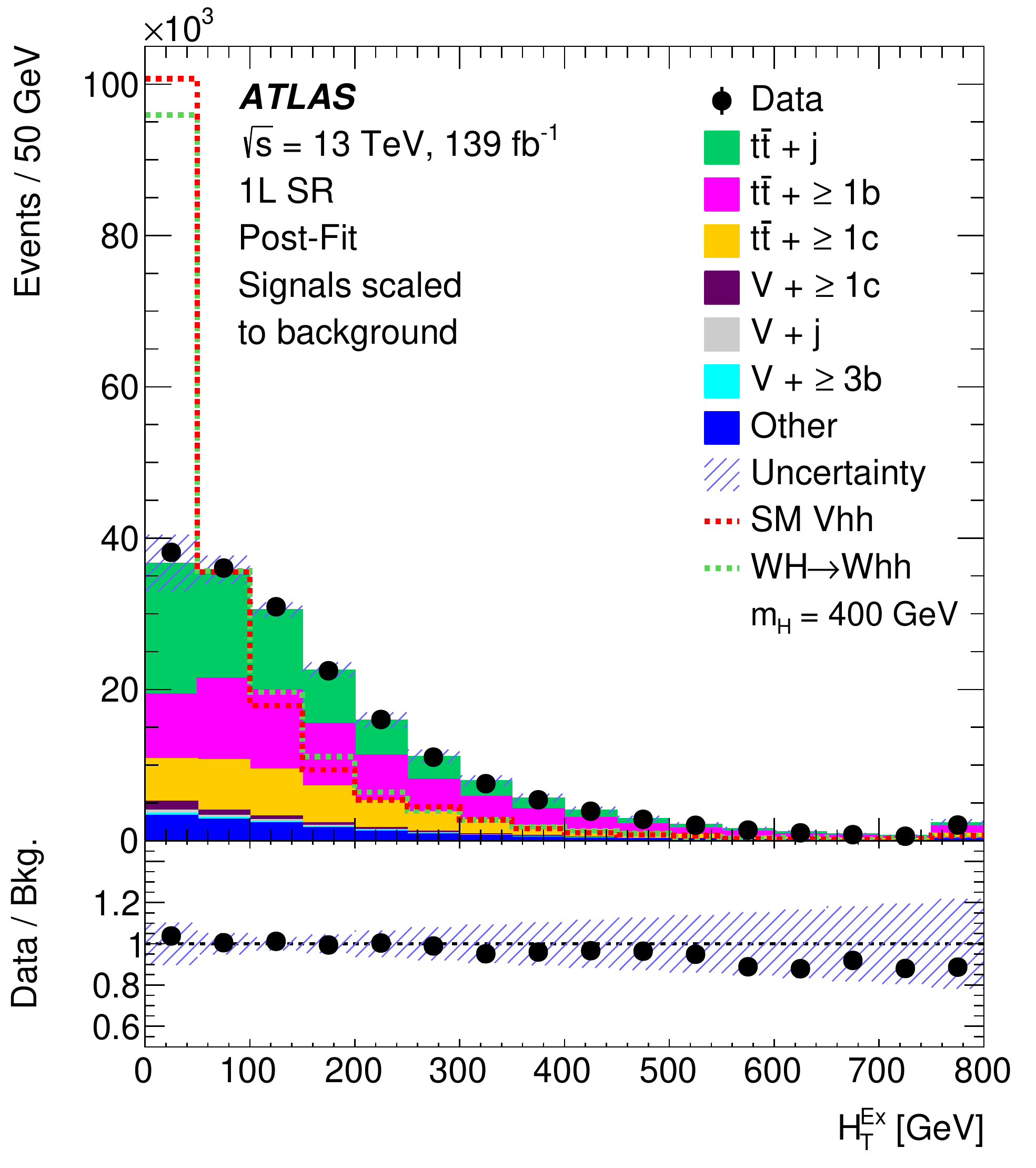}}
\subfigure[]{\includegraphics[width=0.30\textwidth]{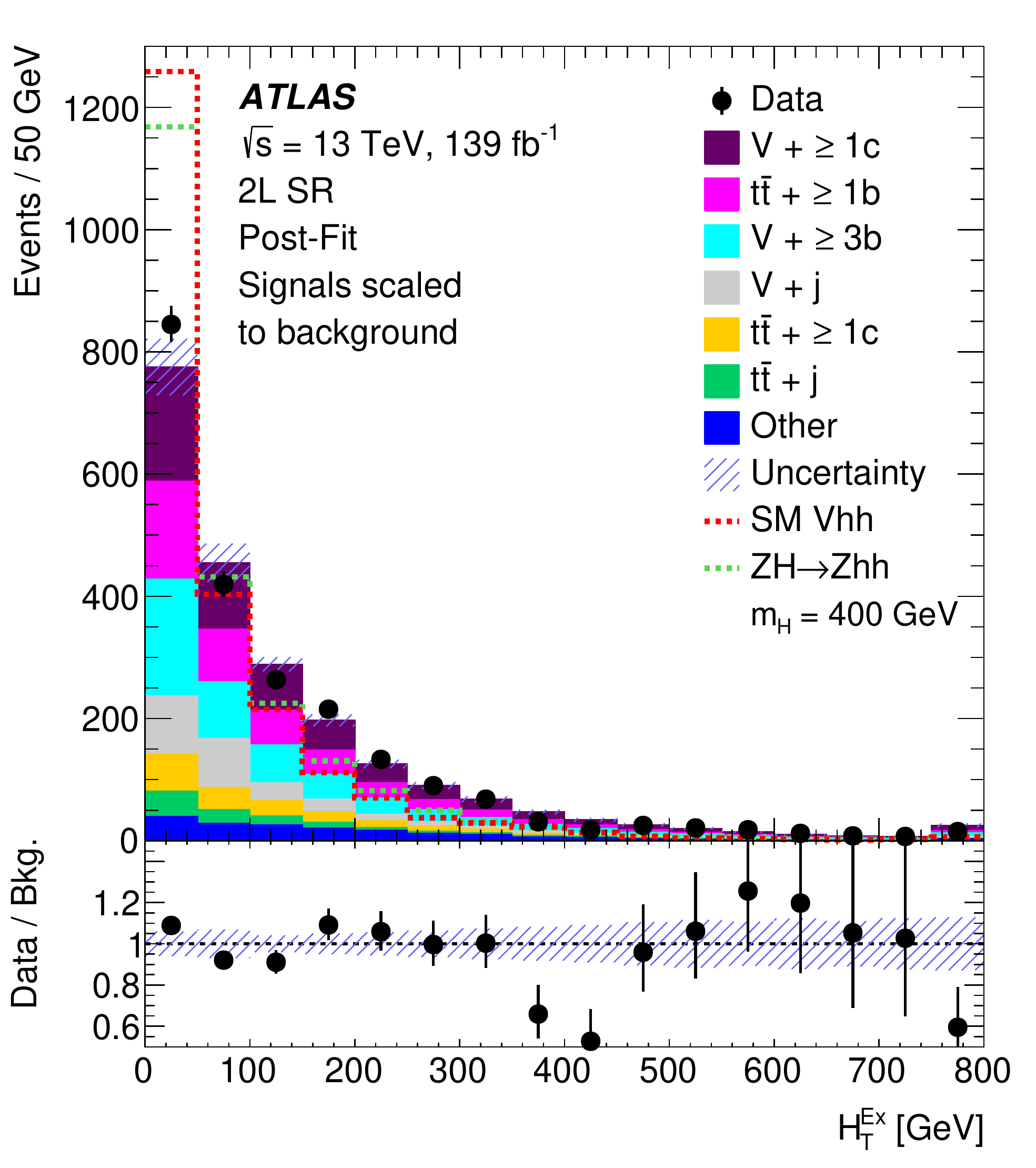}}
\end{center}
\caption{Example distributions of kinematic variables used in the BDT trainings:  the invariant mass \mhh (a, b, c) and the mass difference $m_{h_1}-m_{h_2}$ (d, e, f) of the two Higgs boson candidates, and $\HTex$ (g, h, i) of events passing the 0L (a, d, g),  1L (b, e, h), and 2L (c, f, i) SR selections. The 1L SR combines 1L$+$ and 1L$-$ categories. The expected distributions from example non-resonant and resonant signal models, normalised to the total background expectations, are overlaid. No resonant signal distributions are shown for \mhh, which is used in the BDTs for non-resonant production only. The final bins include overflows.
The bottom panels show the ratios of the data to the total background expectations. The hatched bands represent the combined statistical and systematic uncertainties in the total background predictions.}
\label{fig:BDTVars}
\end{figure}
 
The BDTs are trained in the TMVA~\cite{TMVA} framework. All background processes modelled using simulation are included in the training.
In order to make use of the complete set of simulated MC events
for the BDT training and evaluation in an unbiased way, the MC events are split into two
samples of equal size.
A BDT is trained on each of the two samples and applied to the other, such that the same events are never used for both the BDT training and evaluation.
Similarly, half of the events in data are evaluated with one of the two BDTs, and half are evaluated with the other.
As a result, while significant overtraining is not observed, the effect of any potential overtraining is minimised by following this method.
For each BDT, event weights are included in the training so that the relative importance of each background process is taken into account.
 
For the search for non-resonant $Vhh$ production, the SM, \Chhh-only, and \ChhVV-only signal samples are added together, with equal weights, to form a combined signal sample for the BDT training. The inclusion of the \Chhh and \ChhVV samples in the training improves the sensitivity to their respective couplings. Three different BDTs are trained, one for each leptonic channel.
Little degradation in sensitivity is observed with this strategy relative to training BDTs for each signal sample separately.
 
For the search for resonant $Vhh$ signals, MC samples produced with different values of the resonance mass are combined with equal weights to form the signal sample for the BDT training in each channel.
It results in a BDT that is not strongly correlated with the resonance mass and therefore has good sensitivity to signals with different mass values. This strategy is applied to both the \VH and \AZH searches;  the combination for the latter includes samples with different $m_A$ and $m_H$ values as well as $A$ bosons with narrow and large widths.

% End of text imported from the .//tex/bdt.tex input file

% The next lines are included from the .//tex/signal.tex input file
\subsection{Mass requirements for resonance searches}
\label{sec:signal}

For the resonant \VH and \AZH searches, the Higgs boson pair $hh$ is produced from the decay of a new heavy scalar $H$. Consequently,  the signal \mhh distributions are expected to peak around $m_H$ with a width equal to the natural width of the new scalar convolved with the detector resolution. Therefore, \mhh is a powerful discriminant against the expected continuum SM backgrounds.

Since the lighter Higgs boson $h$ is a narrow resonance with known mass, the \mhh resolution can be improved by constraining the measured masses of the two Higgs boson candidates, $m_{h_1}$ and $m_{h_2}$, to their expected value of 125\,\GeV. This is achieved by scaling the momenta of the $b$-jets from each Higgs boson candidate by the ratio of 125\,\GeV\ to the measured di-$b$-jet mass. Figure~\ref{fig:mhh_ZH} compares the \mhh distributions before and after the rescaling for a few selected $m_H$ values in the 2L channel of the \ZH search. The rescaling improves the relative \mhh resolution from  12.1\% (6.1\%) to 3.5\% (2.6\%) at $m_H=300\,(1000)\,\GeV$. Similar improvements are obtained in the other leptonic channels of both the \VH and \AZH searches. Moreover, in the 2L channel of the \AZH search the rescaling leads to improvement in the mass resolution of $Zhh$ from the decay of a narrow $A$ resonance, as shown in Figure~\ref{fig:mhh_AZH300}.
At $m_H=300\,\GeV$, the relative \Zhh mass resolution improves from 9.4\% (5.4\%) to 2.8\% (2.5\%) at $m_A=400\,(800)\,\GeV$. For LW $A$ bosons, the rescaling has negligible impact on the width of the \Zhh mass distribution. In the following analysis, the rescaled $b$-jet momenta are used solely to calculate the masses of resonance candidates.
 
\FloatBarrier
 
\begin{figure}[!htb]
\begin{center}
\subfigure[]{\includegraphics[width=0.48\textwidth]{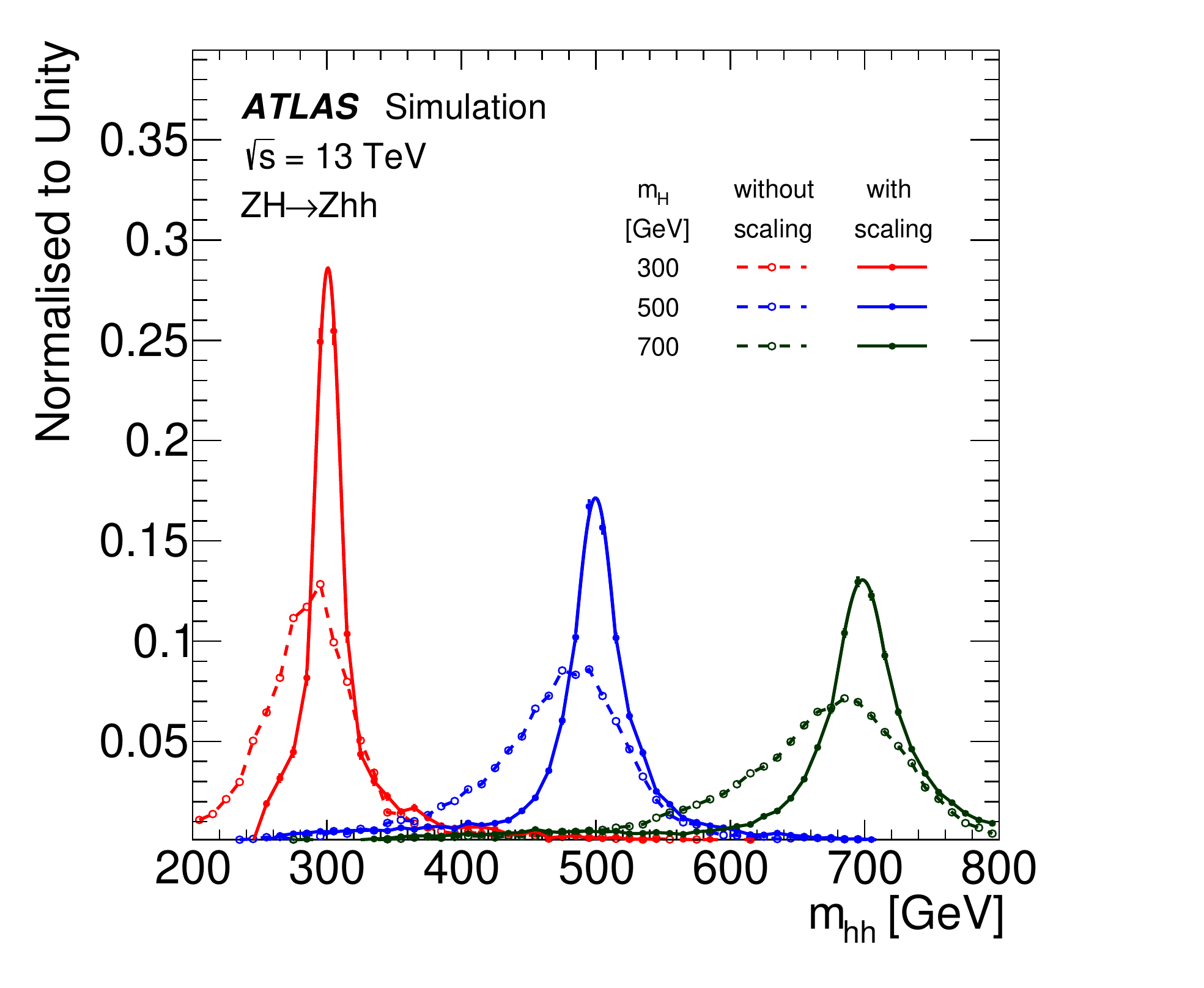}\label{fig:mhh_ZH}}
\subfigure[]{\includegraphics[width=0.48\textwidth]{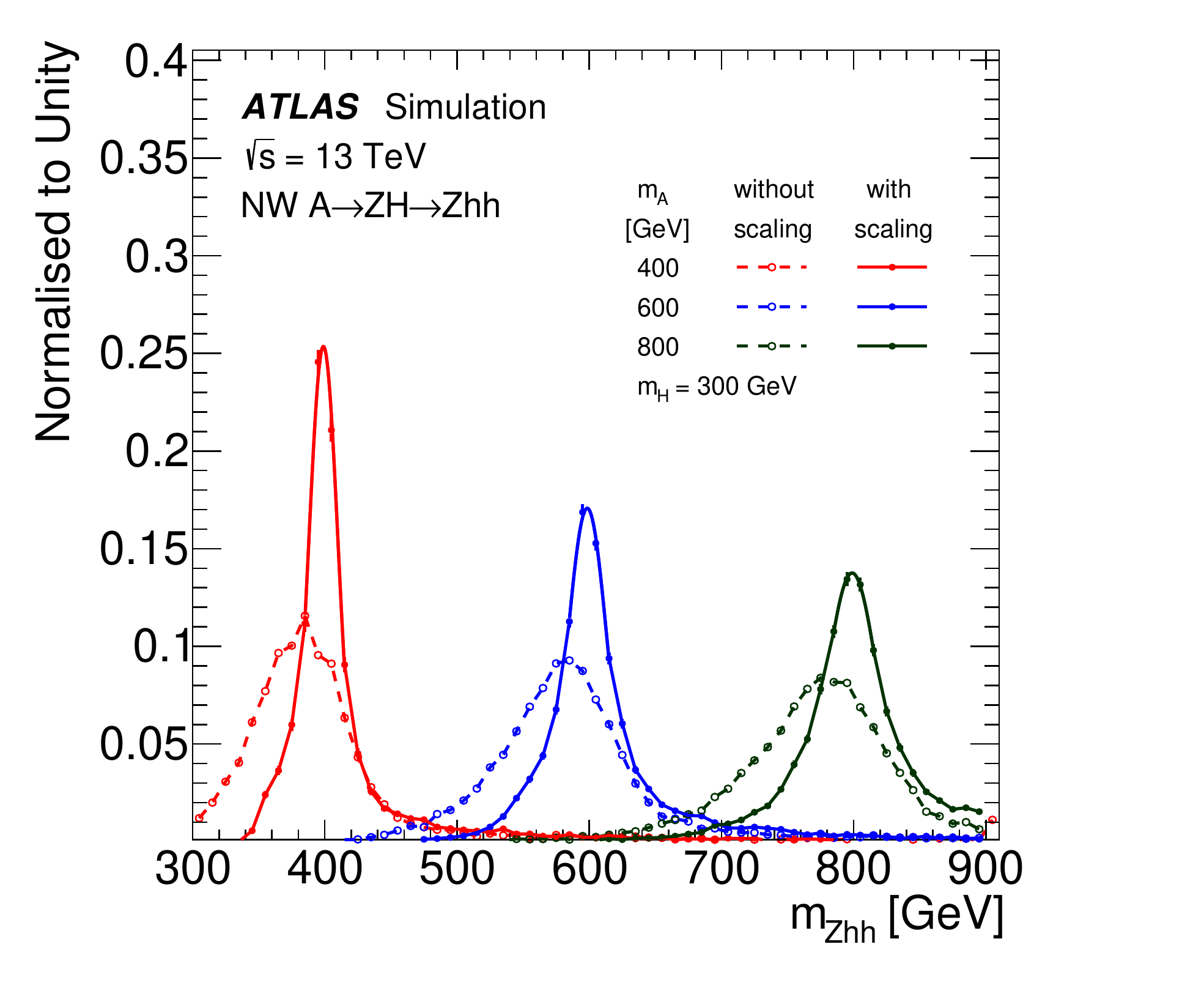}\label{fig:mhh_AZH300}}\\
\end{center}
\caption{Reconstructed invariant mass distributions of (a) the $hh$ system in the \ZH search and (b) the \Zhh system in the \AZH search with a NW $A$ boson for a few representative signal mass points in the 2L channel. Distributions before the $m_h$ rescaling are shown as open circles and dashed lines while those after the rescaling are shown as solid circles and solid lines. All distributions are normalised to unity. A NW $A$ boson has a negligible total decay width compared with the experimental mass resolution.}
\end{figure}

As discussed in Section~\ref{sec:bdt}, the \mhh variable is not used to construct BDTs for the \VH and \AZH searches. Instead, \mhh is required to be in a window around the target $m_H$ value afterwards. The window sizes are optimised to improve the search sensitivities. They vary from 30~GeV at $m_H=260\,\GeV$ to 220~\GeV\ at $m_H=1000\,\GeV$, corresponding to approximately 3--8 times of the expected \mhh resolution after the rescaling.
At high resonance masses the backgrounds are small, so the mass windows are widened (relative to the resolution) to increase signal efficiencies.
The same $m_H$-dependent \mhh windows are used for all channels and for both the \VH and \AZH searches.

The \AZH model has an additional resonance, the  $A$ boson. In the 2L channel, the $A\to ZH\to \ell\ell hh$ decay can be fully reconstructed. For $A$ bosons with a width significantly smaller than the detector mass resolution, i.e.\ the NW case,  the width of the \Zhh mass distribution is dominated by the \mhh resolution as a result of the good lepton momentum resolution and the relatively narrow width of the $Z$ boson. Therefore, the invariant mass of the \Zhh candidate, \mZhh, is required to fall in a window of the same size as the \mhh window for a given $m_H$, but shifted from the $m_H$ value to the targeted $m_A$ value. For $A$ bosons with a width equal to 20\% of the boson mass, i.e.\ the LW case, such requirements reduce the signal efficiencies substantially as the \mZhh distributions are broadened by the $A$ boson width. In this case, $\mZhh>475\,\GeV$ is required only for $m_A\ge 500\,\GeV$ to reduce backgrounds which are present mostly at low \mZhh values. In the 0L channel, the $A\to ZH\to\nu\nu hh$ decay cannot be reconstructed fully, due to the escaping neutrinos. Instead, the transverse mass of the $hh$ and \met system, \mTZhh, is used. However, the \mTZhh distributions are either too broad in the case of LW $A$ bosons or too severely sculpted by the phase-space limitation for NW $A$ bosons with small mass-splittings between the $A$ and $H$ bosons to be effective in discriminating between signal and backgrounds. Thus a requirement on \mTZhh is applied only for NW $A$ bosons with $m_A-m_H\ge 200\,\GeV$. In this case, \mTZhh must be in the window of $[m_A-150,m_A+50]\,\GeV$ for simplicity.
Figure~\ref{fig:Mass} compares the data and expected background distributions of \mhh and \mTZhh in the 0L SR, and those of \mhh and \mZhh in the 2L SR. Expected distributions from an example signal model are overlaid.

\begin{figure}[!htbp]
\begin{center}
\subfigure[]{\includegraphics[width=0.45\textwidth]{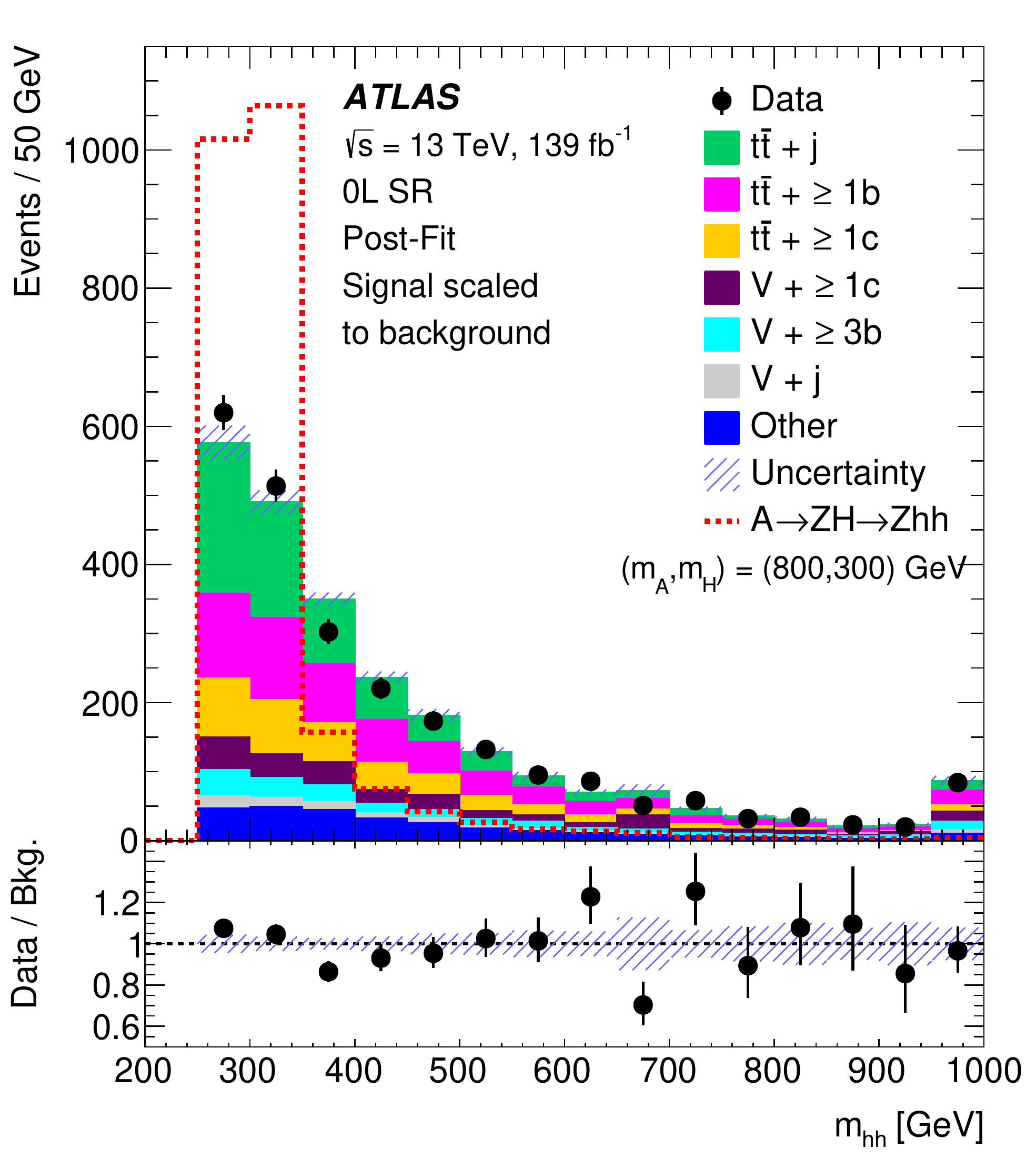}}
\subfigure[]{\includegraphics[width=0.45\textwidth]{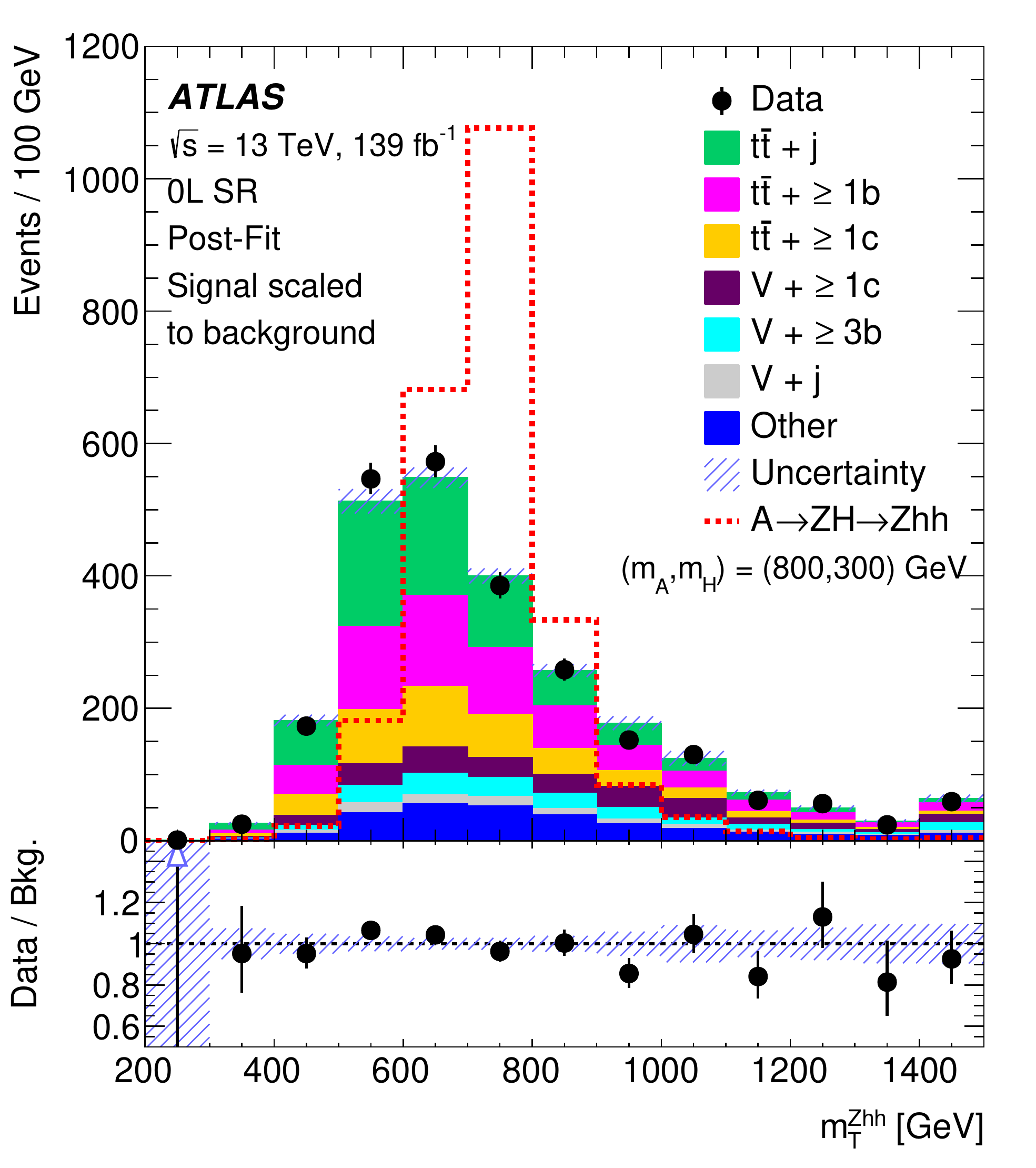}} \\
\subfigure[]{\includegraphics[width=0.45\textwidth]{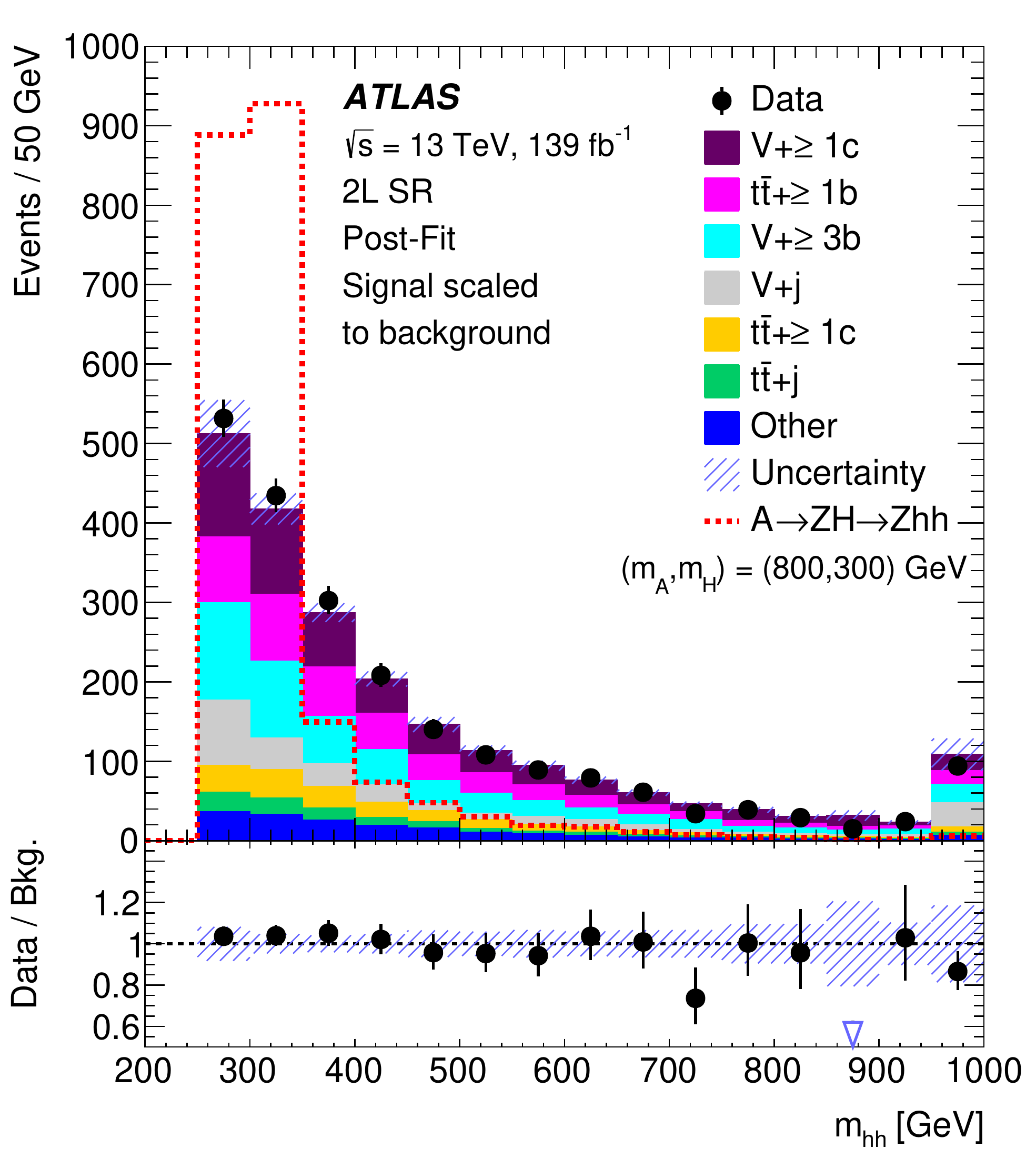}}
\subfigure[]{\includegraphics[width=0.45\textwidth]{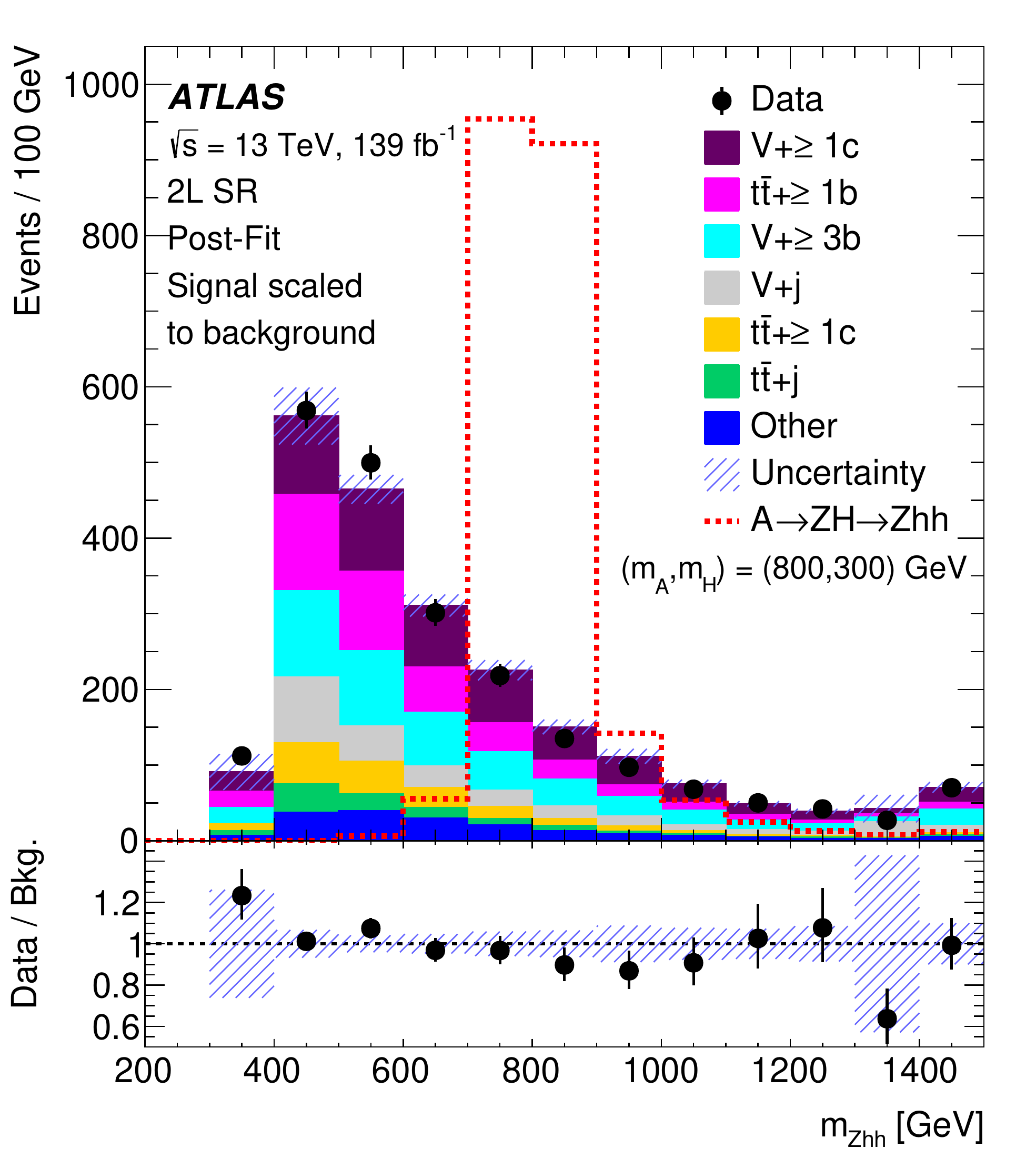}}
\end{center}
\caption{Comparisons between mass distributions of the data and the expected backgrounds:  (a) \mhh and (b) \mTZhh of the 0L SR,  and (c) \mhh and (d) \mZhh of the 2L SR. All masses are calculated after rescaling the measured Higgs boson candidate mass to 125 GeV. The backgrounds are obtained from the background-only fits to the control and signal regions discussed in Section~\ref{sec:results}. Expected distributions from the \AZH signal at $(m_A,m_H)=(800,300)\,\GeV$ for a NW $A$ boson, normalised to the total background expectations, are overlaid. The final bins include overflows.
The bottom panels show the ratios of the data to the total background expectations. The hatched bands represent the combined statistical and systematic uncertainties in the total background predictions.}
\label{fig:Mass}
\end{figure}

% End of text imported from the .//tex/signal.tex input file

% The next lines are included from the .//tex/systematics.tex input file
\section{Systematic uncertainties}
\label{sec:systematics}
 
Systematic uncertainties are divided into four categories: experimental uncertainties, theoretical uncertainties of the overall background normalisations, theoretical uncertainties of acceptances and BDT shapes, and data-driven background modelling uncertainties.
 
\subsection{Experimental systematic uncertainties}
\label{sec:sys-exp}

Jet energy scale (JES) and jet energy resolution (JER) uncertainties comprise the largest group of experimental uncertainties.  The JES uncertainties are primarily determined using data-based $Z$-boson--jet, photon--jet, and multi-jet \pT-balance techniques~\cite{JETM-2018-05}. Additional uncertainties are applied for the energy scale of jets containing \bquarks.  The impact of the JES uncertainties is estimated by scaling the jet energies within their uncertainties.  JER uncertainties are also determined from in situ measurements of $Z$-boson--jet , photon--jet,  and dijet \pT balance~\cite{JETM-2018-05}.  The effect of the JER uncertainties is calculated by increasing the resolution within its uncertainties, smearing the jet energy by the resulting change in resolution, and comparing the result with the nominal shape and normalisation in simulation. 
 
Subdominant experimental uncertainties originate from the $b$-tagging correction factors. The $b$-tagging correction factors, determined from the difference between the efficiencies measured in data and simulation, are evaluated in five DL1r discriminant bins and are derived separately for $b$-jets, $c$-jets, and light-flavour jets~\autocite{FTAG-2018-01, ATLAS-CONF-2018-001, ATLAS-CONF-2018-006}.  All of the correction factors for the three jet flavours have uncertainties estimated from multiple measurements, which are decomposed into uncorrelated components that are then treated independently. 
Extra uncertainties are applied when `truth'-tagging is used (in the highest BDT-score bins for subleading backgrounds), and these are decorrelated between channels and defined as an overall 30\% uncertainty in the yield for these backgrounds.
The `truth'-tagging uncertainties are designed to cover any differences in BDT shapes between background samples when `truth'-tagging is or is not applied.
 
Uncertainties in the reconstruction, identification, isolation, and trigger efficiencies of electrons~\cite{PERF-2017-01} and muons~\autocite{PERF-2015-10} are considered, along with the uncertainty in their energy scale and resolution. These are found to have only a small impact on the results.
The uncertainties in the energy scale and resolution of the jets and leptons are propagated to the calculation of \met, which also has additional uncertainties from the modelling
of the underlying event and the momentum scale, momentum resolution,
and reconstruction efficiency of the tracks used to compute the soft-term (see Section~\ref{sec:objects})~\autocite{PERF-2016-07, ATLAS-CONF-2018-023}.
An uncertainty is assigned to the \met\ trigger correction factors in the 0L channel, defined as the entire difference between the trigger efficiencies for data and simulated events.
The uncertainty in the combined 2015--2018 integrated luminosity is 1.7\%, as described in Section~\ref{sec:samples}.
The average number of interactions per bunch crossing in the simulation is rescaled by 1.03 to improve agreement between the data and the simulation, based on a measurement of the visible cross-section in minimum-bias events~\autocite{STDM-2015-05}, and an uncertainty, as large as the correction, is included.

\subsection{Background normalisation systematic uncertainties}
\label{sec:sys-norm}
 
For all background processes, uncertainties are included in the overall normalisations following a similar approach taken for the ATLAS $t\bar{t}h(\to bb)$ measurement~\cite{HIGG-2020-23}.
These uncertainties are correlated across the three leptonic channels.
The \ttbarB and \VjetsB flavour components are free to float in the fit.
The \ttbarC and \VjetsC flavour components are assigned a large uncertainty, 100\%,
intended to be conservative in case there is significant mismodelling of these processes.
These uncertainties are always constrained by the fits.
The light-flavour \ttbarL and \VjetsL components are assigned a smaller uncertainty, 10\%,
as these are better-measured processes.
In practice, these overall \ttbarL and \VjetsL uncertainties have little impact on the analysis, since the yields from these backgrounds,
particularly at high BDT score, are small.
These uncertainties are all constrained by the \pcbs distributions in \ttbarCR and \VjetsCR, as well as by the low BDT-score regions of the SRs.
 
The above \Vjets uncertainties are correlated between \Gjets, \Zjets, and \Wjets backgrounds.
Extra uncertainties are defined for extrapolations between these backgrounds to account for their kinematic differences.
Each flavour component of \Zjets is assigned an extra  20\% uncertainty, and each component of \Wjets is assigned an extra 30\% uncertainty.
The sizes of the uncertainties are defined so as to cover flavour-composition differences between the \Gjets, \Zjets, and \Wjets backgrounds.
 
The remaining normalisation uncertainties are subdominant in the analysis.
A 100\% uncertainty is applied to the multi-jet backgrounds in both the 1L SRs and \VjetsCR.
The 1L SR multi-jet uncertainty has a small impact due to the size of the background, and
the \VjetsCR multi-jet background is constrained by the fits to have an uncertainty smaller than 100\%.
Uncertainties of 25\% are applied to each of the remaining small SM backgrounds.
Given the small relative yields from these backgrounds, their uncertainties have little impact on the analysis and are thus chosen to be conservative.

\FloatBarrier
 
\subsection{Data-driven background modelling uncertainties}
A set of extra data-driven background uncertainties is included.
The relative differences of the BDT distributions between data and simulation in the validation region define a set of VR non-closure uncertainties, which are determined prior to any fit.
In each of the 0L, 1L (1L$+$ and 1L$-$), and 2L regions, two uncertainties are defined: one for normalisation and one for shape (based on the bin-by-bin differences), each of which is correlated between all backgrounds in the channel.
In practice, each signal region tends to be dominated by a single background (either \ttbar or \Vjets), such that correlating these uncertainties between backgrounds
makes little difference compared to decorrelating the uncertainties between backgrounds.
In \ttbarCR, only the normalisation component of the non-closure is considered, since the individual normalisation factors for each \ttbar flavour component
can cover any shape difference between data and simulation.
In \VjetsCR, no non-closure uncertainty is included, as the normalisation component is redundant because of the multi-jet normalisation uncertainty, which
by design covers any non-closure between data and simulation.
In total, this defines nine components of the validation region non-closure uncertainty, each of which is taken as 100\% of the relative difference in normalisation or shape between data and simulation in the validation regions.
These uncertainties are designed to be conservative, and in practice they tend to be constrained by the fits.
 
For the multi-jet backgrounds in \VjetsCR and the 1L SRs, shape uncertainties are included in addition to the normalisation uncertainties defined in Section~\ref{sec:sys-norm}.
The shape uncertainty included in the \VjetsCR multi-jet template is obtained by increasing or decreasing the \Gjets contamination by 100\% in this region.
Two shape uncertainties are included in the multi-jet template in the 1L SRs: one by increasing or decreasing the prompt-lepton contamination by 30\% when evaluating the template, to account for potential mismodelling of the lepton selection inefficiency, and the other by raising or lowering the \mtW threshold used to determine this template, to account for the potential impact of the selection on the extracted BDT shapes.
 
\subsection{Theoretical uncertainties in BDT shapes and acceptances}
 
In addition to the normalisation uncertainties defined in Section~\ref{sec:sys-norm},
a number of theoretical uncertainties in background and signal modelling are included.
These uncertainties tend to be subdominant relative to the background normalisation and data-driven uncertainties, which are
in turn subdominant relative to the statistical uncertainty of the data.
Each uncertainty is correlated across all analysis regions, and the effects of each uncertainty on the acceptance and BDT shape are correlated.
 
For each component of the \ttbar and \Vjets backgrounds and the single-top-quark background, scale uncertainties are included, defined~\cite{Bothmann:2016nao} using seven variations of the QCD factorisation and renormalisation scales in the matrix elements by factors of 0.5 and 2, avoiding variations in opposite directions.
For each component of the \ttbar background and the single-top-quark background, an extra parton-shower uncertainty is defined by comparing the nominal sample with an alternative sample showered with \Herwig[7.04]~\cite{Bahr:2008pv,Bellm:2015jjp},
and an extra matching uncertainty is defined by comparison with an alternative sample produced with \MGNLO.
For each \ttbar component, initial-state radiation (ISR) and final-state radiation (FSR) variations are included, following the procedure described in Ref.~\cite{ATL-PHYS-PUB-2020-023}.
For the $Wt$ component of the single-top-quark background, a comparison between the nominal sample, which uses the diagram removal scheme~\cite{Frixione:2008yi}, and an alternative sample, which uses the diagram subtraction scheme~\cite{Frixione:2007vw}, defines an extra uncertainty.
For \Vjets, variations of the CKKW parameter for merging/matching the matrix element with the parton shower are included, as are variations of the resummation scale.
 
Modelling of the \ttbar background is improved by correcting the top-quark \pT distribution to that predicted by calculations of top-quark-pair differential distributions at NNLO QCD and NLO EW accuracy~\cite{ttbarnnlo}.
Previous studies have seen improved agreement between data and prediction in \ttbar events, particularly for the top-quark \pT distribution, when comparing the data with the NNLO calculations~\cite{TOPQ-2015-06}.
For each \ttbar component, the change in the BDT shape from the NNLO correction is taken as an uncertainty.
 
For the signal samples, acceptance uncertainties evaluated at the generator level are included to account for scale variations, PDF variations, and variations of the ISR, FSR, multi-parton interaction (MPI), and colour reconnection parameters~\cite{ATL-PHYS-PUB-2014-021}.
For the non-resonant \Zhh signal only, an extra uncertainty is included to cover potential mismodelling of $gg\rightarrow Zhh$.
An uncertainty is included in the $Zhh$ normalisation and BDT shape, covering the difference between the $qq\rightarrow Zhh$ process alone and the sum of the $gg\rightarrow Zhh$ and $qq\rightarrow Zhh$ processes (the effects on the normalisation and shape are correlated).
No other uncertainties are included in the overall cross-section for any signal sample.

\subsection{Impact of systematic uncertainties}
 
The effects of the statistical and systematic uncertainties on the search sensitivities are studied for hypothesised signals following the procedure discussed in Section~\ref{sec:results}. Table~\ref{tab:systematics} lists the leading sources of uncertainty and shows, for a few selected signal models, the expected relative uncertainties in the fitted signal-strength parameter $\mu$, a factor multiplying the predicted cross-section for the hypothesised signal.
In the resonant case, the mass values are chosen to be those with the largest excesses observed in the data as discussed in Section~\ref{sec:results}. The cross-sections used for signal normalisation correspond approximately to their expected upper limits. In each case, the leading source of uncertainty is the statistical uncertainty of the data.

\begin{table}[!htb]
\begin{center}
\caption{Dominant uncertainties in the best-fit signal-strength parameter $\mu$ for hypothesised signals.  For this study, the resonance masses are chosen to be those with the largest excesses in the data for illustration: $m_H=315\,\GeV$ for \WH, $m_H=550\,\GeV$ for \ZH, $(m_A,m_H)=(790,300)\,\GeV$ for NW \AZH, and $(m_A,m_H)=(420,320)\,\GeV$ for LW \AZH.  The hypothesised signals are normalised to reference cross-sections approximately equal to their expected upper limits.}\vspace*{-0.2cm}
\label{tab:systematics}
{\footnotesize
\begin{tabular}{l r@{,\ssp}l r@{,\ssp}lc  r@{,\ssp}lc r@{,\ssp}l r@{,\ssp}l}
\toprule
Model             & \multicolumn{2}{c}{\Vhh like in SM}  & \multicolumn{3}{c}{\WH}  & \multicolumn{3}{c}{\ZH}  & \multicolumn{2}{l}{NW \AZH}  & \multicolumn{2}{l}{LW \AZH}  \\ \midrule
\ \ \ Systematic uncertainty source & \multicolumn{12}{c}{ $\Delta\mu/\mu$ [\%] } \\ \midrule
\hsp Background modelling  & $+20$ & $-15$  &  $+14$  & $-11$   &&  $+4.7$ & $-3.0$  &&  $+17$  & $-13$  &  $+20$ & $-18$  \\
\hsp MC statistics         & $+12$ & $-9.1$ &  $+13$  & $-7.8$  &&  $+4.8$ & $-2.2$  &&  $+7.2$ & $-4.1$ &  $+10$ & $-8.3$ \\
\hsp Objects               & $+12$ & $-8.6$ &  $+8.0$ & $-5.2$  &&  $+4.5$ & $-2.2$  &&  $+19$  & $-11$  &  $+16$ & $-12$  \\
\hsp Signal modelling      & $+10$ & $-4.7$ &  $+12$  &  $-4.9$ &&  $+8.6$ & $-3.0$  &&  $+14$  & $-5.1$ &  $+17$ & $-7.6$ \\
\hsp VR non-closure        & $+14$ & $ -11$ &  $+11$  & $ -9.4$ &&  $+4.4$ & $- 3.0$ &&  $+4.9$ & $-3.7$ &  $+12$ & $-10$  \\ \midrule
\ \ \ Total systematic uncertainty     & $+30$ & $ -22$ &  $+27$  & $ -18$  &&  $+12$  & $-5.8$  &&  $+30$  & $-18$  &  $+33$ & $-24$ \\
\ \ \ Statistical uncertainty         & $+44$ & $-39$  &  $+52$  & $ -43$  &&  $+68$  & $-49$   &&  $+59$  & $-47$  &  $+42$ & $-37$ \\
Total                      & $+52$ & $ -44$ &  $+59$  & $ -47$  &&  $+69$  & $-49$   &&  $+66$  & $-50$  &  $+53$ & $-45$ \\
\toprule
\end{tabular}}
\end{center}
\end{table}

\FloatBarrier

% End of text imported from the .//tex/systematics.tex input file

% The next lines are included from the .//tex/results.tex input file
\section{Results and interpretations}
\label{sec:results}

Potential signal contributions in the data are determined through maximum-likelihood fits to the BDT distributions in the SRs and the \pcbs distributions in \ttbarCR and \VjetsCR. The procedure is based on the framework described in Refs.~\cite{HIGG-2012-17, Moneta:2010pm, Verkerke:2003ir}. A profile-likelihood-ratio test statistic is used to test the signal-plus-background hypothesis, with the signal production rate as the parameter-of-interest. All SRs (0L, 1L, and 2L) are included in the fits for the non-resonant \Vhh search and the resonant \VH search, while only the 0L and 2L SRs are included for the resonant \AZH search.
The BDT distributions are divided into four bins, determined through optimisations of signal sensitivities while maintaining a reasonable number of background MC events in each bin. The binning boundaries are optimised for each channel separately, but are kept the same for different signal models for simplicity.

The \ttbar and \Vjets backgrounds in the SRs from MC simulation are decomposed into three jet-flavour categories in the same way as those in the \ttbar and \Vjets CRs discussed in Section~\ref{sec:backgrounds}. These flavour-dependent contributions share the same normalisation factors~(NFs) as their corresponding components in the CRs. In the fits, the NFs are unconstrained for the \ttbarB and \VjetsB components and are constrained to unity within their estimated uncertainties for other components.
Systematic uncertainties, described in Section~\ref{sec:systematics}, are incorporated as additional multiplicative terms, parameterised with nuisance parameters, in the likelihood calculations, where each nuisance parameter is given a prior distribution based on individual studies.

To test the overall compatibility of the data with the background expectations, fits are first performed for the background-only hypothesis.
Since BDTs are model dependent, a fit to the BDT and \pcbs distributions for the non-resonant search is chosen to illustrate the background modelling. Figure~\ref{fig:BDTVars} compares the data with the background expectations from the fit for a few selected kinematic variables used in training the BDTs.  Overall, the post-fit backgrounds are found to reproduce the data well.
The fitted NF values of the heavy-flavour components of the \ttbar and \Vjets backgrounds are listed in Table~\ref{tab:NFs}, along with those from fits to the different signal-plus-background hypotheses discussed below.
 
The fits are repeated for the signal-plus-background hypotheses for each signal model and resonance mass assumption. The NF values shown in Table~\ref{tab:NFs} from different fits are stable and consistent.  Upper limits on signal production cross-sections are calculated with the CL$_\mathrm{s}$ method~\cite{Read:2002hq}, using the $\tilde{q}_\mu$ test statistic in the asymptotic approximation~\cite{Cowan:2010js}.
 
Global significances are calculated following the procedure given in Ref.~\cite{globalsig}.
Pseudo-experiments are generated for the background-only hypothesis.
For each pseudo-experiment, a scan over all considered mass points is performed to determine the largest local significance for that pseudo-experiment.
The fraction of pseudo-experiments with a local significance greater than the maximum local significance found in the scan over the real data defines
the global $p$-value, which in turn defines the global significance.

\begin{table}[!htb]
\begin{center}
\caption{Values of the normalisation factors of the heavy-flavour background components from the simultaneous fits of the signal and control regions to the background-only and different signal-plus-background hypotheses.
For this study, the resonance masses with the largest excesses in the data are chosen: $m_H=315\,\GeV$ for \WH, $m_H=550\,\GeV$ for \ZH, $(m_A,m_H)=(790,300)\,\GeV$ for NW \AZH, and $(m_A,m_H)=(420,320)\,\GeV$ for LW \AZH.}
\label{tab:NFs}
{\footnotesize
\begin{tabular}{cccccccccc}\toprule
Background   & Normalisation & Bkg-only    & \Vhh & \multicolumn{2}{c}{\VH}  & & \multicolumn{2}{c}{\AZH} \\ \cline{5-6}\cline{8-9} \\[-3mm]
source  & constraint   & fit &  like in SM    & \WH  & \ZH & & NW  & LW \\ \midrule \\[-2mm]
\ttbarB        & free-float  & $1.21^{+0.19}_{-0.16}$ & $1.21^{+0.19}_{-0.17}$ & $1.20^{+0.21}_{-0.18}$  & $1.18^{+0.22}_{-0.18}$ & & $1.29^{+0.30}_{-0.23}$ & $1.27^{+0.26}_{-0.22}$ \\[2mm]
\ttbarC        & $\pm 100\%$ & $1.54^{+0.62}_{-0.54}$ & $1.51^{+0.64}_{-0.53}$ & $1.56^{+0.68}_{-0.57}$  & $1.44^{+0.61}_{-0.49}$ & & $1.65^{+0.70}_{-0.63}$ & $1.85^{+0.57}_{-0.74}$ \\[2mm]
 
\VjetsB        & free-float  & $0.98^{+0.28}_{-0.30}$ & $0.86^{+0.33}_{-0.30}$ & $0.83^{+0.35}_{-0.29}$  & $0.90^{+0.31}_{-0.30}$ & & $0.76^{+0.38}_{-0.30}$ & $0.72^{+0.38}_{-0.28}$ \\[2mm]
\VjetsC        & $\pm 100\%$ & $1.29^{+0.55}_{-0.40}$ & $1.17^{+0.60}_{-0.35}$ & $1.09^{+0.61}_{-0.33}$  & $1.24^{+0.59}_{-0.39}$ & & $1.01^{+0.63}_{-0.31}$ & $0.97^{+0.60}_{-0.30}$ \\[2mm]
\toprule
\end{tabular}}
\end{center}
\end{table}

\subsection{Search for non-resonant $Vhh$ production}
The BDTs for non-resonant production are used to search for \Vhh signals from three production scenarios: `SM', \Chhh, and \ChhVV. The `SM' scenario assumes SM kinematics but with its cross-section scaled by a signal-strength parameter $\mu$. The \Chhh scenario tests for an anomalous tri-linear $hhh$ coupling, assuming SM couplings for the rest. Similarly, the \ChhVV scenario tests for an anomalous quartic $hhVV$ coupling. For the \Chhh and \ChhVV scenarios, both the event kinematics and production cross-section depend on their respective coupling modifier.

Constraints on non-resonant \Vhh production are obtained through fits to the signal-plus-background hypotheses described above, assuming the SM value for the $h\to bb$ decay branching ratio~\cite{deFlorian:2016spz}. The three non-resonant scenarios have the same data BDT distribution, but differ in the signal BDT distributions. The background BDT distributions are largely the same, barring small variations in the post-fit background contributions.
For `SM' \Vhh production, a 95\% confidence-level (CL) upper limit of \VhhObs\, on $\mu$ is observed compared with \VhhExpErr\, expected. The corresponding post-fit BDT distribution is shown in Figure~\ref{fig:SMBDT}.
For the \Chhh and \ChhVV scenarios, \Vhh cross-section upper limits are derived from the fits for different values of the coupling modifiers. These limits lead to the observed (expected) 95\% CL intervals of $\ChhhLObs<\Chhh<\ChhhUObs$ ($\ChhhLExp<\Chhh<\ChhhUExp$) and $\ChhVVLObs<\ChhVV<\ChhVVUObs$ ($\ChhVVLExp<\ChhVV<\ChhVVUExp$) for the two coupling modifiers.
The observed bounds are weaker than the expectations largely because of small excesses of data in the highest BDT bins. 
These are the first limits derived from the \Vhh process, and are considerably weaker than those obtained from the $hh$ searches focused on the ggF and VBF processes~\cite{HDBS-2018-58,CMS-HIG-20-005,CMS-HIG-17-030,HDBS-2018-18,HDBS-2018-34,CMS-HIG-19-018}.
In addition, this analysis can search for deviations of the $WWhh$ and $ZZhh$ couplings from their SM values, parameterised by the respective coupling modifiers \ChhWW and \ChhZZ (in the SM, $\ChhWW = 1$, and $\ChhZZ = 1$). Cross-section upper limits are derived from the fits for different values of these coupling modifiers, leading to limits for the observed (expected) 95\% CL intervals of $\ChhWWLObs<\ChhWW<\ChhWWUObs$ ($\ChhWWLExp<\ChhWW<\ChhWWUExp$) and $\ChhZZLObs<\ChhZZ<\ChhZZUObs$ ($\ChhZZLExp<\ChhZZ<\ChhZZUExp$) for the two coupling modifiers. Higgs boson couplings other than the one being tested are set to their SM values.

\begin{figure}[!htb]
\begin{center}
\subfigure[]{\includegraphics[width=0.7\textwidth]{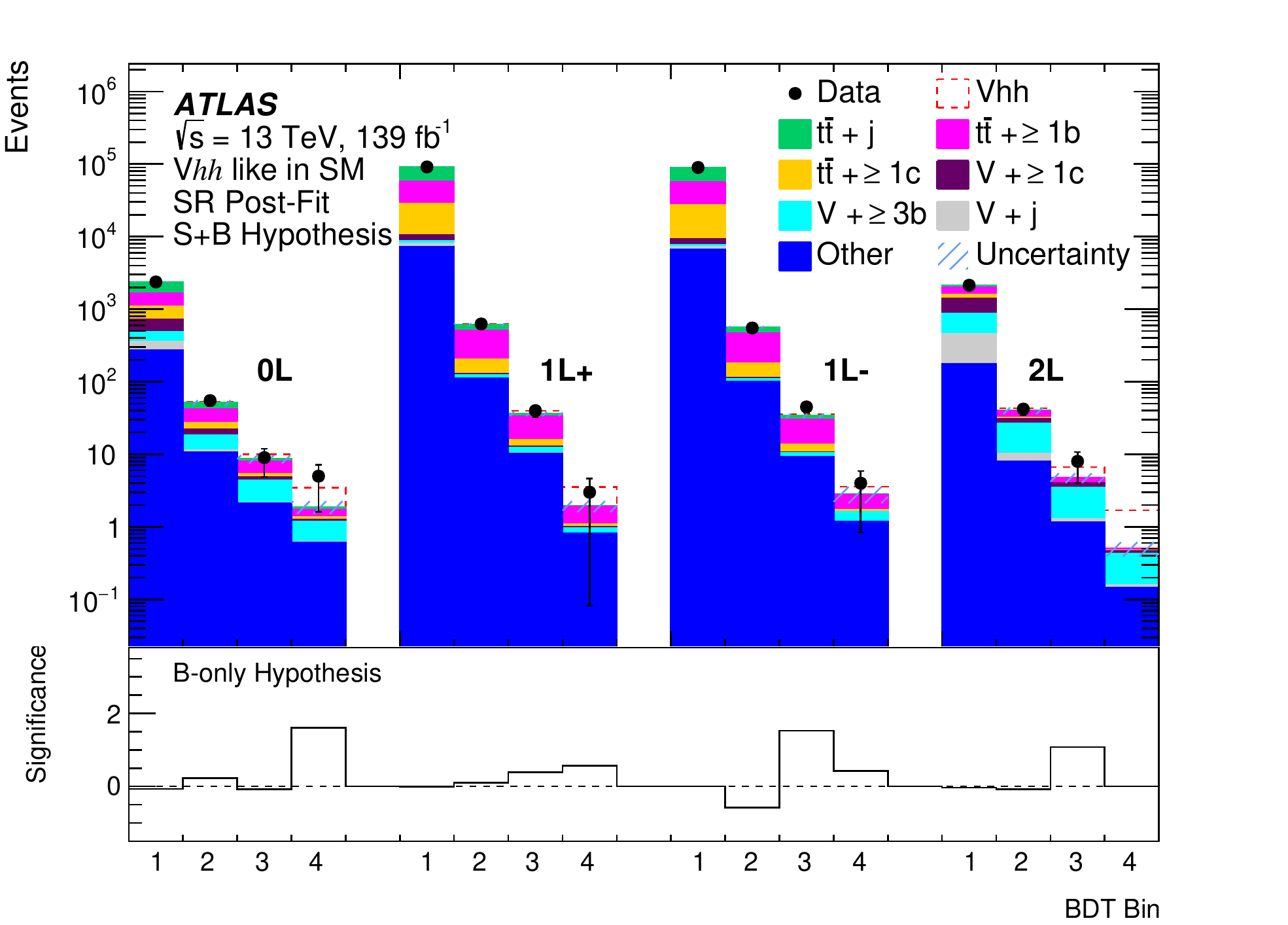}\label{fig:SMBDT}} \\
\subfigure[]{\includegraphics[width=0.48\textwidth]{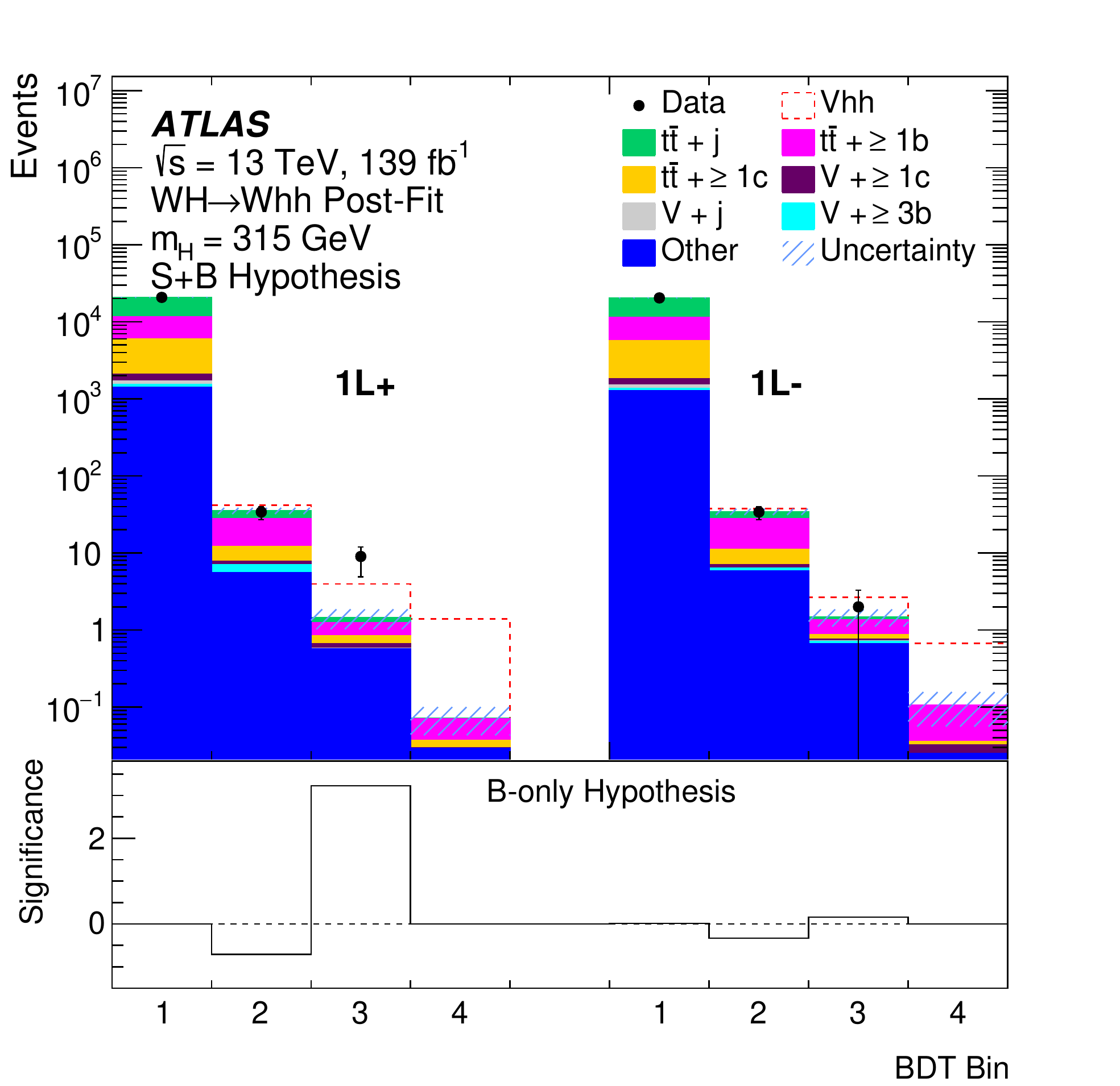}\label{fig:WHBDT}}
\subfigure[]{\includegraphics[width=0.48\textwidth]{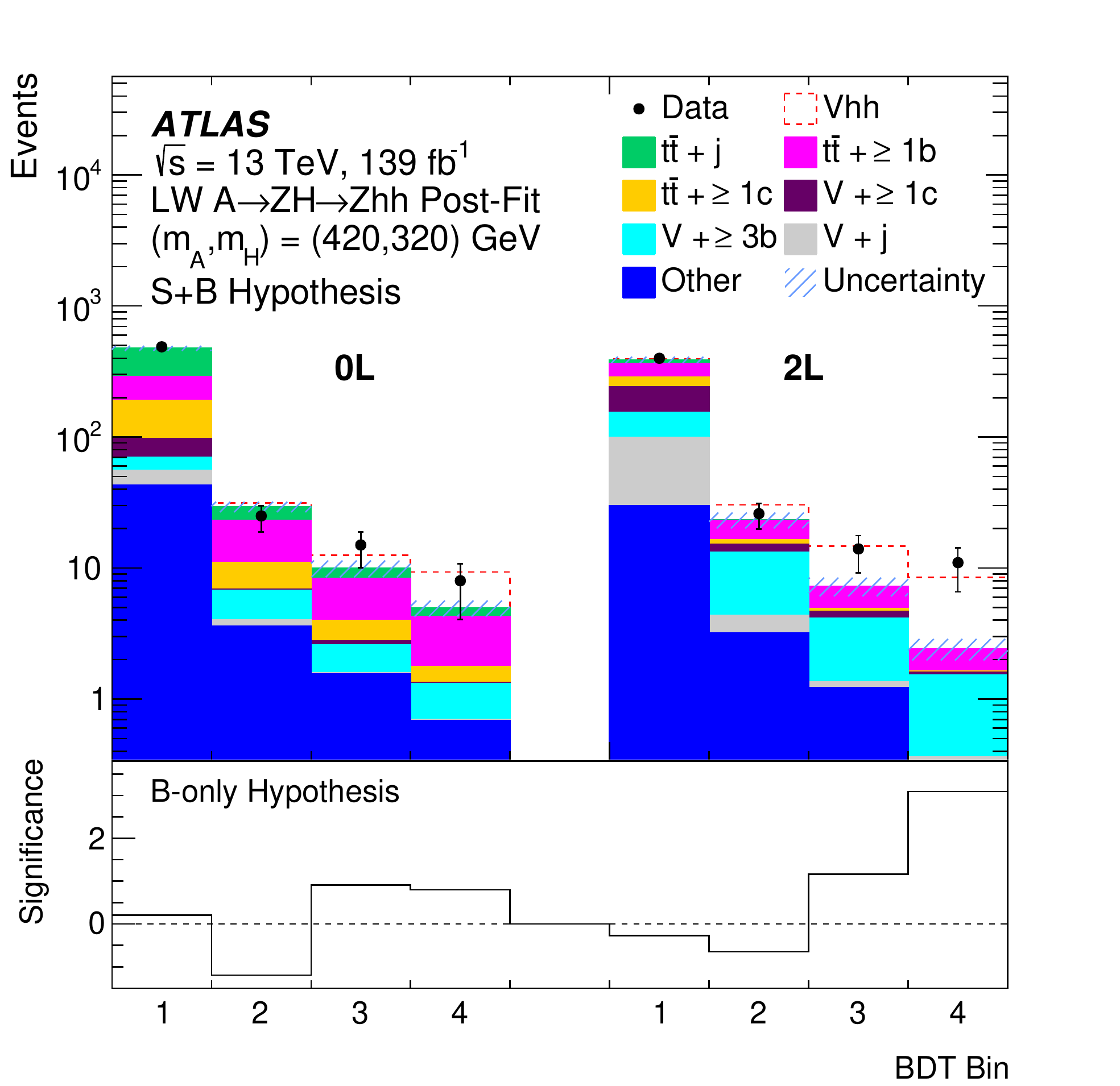}\label{fig:AZHLWBDT}}
\end{center}
\caption{The post-fit BDT distributions of the signal-plus-background hypotheses from the searches for (a) non-resonant \Vhh production like in the SM, (b) the resonant \WH process at $m_H=315\,\GeV$, and (c) the resonant \AZH process at $(m_A,m_H)=(420,320)\,\GeV$ for a LW $A$ boson.
The chosen mass points in (b) and (c) correspond to the most significant excesses of data over the background expectations in their respective search. The fitted signal and background contributions are shown as stacked histograms. Bins without data points have zero observed events. The hatched bands represent the combined statistical and systematic uncertainties in the total background predictions. The bottom panels show the observed bin-by-bin significances for the background-only hypotheses, calculated following the prescription of Ref.~\cite{ATL-PHYS-PUB-2020-025}. The LW $A$ boson has a total decay width equal to 20\% of its mass.
}
\end{figure}

\subsection{Searches for $VH\to Vhh$ production}
Constraints on the production of a heavy narrow scalar resonance $H$ in association with a $V$ boson are determined through the fits of the BDT distributions for resonant \VH production to the signal-plus-background hypothesis in 5~\GeV\ $m_H$ steps. The step size is chosen to be comparable to, or smaller than, the experimental \mhh mass resolution. For each tested $m_H$ value, the BDT distributions are obtained after imposing the \mhh mass window requirement discussed in Section~\ref{sec:signal}. For $m_H$ values with no corresponding MC signal sample, the BDT distributions are linearly interpolated from those of the two closest neighbouring mass points with MC signal samples.
To validate this interpolation, the results of fits performed at $m_H$ points with a MC signal sample are compared with the results obtained when the BDT distribution is an interpolation between those of neighbouring points with MC signal samples.
 
Fits are performed separately for the $WH\to \ell\nu hh$ and $ZH\to (\ell\ell/\nu\nu)hh$ searches. All three channels are included in the fits for both searches. The \WH signal contributes mostly to the 1L channel, but with a sizeable contribution in the 0L channel due to an inefficiency in lepton identification. Its contribution to the 2L channel is negligible. Inclusion of the 2L channel in the fit effectively makes the channel an additional CR for the $WH\to\ell\nu\, hh$ search, further constraining the \Zjets background. Similarly, the \ZH signal contributes mostly to the 0L and 2L channels. Inclusion of the 1L channel helps to constrain the \Wjets background.

The data are found to be consistent with the estimated background contributions. The largest upward deviations from the background expectations are at $m_H=315\,\GeV$ with a local (global) significance of $\WHSigLoc\, (\WHSigGlo)$ standard deviations ($\sigma$) in the \WH search and at $m_H=550\,\GeV$ with a local (global) significance of $\ZHSigLoc\sigma~(\ZHSigGlo\sigma)$ in the \ZH search. These small excesses are largely correlated with those observed in the non-resonant \Vhh search.
The post-fit BDT distribution in the \WH search at $m_H=315\,\GeV$ is illustrated in Figure~\ref{fig:WHBDT}. The heavy-flavour background NFs from the fits are compared  with those from other fits in Table~\ref{tab:NFs}.
 
The observed and expected 95\% CL upper limits on the cross-section $\sigma(VH)\times B(H\to hh\to bbbb)$ as a function of $m_H$ are shown in Figure~\ref{fig:VHLimits}.
The resonant $VH$ search is sensitive to the  $HWW$ and $HZZ$ couplings separately. Compared with the search in the VBF channel~\cite{HDBS-2018-18}, which is sensitive to the combination of the two couplings, the $VH$ search has better sensitivity for $m_H$ up to $\sim 450\,\GeV$, assuming that SU(2) custodial symmetry (like that in the SM) applies to the $HWW$ and $HZZ$ couplings~\cite{Sikivie:1980hm}.

\begin{figure}[!htb]
\begin{center}
\subfigure[]{\includegraphics[width=0.48\textwidth]{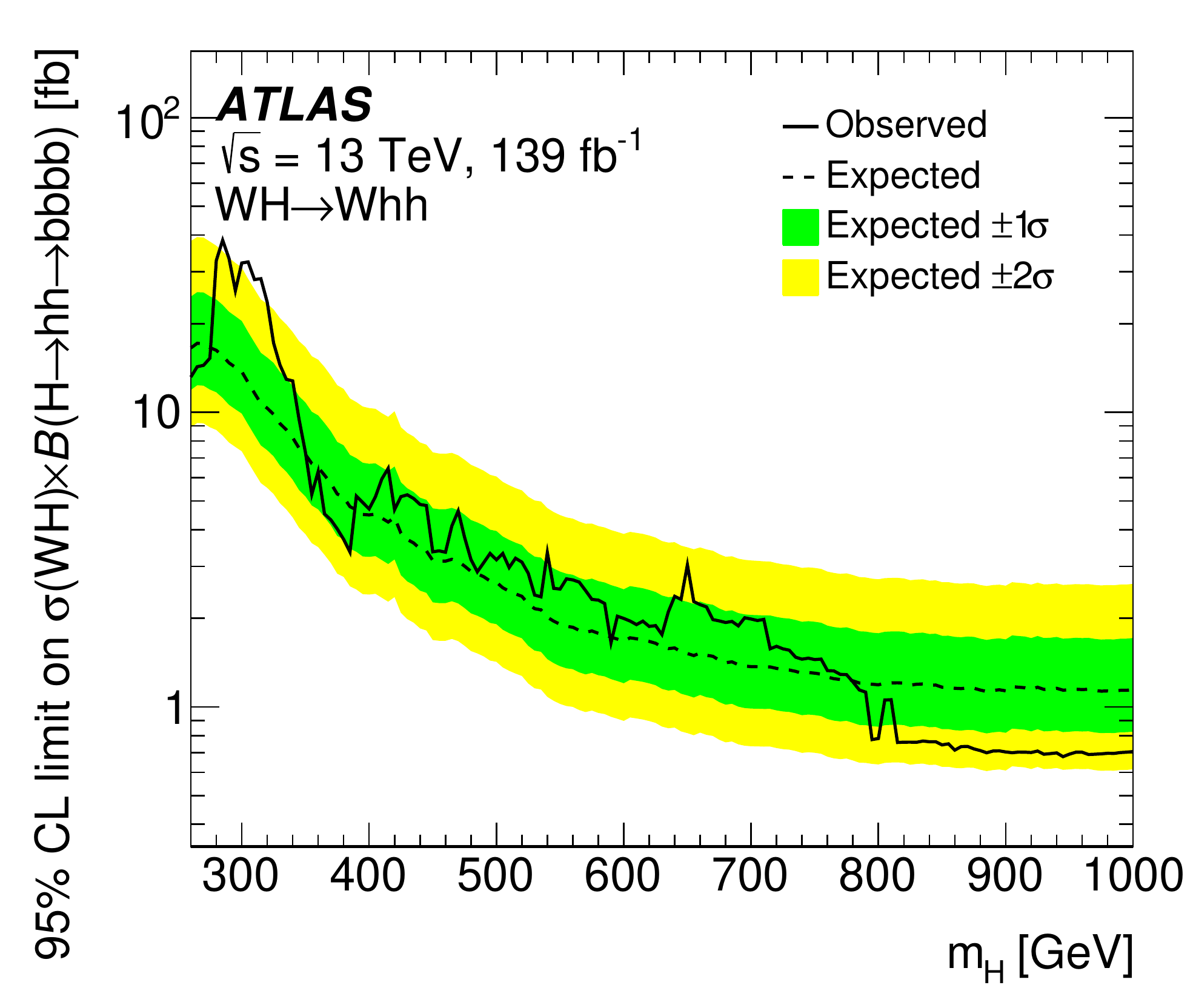}}
\subfigure[]{\includegraphics[width=0.48\textwidth]{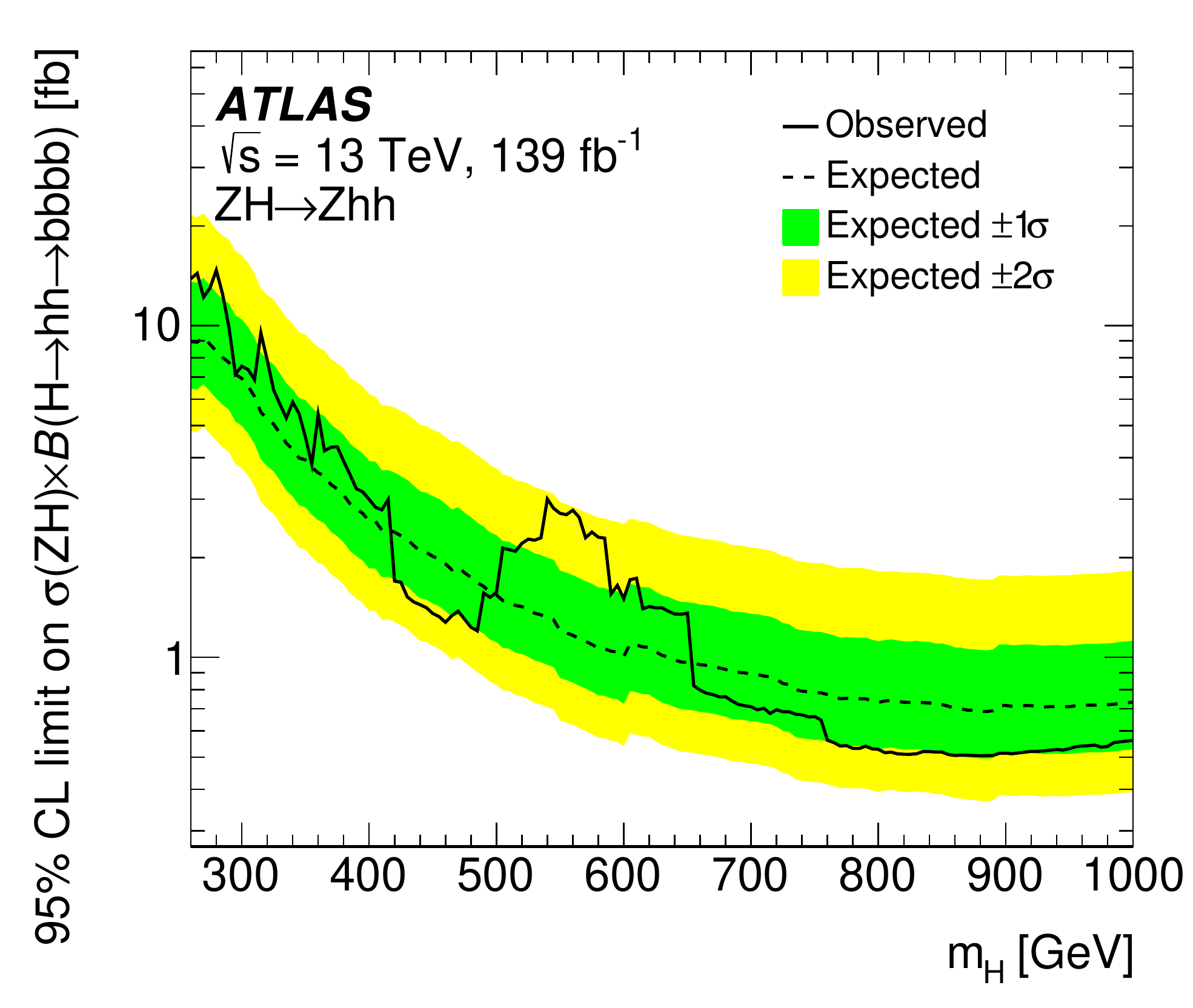}}
\end{center}
\caption{Observed (black solid curve) and expected (black dashed curve) 95\% CL upper limits on the production cross-section at $\sqrt{s}=13\,\TeV$ of a heavy narrow scalar resonance $H$ in the decay mode $H\to hh\to bbbb$ in association with (a) a $W$ boson and (b) a $Z$ boson as a function of the resonance mass. The green (inner) and yellow (outer) bands represent $\pm 1\sigma$ and $\pm 2\sigma$ uncertainty in the expected limits.}
\label{fig:VHLimits}
\end{figure}

\subsection{Search for \AZH production}
The \AZH search follows a strategy similar to that used in the \VH search. The \AZH BDT distributions in the 0L and 2L channels, after applying the mass requirements discussed in Section~\ref{sec:signal}, are used to constrain $gg\to A\to ZH\to Zhh\to Zbbbb$ production for each $(m_A,m_H)$ hypothesis with 10~\GeV~steps in both masses, separately for NW and LW $A$ bosons. The BDT distributions for $(m_A,m_H)$ hypotheses without MC samples are linearly interpolated from those of the four closest neighbouring mass points with  MC samples.
For a NW $A$ boson, the expected and observed upper limits in the $(m_A, m_H)$ plane are shown in Figure~\ref{fig:exclusion_azh_expected_nw} and Figure~\ref{fig:exclusion_azh_observed_nw} respectively. The upper limits for a LW $A$ boson, with a width of 20\%, are shown in Figure~\ref{fig:exclusion_azh_expected_lw} (expected) and Figure~\ref{fig:exclusion_azh_observed_lw} (observed). The ratios of the observed and the expected upper bounds in the $(m_A,m_H)$ plane are shown in Figure~\ref{fig:AZHLimitRatios} for the two width scenarios.

\begin{figure}[!htb]
\begin{center}
\subfigure[]{\includegraphics[width=0.47\textwidth]{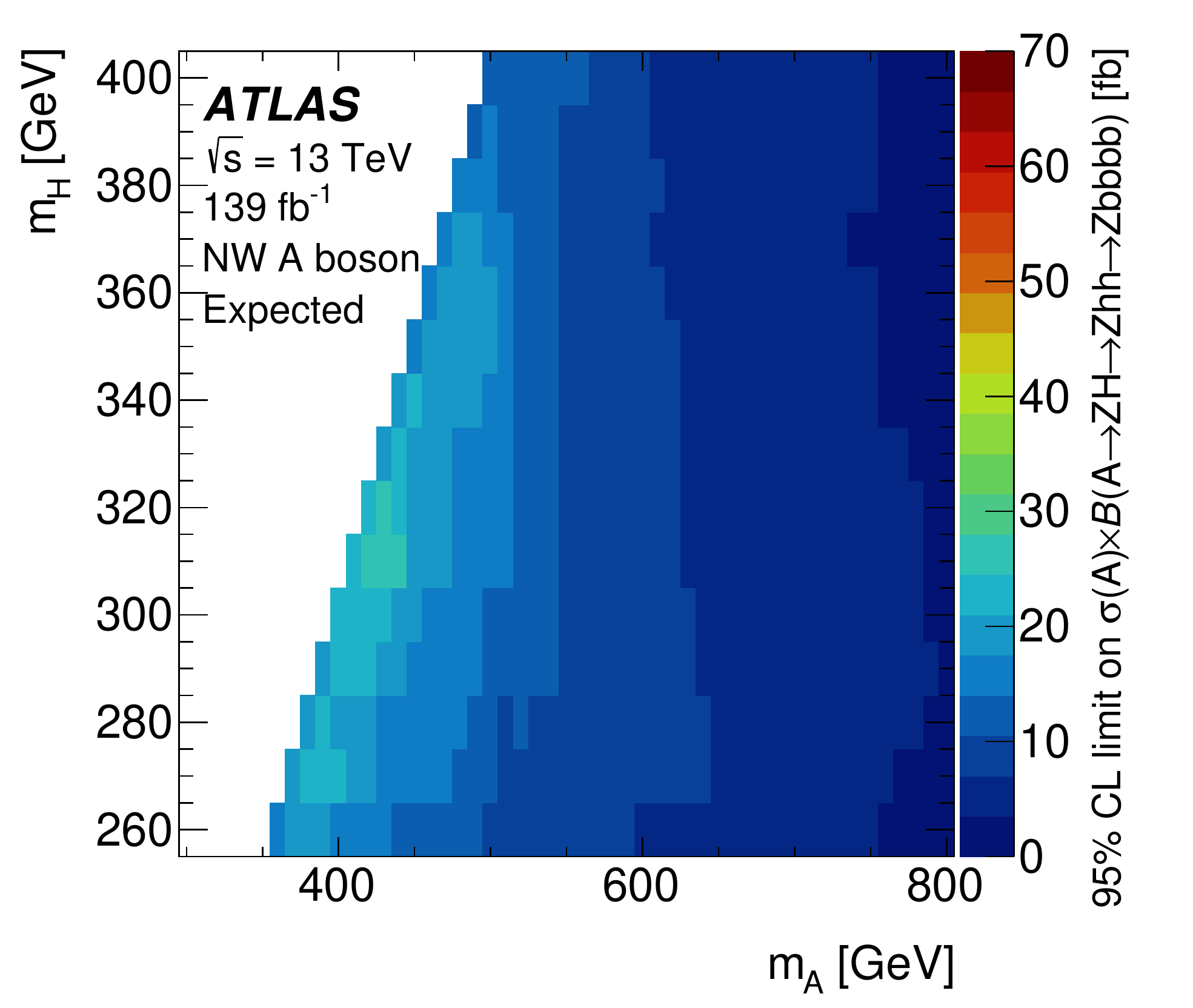}\label{fig:exclusion_azh_expected_nw}}
\subfigure[]{\includegraphics[width=0.47\textwidth]{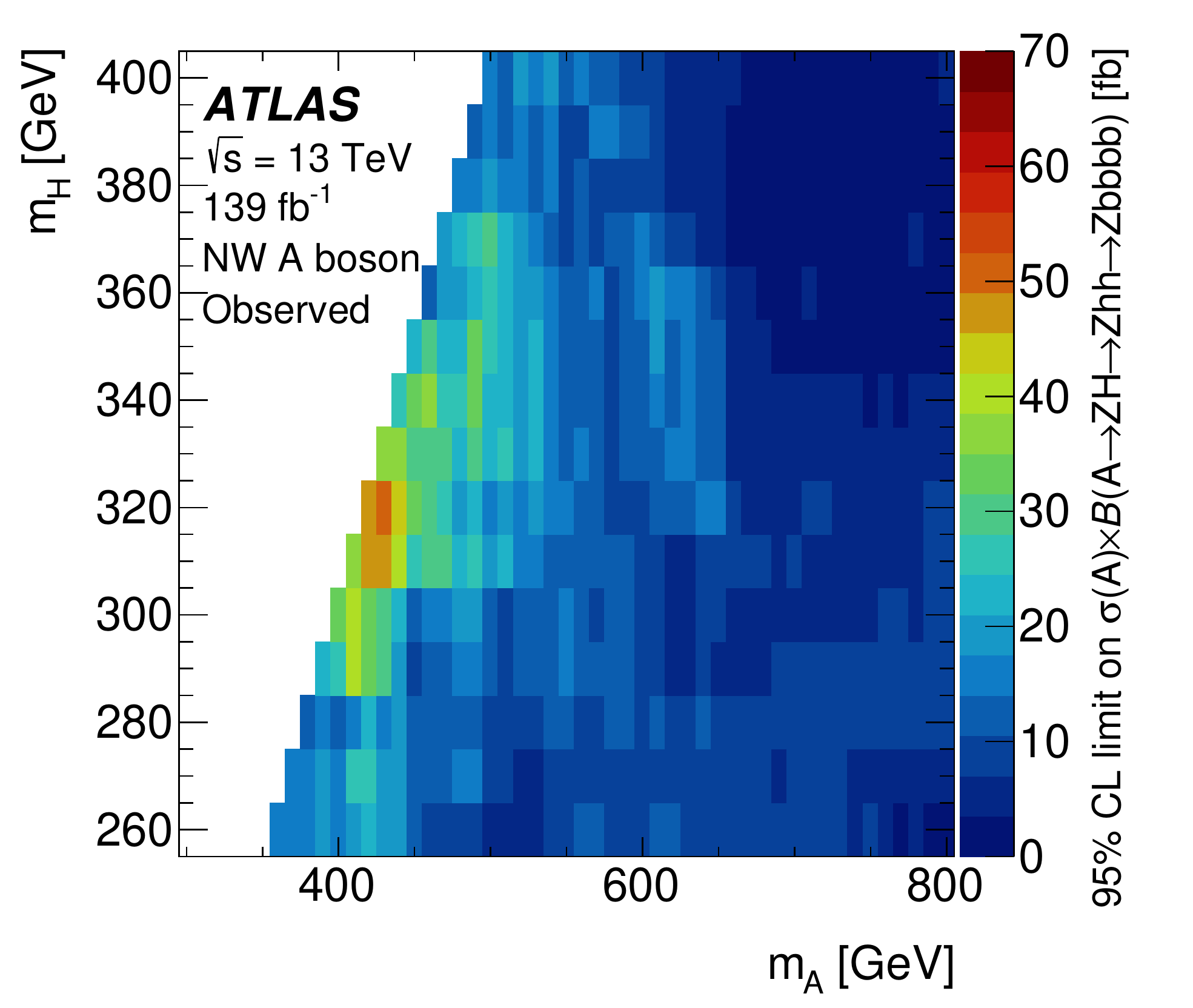}\label{fig:exclusion_azh_observed_nw}}\\
\subfigure[]{\includegraphics[width=0.47\textwidth]{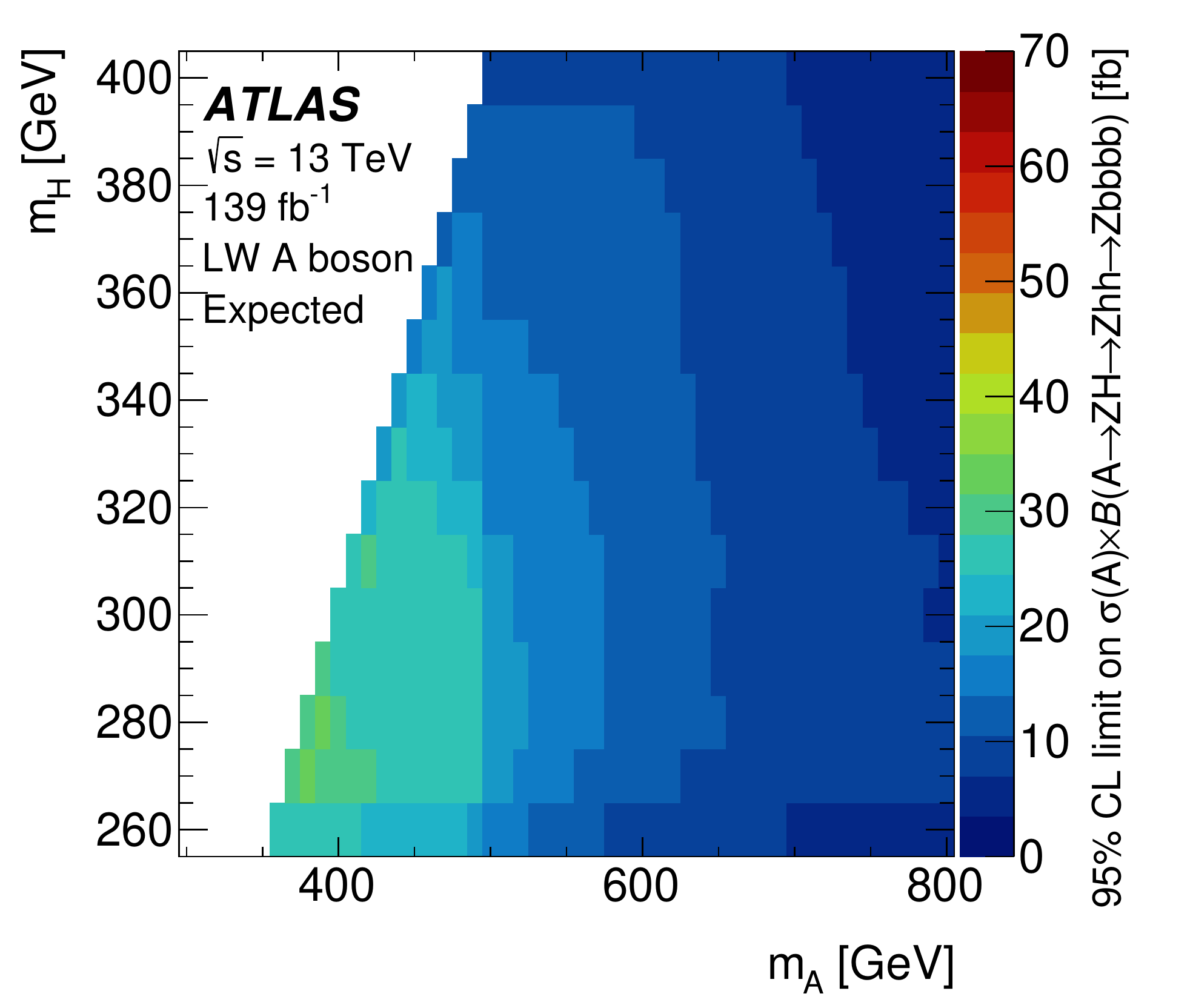}\label{fig:exclusion_azh_expected_lw}}
\subfigure[]{\includegraphics[width=0.47\textwidth]{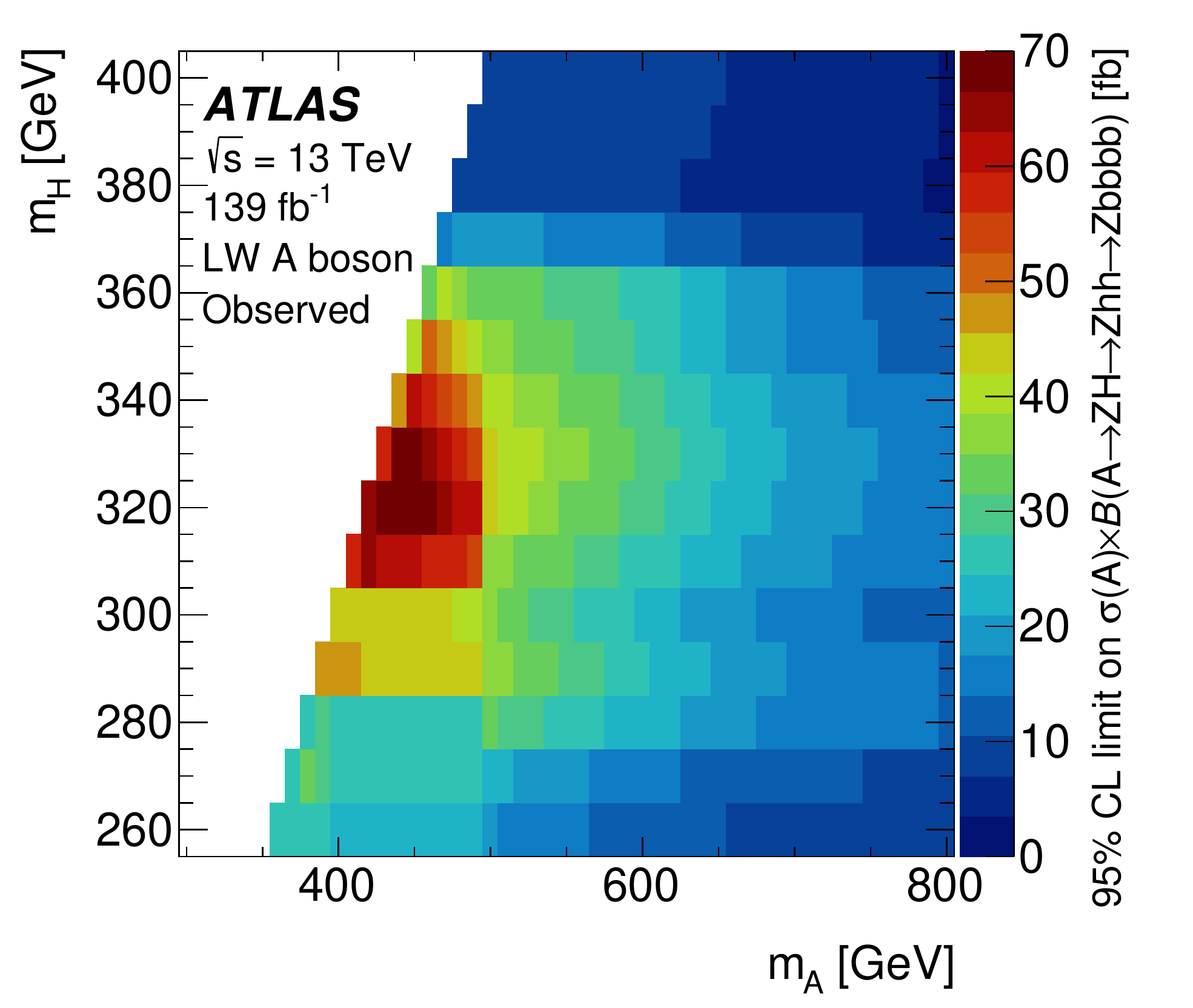}\label{fig:exclusion_azh_observed_lw}}
\end{center}
\caption{Upper bounds at 95\% CL on $\sigma(A)\times B(A\to ZH\to Zhh\to Zbbbb)$ in the $(m_A, m_H)$ plane for  (a,\,b) a NW $A$ boson and (c,\,d) a LW $A$ boson. The expected upper limits are shown in (a) and (c) while the observed limits are shown in (b) and (d). The $A$ boson has a total decay width that is negligible compared to the experimental mass resolution in the NW case and is 20\% of its mass in the LW case. }
\end{figure}

The most significant excesses are observed at $(m_A,m_H)=(790,300)\,\GeV$ with a local (global) significance of $\AZHNWSigLoc\sigma~(\AZHNWSigGlo\sigma)$ in the NW scenario and at $(m_A,m_H)=(420,320)\,\GeV$ with a local (global) significance of $\AZHLWSigLoc\sigma~(\AZHLWSigGlo\sigma)$ in the LW scenario. Figure~\ref{fig:AZHLWBDT} shows the post-fit BDT distributions in the LW scenario from the search at $(m_A,m_H)=(420,320)\,\GeV$.
Due to the selection for the LW scenario, there is little sensitivity to either the $A$ boson mass or width,
and a real LW signal would likely produce a higher-than-otherwise-expected cross-section upper limit over a wide range of probed $A$ boson mass values.
Thus the broad excess in the $A$ boson mass distribution in the LW scenario is consistent with the expectation for a signal.
 
\begin{figure}[!htb]
\begin{center}
\subfigure[]{\includegraphics[width=0.47\textwidth]{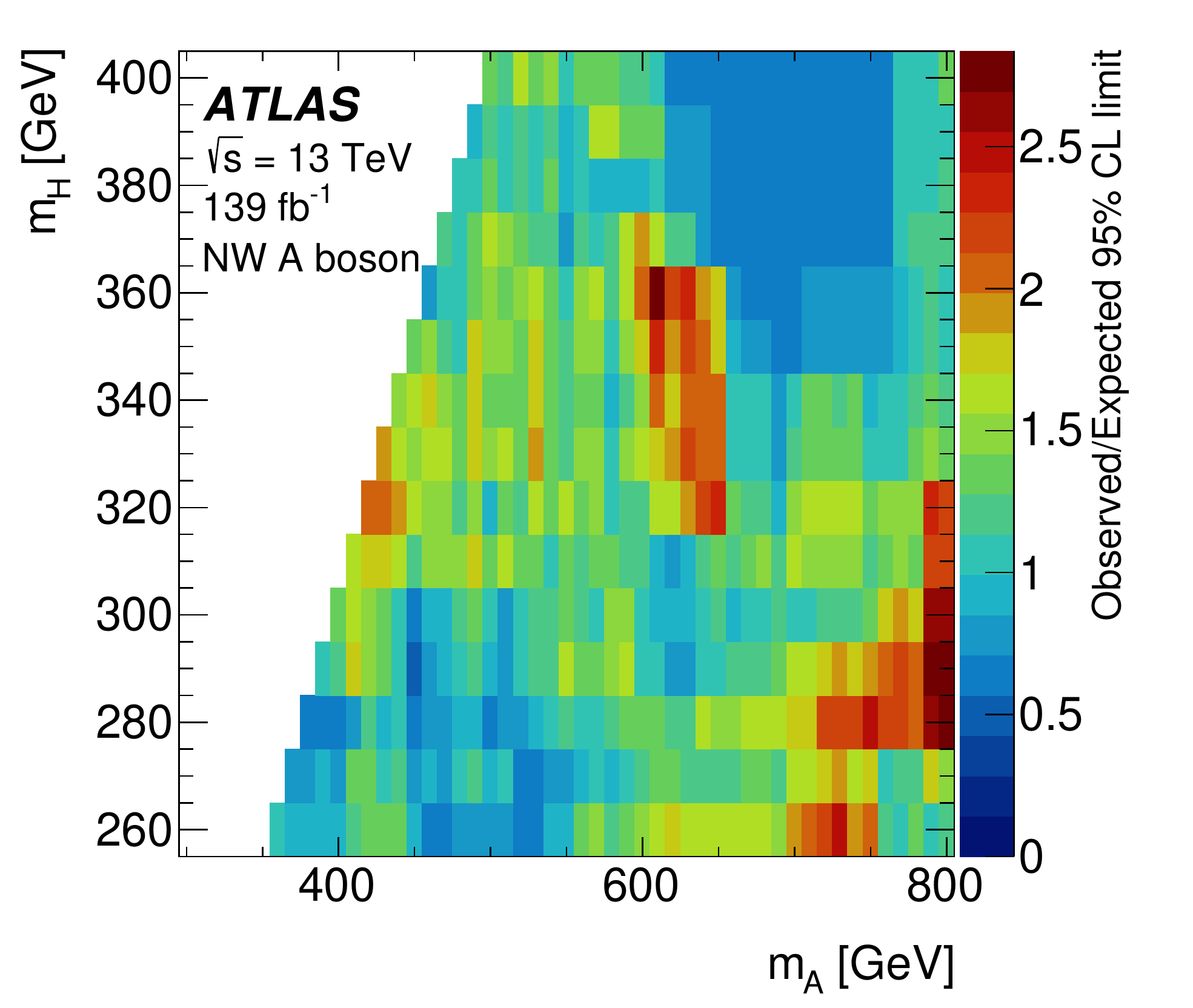}}
\subfigure[]{\includegraphics[width=0.47\textwidth]{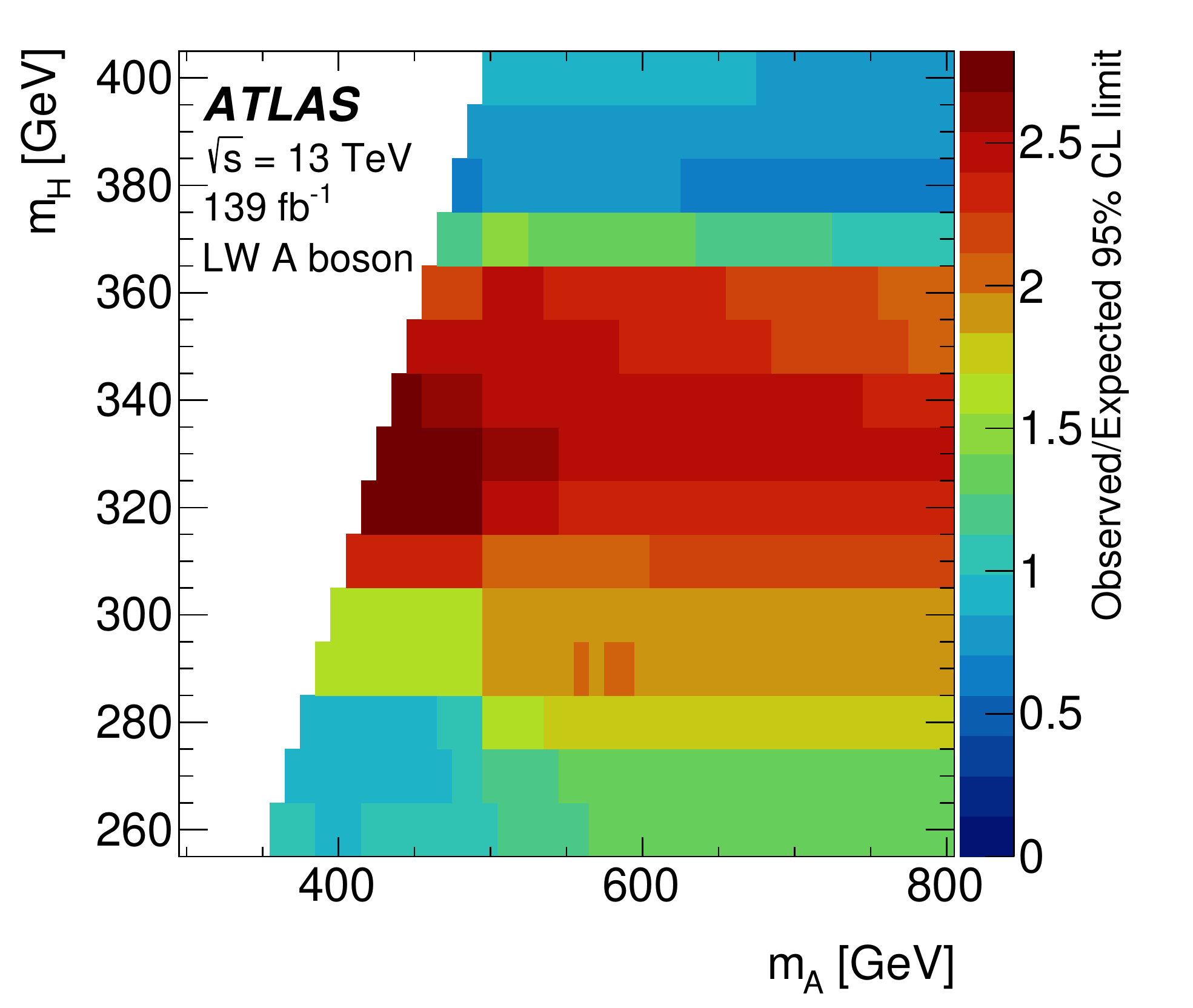}}
\end{center}
\caption{Ratios of the observed and the expected 95\% CL upper bounds on $\sigma(A)\times B(A\to ZH\to Zhh\to Zbbbb)$ in the $(m_A,m_H)$ plane of the \AZH search with a (a) NW and (b) LW $A$ boson. The $A$ boson has a total decay width that is negligible compared to the experimental mass resolution in the NW case and is 20\% of its mass in the LW case.}
\label{fig:AZHLimitRatios}
\end{figure}

The 2HDM benchmark used for interpretations has four free parameters: $m_A$, $m_H$, $\tan\beta$,
and $\cos(\beta-\alpha)$, where $\tan\beta$ is the ratio of the vacuum
expectation values of the two doublets and $\alpha$ is the mixing angle
of the CP-even Higgs bosons. The limiting case of $\cos(\beta-\alpha)\to 0$ corresponds to the
2HDM weak decoupling limit~\cite{Gunion:2002zf} in which
the lightest CP-even Higgs boson has the same couplings as the SM Higgs boson at the lowest order.
The \AZH search results are interpreted as constraints in the plane defined by
$\cos(\beta-\alpha)$ and $m_A$ for given $m_H$ and $\tan\beta$ values.
For the remaining 2HDM parameters, the mass of the charged Higgs boson
is set to be equal to $m_A$ and the potential parameter $m^2_{12}$ is set to
$m^2_{A}\tan\beta/(1+\tan^2\beta)$.

For regions relevant to the sensitivity of this search, the natural width of the $H$ boson remains narrow, especially where
$\cos(\beta-\alpha)$ is close to zero. It is estimated that the cross-section upper limits should be valid as long as the natural width of the $H$ boson is less than 1\% and, therefore, the search constrains only parts of the 2HDM parameter space that conform with this requirement. On the other hand, the natural
width of the $A$ boson varies from narrow to about 20\%.
This means that cross-section limits have to be calculated for a range of $A$ boson natural widths in the
$(m_A, m_H)$ plane, which are subsequently interpreted in the 2HDM planes
discussed previously. The signal hypothesis is tested for several values
of the $A$ boson natural width, and cross-section upper limits for those widths are derived as
a function of $m_A$. Linear interpolation is used to derive
the limit for any natural width between a pair of tested width values.

To interpret these results in the 2HDM, the upper limits
on the cross-section are compared with the theoretical predictions of the
model. In the type-I and lepton-specific 2HDMs, only gluon--gluon fusion production
of the $A$ boson, $gg\to A$, is relevant, and it is calculated with corrections
at up to NNLO in QCD as implemented in
{\texttt SusHi}~\cite{Harlander:2012pb,Harlander:2005rq,Harlander:2003ai,Harlander:2002wh}.
The  widths and branching
ratios of the Higgs bosons ($A$, $H$, and $h$) are calculated using the 2HDMC code~\cite{Eriksson:2009ws}. The procedure used to calculate the cross-sections
and branching ratios, as well as to choose the 2HDM parameter values, follows Ref.~\cite{deFlorian:2016spz}.
The upper limits are shown for some representative values in Figure~\ref{fig:2HDMLimits} for the type-I 2HDM and in Figure~\ref{fig:2HDMLimits-t4} for the lepton-specific 2HDM.
The $Hhh$ coupling vanishes at $\cos(\beta-\alpha)=0$, a feature which is reflected by the inability of this analysis to exclude this region of the $(\cos(\beta-\alpha), m_A)$ plane.
For $\tan\beta=1$ the sensitivity is the same for both 2HDM scenarios and therefore the corresponding results are omitted from Figure~\ref{fig:2HDMLimits-t4}. The sensitivity of this search is complementary to that of the $A\to ZH\to \ell\ell bb/\ell\ell WW$ search~\cite{HDBS-2018-13} and to the constraints from the Higgs boson coupling measurements~\cite{HIGG-2021-23}.

\begin{figure}[!htb]
\begin{center}
\subfigure[]{\includegraphics[width=0.48\textwidth]{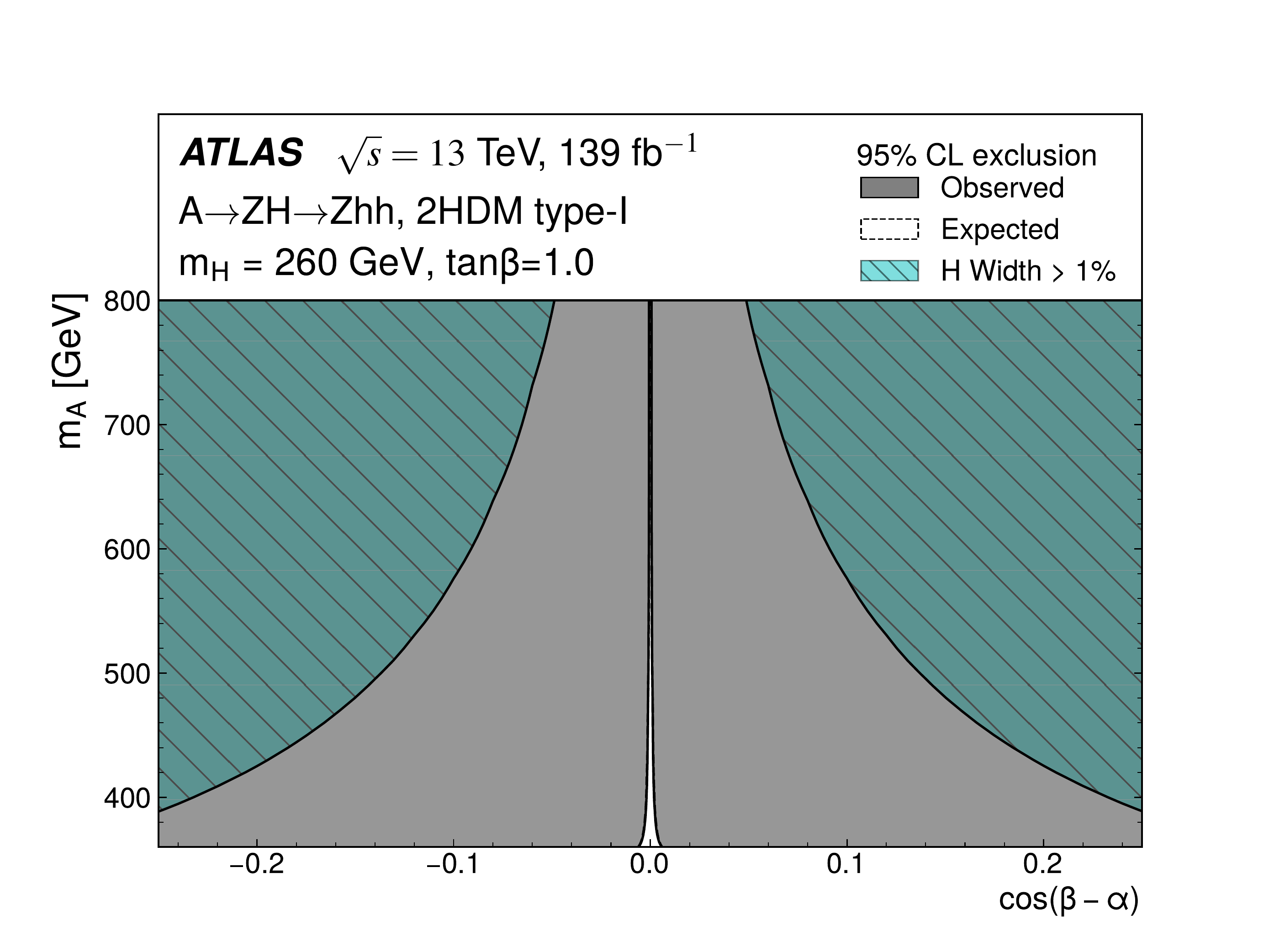}}
\subfigure[]{\includegraphics[width=0.48\textwidth]{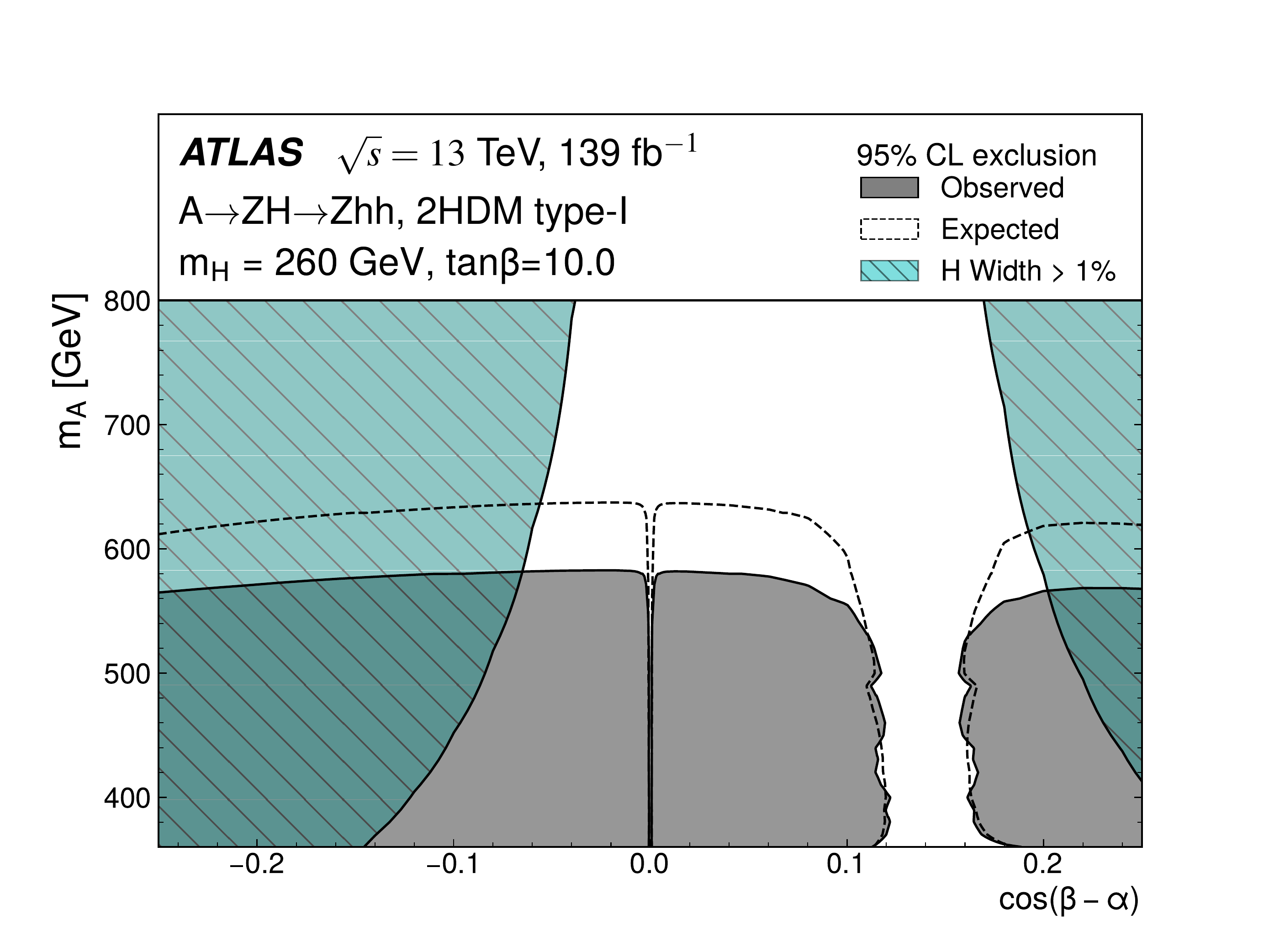}}\\
\subfigure[]{\includegraphics[width=0.48\textwidth]{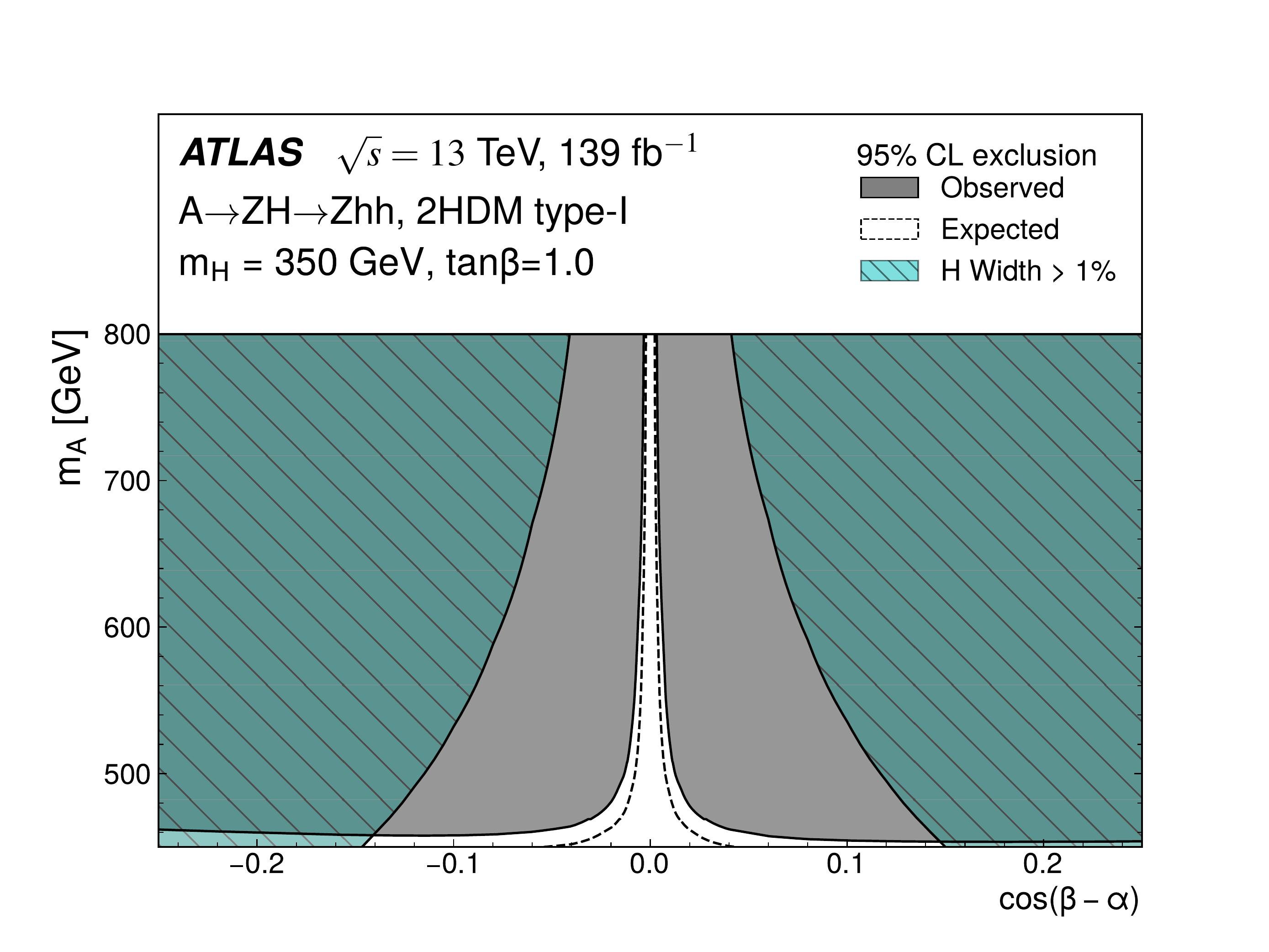}}
\subfigure[]{\includegraphics[width=0.48\textwidth]{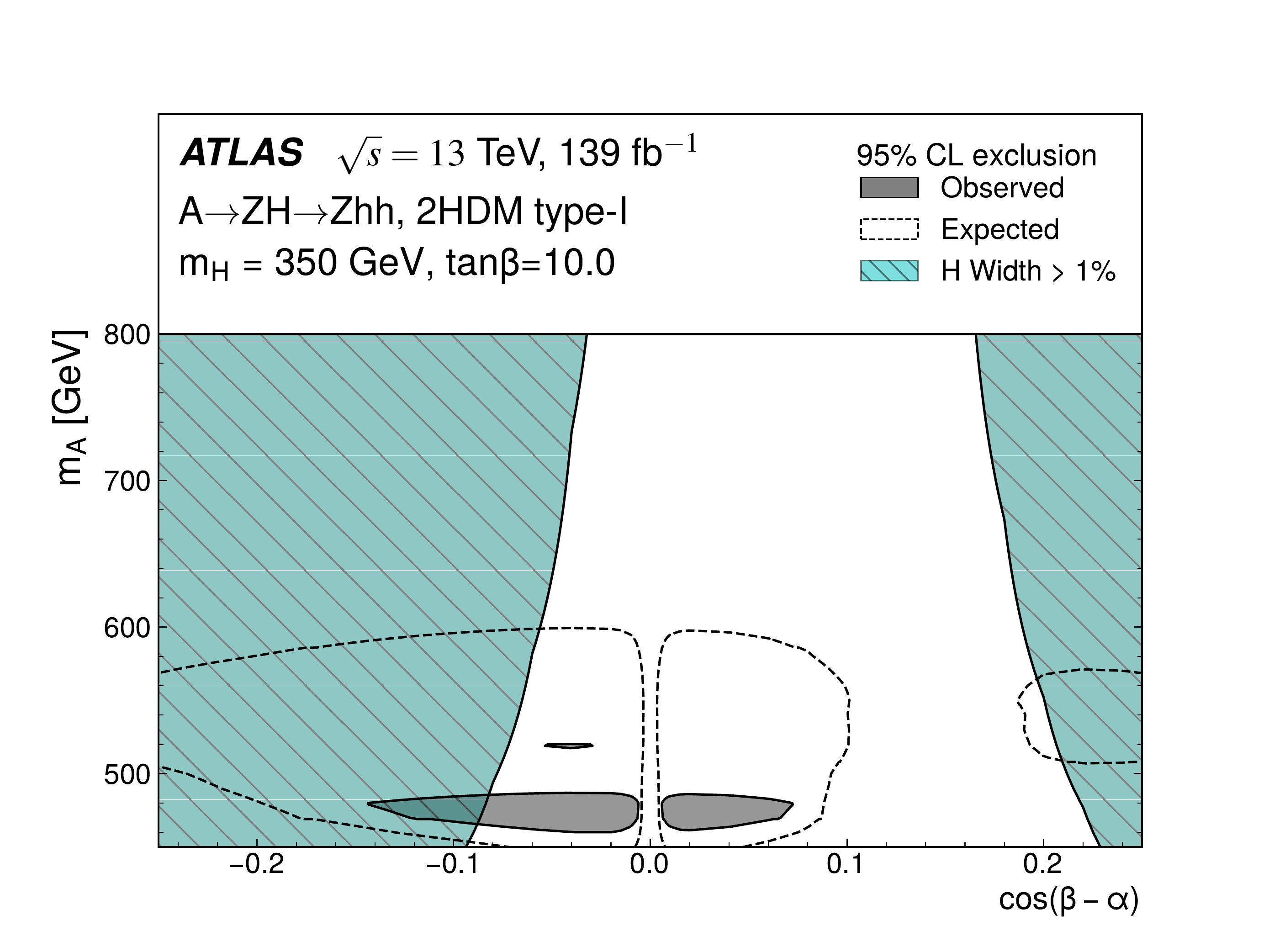}}
\end{center}
\caption{Interpretation of the upper limits on $\sigma(A)\times B(A\to ZH\to Zhh\to Zbbbb)$ in the parameter space of the type-I 2HDM. The shaded areas correspond to 95\% CL exclusion regions in the $(\cos(\beta-\alpha), m_A)$ plane for a given $m_H$ and $\tan\beta$. The hatched area corresponds to natural widths of the $H$ boson for which the upper limits are not valid. Panels (a) and (b) refer to $m_H=260$~\GeV\ and to $\tan\beta= 1$ and 10, respectively; (c) and (d) refer to $m_H=350$~\GeV\ and to $\tan\beta= 1$ and 10, respectively. }
\label{fig:2HDMLimits}
\end{figure}
 
\begin{figure}[!htb]
\begin{center}
\subfigure[]{\includegraphics[width=0.48\textwidth]{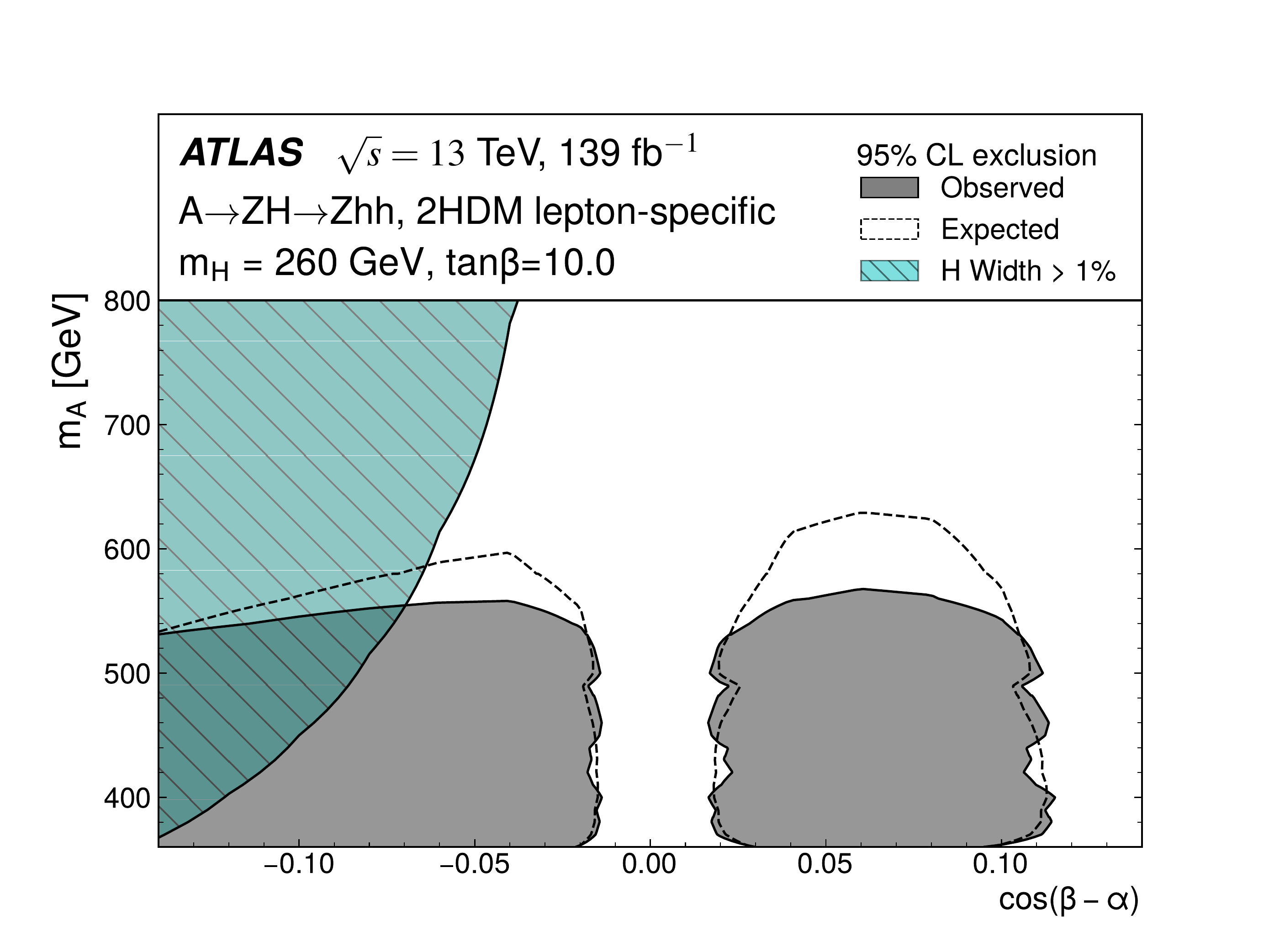}}
\subfigure[]{\includegraphics[width=0.48\textwidth]{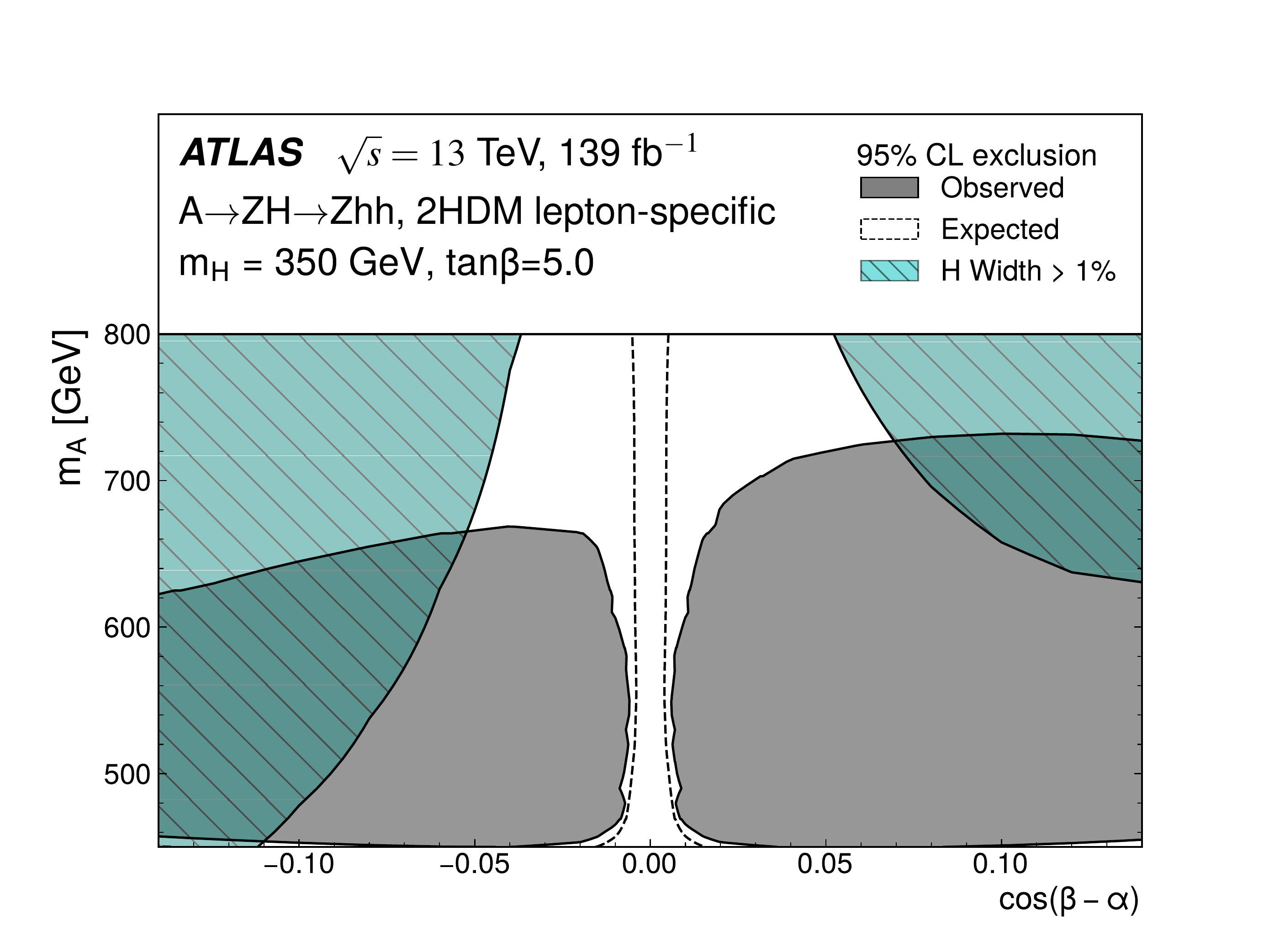}}
\end{center}
\caption{Interpretation of the upper limits on $\sigma(A)\times B(A\to ZH\to Zhh\to Zbbbb)$ in the parameter space of the lepton-specific 2HDM. The same notation
as in Figure~\ref{fig:2HDMLimits} is used.
Panel (a) refers to $m_H=260$~\GeV, $\tan\beta=10$ and (b)
refers to $m_H=350$~\GeV, $\tan\beta=5$. }
\label{fig:2HDMLimits-t4}
\end{figure}

\FloatBarrier

% End of text imported from the .//tex/results.tex input file

% The next lines are included from the .//tex/summary.tex input file
\section{Summary}
 
Searches for Higgs boson pair production in association with a vector boson in $pp$ collisions at $\sqrt{s}=13\,\TeV$ are performed using a data sample corresponding to an integrated luminosity of \lumi\ifb, recorded by the ATLAS experiment between 2015 and 2018 at the LHC. The Higgs bosons are identified via their decays into a pair of $b$-quarks and the vector bosons are required to decay into leptons, leading to final states with zero, one or two charged leptons along with four $b$-jets. The searches target both SM-inspired non-resonant $hh$ production and BSM-motivated resonant $hh$ production. The non-resonant $Vhh$ search is carried out for scenarios with either SM kinematics but an enhanced production cross-section or modified Higgs boson couplings to vector bosons or itself. The resonant $Vhh$ searches are designed for the production of a vector boson along with a heavy neutral scalar Higgs boson $H$ decaying into $hh$, either directly or indirectly from the decay of another heavier neutral pseudoscalar Higgs boson $A$.
 
In general, the data are found to be in good agreement with the estimated background contributions, except for a few notable excesses. The most significant global excess is observed in the $gg\to A\to ZH\to Zhh$  search for a large-width $A$ boson at $(m_A,m_H)=(420, 320)\,\GeV$, where the local (global) significance is $\AZHLWSigLoc\, (\AZHLWSigGlo)$ standard deviations. More data are needed to ascertain the nature of this excess.
Upper bounds on the $Vhh$ production cross-sections are derived. For non-resonant production with SM kinematics, a 95\% CL upper limit of \VhhObs\, (\VhhExp) is observed (expected) for the $Vhh$ cross-section relative to its SM prediction. 
For resonant production, the observed (expected) upper limits are presented as a function of $m_H$ in the range 260--1000\,\GeV\ for \WH and \ZH separately, and in the $(m_A,m_H)$ plane for \AZH, covering the $m_A$ range 360--800\,\GeV\ and $m_H$ range 260--400\,\GeV. The constraints on \AZH production are also interpreted in the $(\cos(\beta-\alpha), m_A)$ parameter space of type-I and lepton-specific two-Higgs-doublet models.

% End of text imported from the .//tex/summary.tex input file

\section*{Acknowledgements}

% The next lines are included from the .//acknowledgements/Acknowledgements.tex input file

We thank CERN for the very successful operation of the LHC, as well as the
support staff from our institutions without whom ATLAS could not be
operated efficiently.
 
We acknowledge the support of
ANPCyT, Argentina;
YerPhI, Armenia;
ARC, Australia;
BMWFW and FWF, Austria;
ANAS, Azerbaijan;
CNPq and FAPESP, Brazil;
NSERC, NRC and CFI, Canada;
CERN;
ANID, Chile;
CAS, MOST and NSFC, China;
Minciencias, Colombia;
MEYS CR, Czech Republic;
DNRF and DNSRC, Denmark;
IN2P3-CNRS and CEA-DRF/IRFU, France;
SRNSFG, Georgia;
BMBF, HGF and MPG, Germany;
GSRI, Greece;
RGC and Hong Kong SAR, China;
ISF and Benoziyo Center, Israel;
INFN, Italy;
MEXT and JSPS, Japan;
CNRST, Morocco;
NWO, Netherlands;
RCN, Norway;
MEiN, Poland;
FCT, Portugal;
MNE/IFA, Romania;
MESTD, Serbia;
MSSR, Slovakia;
ARRS and MIZ\v{S}, Slovenia;
DSI/NRF, South Africa;
MICINN, Spain;
SRC and Wallenberg Foundation, Sweden;
SERI, SNSF and Cantons of Bern and Geneva, Switzerland;
MOST, Taiwan;
TENMAK, T\"urkiye;
STFC, United Kingdom;
DOE and NSF, United States of America.
In addition, individual groups and members have received support from
BCKDF, CANARIE, Compute Canada and CRC, Canada;
PRIMUS 21/SCI/017 and UNCE SCI/013, Czech Republic;
COST, ERC, ERDF, Horizon 2020 and Marie Sk{\l}odowska-Curie Actions, European Union;
Investissements d'Avenir Labex, Investissements d'Avenir Idex and ANR, France;
DFG and AvH Foundation, Germany;
Herakleitos, Thales and Aristeia programmes co-financed by EU-ESF and the Greek NSRF, Greece;
BSF-NSF and MINERVA, Israel;
Norwegian Financial Mechanism 2014-2021, Norway;
NCN and NAWA, Poland;
La Caixa Banking Foundation, CERCA Programme Generalitat de Catalunya and PROMETEO and GenT Programmes Generalitat Valenciana, Spain;
G\"{o}ran Gustafssons Stiftelse, Sweden;
The Royal Society and Leverhulme Trust, United Kingdom.
 
The crucial computing support from all WLCG partners is acknowledged gratefully, in particular from CERN, the ATLAS Tier-1 facilities at TRIUMF (Canada), NDGF (Denmark, Norway, Sweden), CC-IN2P3 (France), KIT/GridKA (Germany), INFN-CNAF (Italy), NL-T1 (Netherlands), PIC (Spain), ASGC (Taiwan), RAL (UK) and BNL (USA), the Tier-2 facilities worldwide and large non-WLCG resource providers. Major contributors of computing resources are listed in Ref.~\cite{ATL-SOFT-PUB-2021-003}.

% End of text imported from the .//acknowledgements/Acknowledgements.tex input file

\iffalse
The \texttt{atlaslatex} package contains the acknowledgements that were valid
at the time of the release you are using.
These can be found in the \texttt{acknowledgements} subdirectory.
When your ATLAS paper or PUB/CONF note is ready to be published,
download the latest set of acknowledgements from:\\
\url{https://twiki.cern.ch/twiki/bin/view/AtlasProtected/PubComAcknowledgements}
\fi

\clearpage

\printbibliography

\clearpage
% ATLAS Collaboration author list
% Reference date of HDBS-2019-31 is 2022-06-02
% Author list last updated on date 21-APR-23
% Data extracted on 21-Apr-2023 for paper reference HDBS-2019-31
% at 2:22pm
 
\begin{flushleft}
\hypersetup{urlcolor=black}
{\Large The ATLAS Collaboration}

\bigskip

\AtlasOrcid[0000-0002-6665-4934]{G.~Aad}$^\textrm{\scriptsize 101}$,
\AtlasOrcid[0000-0002-5888-2734]{B.~Abbott}$^\textrm{\scriptsize 119}$,
\AtlasOrcid[0000-0002-7248-3203]{D.C.~Abbott}$^\textrm{\scriptsize 102}$,
\AtlasOrcid[0000-0002-1002-1652]{K.~Abeling}$^\textrm{\scriptsize 55}$,
\AtlasOrcid[0000-0002-8496-9294]{S.H.~Abidi}$^\textrm{\scriptsize 29}$,
\AtlasOrcid[0000-0002-9987-2292]{A.~Aboulhorma}$^\textrm{\scriptsize 35e}$,
\AtlasOrcid[0000-0001-5329-6640]{H.~Abramowicz}$^\textrm{\scriptsize 150}$,
\AtlasOrcid[0000-0002-1599-2896]{H.~Abreu}$^\textrm{\scriptsize 149}$,
\AtlasOrcid[0000-0003-0403-3697]{Y.~Abulaiti}$^\textrm{\scriptsize 116}$,
\AtlasOrcid[0000-0003-0762-7204]{A.C.~Abusleme~Hoffman}$^\textrm{\scriptsize 136a}$
% ...
\bigskip
\\

% ...
$^{29}$Physics Department, Brookhaven National Laboratory, Upton NY; United States of America.\\
% ...
$^{35}$$^{(a)}$Facult\'e des Sciences Ain Chock, R\'eseau Universitaire de Physique des Hautes Energies - Universit\'e Hassan II, Casablanca;$^{(b)}$Facult\'{e} des Sciences, Universit\'{e} Ibn-Tofail, K\'{e}nitra;$^{(c)}$Facult\'e des Sciences Semlalia, Universit\'e Cadi Ayyad, LPHEA-Marrakech;$^{(d)}$LPMR, Facult\'e des Sciences, Universit\'e Mohamed Premier, Oujda;$^{(e)}$Facult\'e des sciences, Universit\'e Mohammed V, Rabat;$^{(f)}$Institute of Applied Physics, Mohammed VI Polytechnic University, Ben Guerir; Morocco.\\
% ...
$^{55}$II. Physikalisches Institut, Georg-August-Universit\"{a}t G\"ottingen, G\"ottingen; Germany.\\
% ...
$^{101}$CPPM, Aix-Marseille Universit\'e, CNRS/IN2P3, Marseille; France.\\
$^{102}$Department of Physics, University of Massachusetts, Amherst MA; United States of America.\\
% ...
$^{116}$Department of Physics, New York University, New York NY; United States of America.\\
% ...
$^{119}$Homer L. Dodge Department of Physics and Astronomy, University of Oklahoma, Norman OK; United States of America.\\
% ...
$^{136}$$^{(a)}$Departamento de F\'isica, Pontificia Universidad Cat\'olica de Chile, Santiago;$^{(b)}$Millennium Institute for Subatomic physics at high energy frontier (SAPHIR), Santiago;$^{(c)}$Instituto de Investigaci\'on Multidisciplinario en Ciencia y Tecnolog\'ia, y Departamento de F\'isica, Universidad de La Serena;$^{(d)}$Universidad Andres Bello, Department of Physics, Santiago;$^{(e)}$Instituto de Alta Investigaci\'on, Universidad de Tarapac\'a, Arica;$^{(f)}$Departamento de F\'isica, Universidad T\'ecnica Federico Santa Mar\'ia, Valpara\'iso; Chile.\\
% ...
$^{149}$Department of Physics, Technion, Israel Institute of Technology, Haifa; Israel.\\
$^{150}$Raymond and Beverly Sackler School of Physics and Astronomy, Tel Aviv University, Tel Aviv; Israel.\\
% ...
$^{a}$ Also Affiliated with an institute covered by a cooperation agreement with CERN.\\
$^{b}$ Also at Borough of Manhattan Community College, City University of New York, New York NY; United States of America.\\
$^{c}$ Also at Bruno Kessler Foundation, Trento; Italy.\\
$^{d}$ Also at Center for High Energy Physics, Peking University; China.\\
$^{e}$ Also at Centro Studi e Ricerche Enrico Fermi; Italy.\\
$^{f}$ Also at CERN, Geneva; Switzerland.\\
% ...

\end{flushleft}

% Created with Glance <Atlas.Glance@cern.ch>

\end{document}